%% file: main_with_supp.tex
\documentclass[12pt]{article}

\usepackage[sectionbib]{chapterbib}
\usepackage[round,authoryear]{natbib}
\usepackage{authblk}
\input{preamble}

\usepackage{parskip}
\usepackage{ltablex}
\keepXColumns
\usepackage{caption}
\usepackage{fvextra}
\usepackage{enumitem}
\hypersetup{hidelinks}

\DefineVerbatimEnvironment{WrappedVerbatim}{Verbatim}{
  breaklines,
  breakanywhere
}

\makeatletter
\renewcommand\AB@affilsepx{ \protect\\ }
\makeatother

\title{Estimating heritability of survival traits using censored multiple variance component model}

\author[1,$\ast$]{Do Hyun Kim}
\author[1,$\ast$]{Hua Zhou}
\author[1]{Brendon Chau}
\author[1]{Aubrey Jensen}
\author[2]{Judong Shen}
\author[2]{Devan Mehrotra}
\author[1]{Gang Li}
\author[1,$\ast$]{Jin J. Zhou}
\affil[1]{Department of Biostatistics, UCLA, Los Angeles, CA, USA}
\affil[2]{Biostatistics and Research Decision Sciences, Merck \& Co., Inc., Rahway, NJ, USA}
\affil[$\ast$]{Corresponding authors: \href{mailto:dkim116@ucla.edu}{dkim116@ucla.edu}; \href{mailto:huazhou@ucla.edu}{huazhou@ucla.edu}; \href{mailto:jinjinzhou@ucla.edu}{jinjinzhou@ucla.edu}}
\date{}

\begin{document}
\maketitle

\baselineskip=20.5pt

\begin{abstract}
Characterizing the genetic basis of survival traits, such as age at disease onset, is critical for risk stratification, early intervention, and elucidating biological mechanisms that can inform therapeutic development. However, time‐to‐event outcomes in human cohorts are frequently right‐censored, complicating both the estimation and partitioning of total heritability. Modern biobanks linked to electronic health records offer the unprecedented power to dissect the genetic basis of age-at-diagnosis traits at large scale. Yet, few methods exist for estimating and partitioning the total heritability of censored survival traits. Existing methods impose restrictive distributional assumptions on genetic and environmental effects and are not scalable to large biobanks with a million subjects. We introduce a censored multiple variance component model to robustly estimate the total heritability of survival traits under right-censoring. We demonstrate through extensive simulations that the method provides accurate total heritability estimates of right-censored traits at censoring rates up to $80\%$ given sufficient sample size. The method is computationally efficient in estimating one hundred genetic variance components of a survival trait using large-scale biobank genotype data consisting of a million subjects and a million SNPs in nine hours, including uncertainty quantification. We apply our method to estimate the total heritability of four age-at-diagnosis traits from the UK Biobank study. Our results establish a scalable and robust framework for heritability analysis of right-censored survival traits in large-scale genetic studies.

\end{abstract}

\include{main_content}

\appendix
\setkeys{Gin}{height=0.68\textheight,keepaspectratio}
\captionsetup{font=small}
\include{supp_content}

\end{document}

%% file: preamble.tex
\usepackage{amsbsy, amstext, amssymb, amsthm, amsmath,bm,wasysym}
\usepackage{dsfont}
\usepackage{algorithmic,graphicx,booktabs,rotating}
\usepackage[letterpaper, margin=1in]{geometry}
\usepackage{indentfirst}
\usepackage{listings}
\usepackage[lined,boxed,linesnumbered]{algorithm2e}
\usepackage{array,stmaryrd}
\usepackage{tabularx}
\usepackage{pdflscape}
\usepackage{adjustbox}
\usepackage{verbatimbox}
\usepackage{subfig}
\usepackage{float}
\usepackage{multirow}

\usepackage{color}

\usepackage{blkarray}
\usepackage{hyperref}

\newcommand{\inv}{{^{-1}}}

\newcommand{\bmat}[1]{\begin{bmatrix}#1\end{bmatrix}}

\newcommand{\E}{\mathop{\mathbb{E}}} 
\newcommand{\V}{\operatorname{\mathbb{V}ar}} 

\newcommand{\tr}{\operatorname{tr}}
\newcommand{\indep}{\;\, \rule[0em]{.03em}{.6em} \hspace{-.25em}
\rule[0em]{.65em}{.03em} \hspace{-.25em}
\rule[0em]{.03em}{.6em}\;\,}

\newcommand{\I}{\mathds{1}}

\newcommand{\diag}{\,\operatorname{diag}}

\newcommand{\gbar}{\bar{g}}

\newcommand{\ybar}{\bar{y}}

\newcommand{\bHbar}{\bar{{\bm H}}}

\newcommand{\ystar}{y^{*}}

\newcommand{\bystar}{\bm{y}^{*}}
\newcommand{\bYstar}{\bm{Y}^{*}}

\newcommand{\muhat}{\hat{\mu}}

\newcommand{\sigmahat}{\hat{\sigma}}

\newcommand{\hhat}{\hat{h}}

\newcommand{\Ghat}{\hat{G}}

\newcommand{\bbeta}{\bm{\beta}}

\newcommand{\bsigmahat}{\hat{\bm{\sigma}}}

\newcommand{\Dcal}{\mathcal D}

\newcommand{\Ncal}{\mathcal N}

\newcommand{\Rcal}{\mathcal R}

\newcommand{\bA}{{\bm A}}

\newcommand{\bD}{{\bm D}}

\newcommand{\bF}{{\bm F}}
\newcommand{\bG}{{\bm G}}
\newcommand{\bH}{{\bm H}}
\newcommand{\bI}{{\bm I}}

\newcommand{\bL}{{\bm L}}
\newcommand{\bK}{{\bm K}}

\newcommand{\bP}{{\bm P}}
\newcommand{\bQ}{{\bm Q}}
\newcommand{\bR}{{\bm{R}}}

\newcommand{\bT}{{\bm T}}
\newcommand{\bU}{{\bm U}}
\newcommand{\bV}{{\bm V}}
\newcommand{\bW}{{\bm W}}
\newcommand{\bX}{{\bm{X}}}

\newcommand{\bZ}{{\bm Z}}

\newcommand{\bb}{{\bm b}}
\newcommand{\bc}{{\bm c}}

\newcommand{\bg}{{\bm g}}

\newcommand{\br}{{\bm r}}

\newcommand{\bw}{{\bm w}}
\newcommand{\bx}{{\bm{x}}}
\newcommand{\by}{{\bm y}}
\newcommand{\bz}{{\bm z}}
\newcommand{\balpha}{\bm{\alpha}}

\newcommand{\bepsilon}{\bm{\varepsilon}}

\newcommand{\bmu}{\bm{\mu}}

\newcommand{\bsigma}{\bm{\sigma}}

\newcommand{\bSigma}{\bm{\Sigma}}

\newcommand{\bztilde}{\Tilde{\bm{z}}}
\newcommand{\bAtilde}{\tilde{{\bm A}}}

\newcommand{\bFtilde}{\tilde{{\bm F}}}

\newcommand{\bHtilde}{\tilde{{\bm H}}}

\newcommand{\bLtilde}{\tilde{{\bm L}}}

\newcommand{\bQtilde}{\tilde{{\bm Q}}}
\newcommand{\bRtilde}{\tilde{{\bm R}}}

\newcommand{\bUtilde}{\tilde{{\bm U}}}
\newcommand{\bVtilde}{\tilde{{\bm V}}}

\newcommand{\bZtilde}{\tilde{{\bm Z}}}

\newcommand{\bzero}{{\mathbf 0}}
\newcommand{\bone}{{\mathbf 1}}

\newcommand{\iter}[1]{^{(#1)}}

\newcommand\norm[1]{\left\lVert#1\right\rVert}

\newcommand{\tp}{{^\mathrm{T}}}

%% file: main_content.tex
\newpage
\section{Introduction}
Modern biobanks, such as the UK Biobank, the Million Veteran Program, and the National Institutes of Health (NIH) All of Us Research Program~\citep{sudlow_uk_2015, gaziano_million_2016, sankar_precision_2017}, provide extensive longitudinal and genomic data. These resources enable large-scale studies of the genetic underpinnings of time-to-event outcomes, including chronic diseases, cancer, and neurodegenerative diseases.

These data facilitate genome-wide association studies (GWASs) that identify single nucleotide polymorphisms (SNPs) or genes associated with survival traits via single-variant~\citep{dey_efficient_2022, bi_fast_2020} or multi-marker analyses~\citep{choi_variancecomponents_2024, sun_inference_2023, wu_multimarker_2021, chen_sequence_2014}. Indeed, recent GWASs have implicated  genetic variants in survival traits such as type 2 diabetes, Alzheimer's disease, heart disease, and fractures~\citep{dey_efficient_2022, bi_fast_2020, choi_variancecomponents_2024, sun_inference_2023}. An implicit but essential assumption in GWAS is that the disease has a genetic basis, i.e., heritability. Heritability is defined as the proportion of phenotypic variation in a population attributable to genetic variation~\citep{falconer_introduction_2009, visscher_heritability_2008}, and accurate estimation provides valuable insight into the polygenicity of complex traits.

In animal breeding, heritability estimation for survival traits has been used to select and breed cows for extended milk production~\citep{ducrocq1996bayesian, korsgaard1998bayesian, yazdi2002heritability}. These studies used Bayesian parametric hazard models with frailty terms whose variance components were estimated via Markov Chain Monte Carlo (MCMC) methods. While effective in small-scale settings, these approaches are unsuitable for modern human biobank data due to the simplicity of the models with a single genetic component and computational bottlenecks arising from MCMC-based estimation.

In human genetics, methods for investigating the role of genetic and environmental effects on censored survival traits have been developed for pedigree data. One class of methods adopted Bayesian parametric survival models with normally distributed frailty terms~\citep{pitkaniemi_genetic_2007, locatelli_correlated_2007}. These methods not only estimated genetic and family-related environmental variance components but also estimated the cure fraction of the cohort ``immune" to developing the disease. These methods are limited by restrictive distributional assumptions and resource intensive MCMC computation for obtaining posterior parameter estimates. Another class relied on martingale residuals from Cox proportional hazards (Cox PH) models~\citep{yoo_practical_2001, wintrebert_assessing_2006}, but required normality assumptions on residuals or frailty terms. Additional approaches included random effects Cox PH model~\citep{pankratz_randomeffects_2005} and Tobit variance component model~\citep{epstein_tobit_2003}, both involving computationally intensive numerical integration. Overall, these methods are ill-suited for biobank-scale human data: they are designed for family data, computationally expensive, and unable to model multiple genetic variance components that capture different genomic characteristics.

More recently, efforts have been made to estimate the heritability of survival traits using population-level human genetic data~\citep{gorfine_heritability_2017, ojavee_genomic_2021}. Heritability Estimation using a Regularized Regression Approach (HERRA)~\citep{gorfine_heritability_2017} used an accelerated failure time (AFT) model as the base model and selected a subset of SNPs through multiple rounds of variable selection. An inverse probability censoring weighting approach was used to account for right-censoring. However, because HERRA does not estimate multiple genetic variance components, it is less robust to different genetic architectures, potentially introducing biases in heritability estimates~\citep{haplotype_reference_consortium_comparison_2018}.

Another method, BayesW~\citep{ojavee_genomic_2021}, also employed an AFT model but assumed a Weibull distribution for survival traits. It modeled genetic effects using a latent mixture of Normal distributions, allowing for multiple genetic variance components and improved robustness across different underlying genetic architectures. A scalable MCMC technique, leveraging message-passing interfaces (MPIs), was developed to obtain posterior distributions of heritability estimates. While the latent mixture assumption on genetic effects enhanced robustness, the Weibull distributional assumption for survival traits restricts the model’s flexibility. Additionally, despite its scalability, BayesW remains computationally intensive, requiring several days to analyze biobank-scale data.

In this paper, we introduce Censored Variance Component (CVC) model for estimating the narrow-sense heritability $h^2$~\citep{visscher_heritability_2008}. Our method extends the scalable method-of-moments approach leveraging randomized trace estimator for analyzing and estimating heritability with biobank-scale genetic  data~\citep{pazokitoroudi_efficient_2020}. The CVC model accommodates right-censored survival traits, is able to estimate partitioned heritability and robust to model misspecification, and scales efficiently to biobank-scale data. A key feature of our approach is the use of synthetic variables~\citep{KoulSusarlaVvanRyzin81RightCensor, Leurgans87RandomCensoring, Zheng87PseudoVariableCensor}, whose moments match the corresponding moments of the uncensored survival trait, enabling unbiased variance component estimation via least squares.

Through simulations on both synthetic and real UK Biobank genotypes, we demonstrate that our method provides accurate heritability estimates even under model misspecification induced by diverse genetic architectures. We also compare the performance of the CVC model to BayesW and liability-threshold (LT) models~\citep{lee_estimating_2011} and show superior performance of the CVC model in estimation accuracy and uncertainty quantification of total and partitioned heritability. We further apply our method to analyze age-at-diagnosis survival traits in UKB participants, and provide total heritability as well as partitioned heritability estimates of those traits.

The remainder of the paper is structured as follows. In Section 2, we introduce the CVC model, first- and second-moment synthetic variables, and discuss randomized trace estimator and jackknife standard error estimator. In Section 3, we focus on the robustness of the CVC model under model misspecification through extensive simulation studies. Results under the correctly specified model, which serve as a baseline validation of the proposed method, are presented in Section~\ref{sec:correct_spec} of Supplementary Material. In Section 4, we compare the statistical performance of the CVC model with that of the BayesW and LT models. In Section 5, we benchmark the running times of the CVC model against those of the BayesW model. In Section 6, we present the total and partitioned heritability estimates of four age-at-diagnosis traits from the UKB study. In Section 7, we provide concluding remarks, limitations and future directions. Additional technical details and simulation results are included in Supplementary Material.

\section{Censored variance component model}
We present the CVC model, which leverages synthetic variables to analyze right-censored survival traits. First, we introduce a multi-component linear random-effects model augmented by synthetic variables, enabling estimation of multiple variance components using a least-squares optimization. Next, we extend this model to a multi-component linear mixed-effects framework, incorporating fixed effects arising from confounders. We also briefly discuss a randomized trace estimator, which significantly improves computational efficiency when fitting the model to biobank-scale genotype data involving up to one million subjects and one million SNPs. Lastly, we describe the column-wise block jackknife method for estimating the standard errors of the heritability estimates.

\subsection{Censored multi-component linear random-effects model with synthetic variables and total heritability}
Suppose $\by$ is an $N$-vector of phenotypes and $\bX$ is an $N \times M$ standardized genotype matrix. We relate $\by$ and $\bX$ through the following linear random-effects model:
\begin{equation}\label{eqn:ranef_model}
\begin{aligned}
    \by &= \sum_k \bX_k\bbeta_k + \bepsilon,\\
    \bepsilon &\sim \Dcal_e(\bzero,\sigma_e^2\bI_N),\\
    \bbeta_k &\sim \Dcal_k\left(\bzero,\frac{\sigma_k^2}{M_k}\bI_{M_k}\right),\ k \in \{1,...,K\},
\end{aligned}
\end{equation} where $\bX_k$ is an $N \times M_k$ standardized genotype matrix for the partition $k$ with $\sum_k M_k = M.$ $\bbeta_k$ is a random genetic effect associated with the $k$th genomic partition, and $\Dcal(\bmu, \bSigma)$ refers to an arbitrary distribution parameterized by mean $\bmu$ and variance $\bSigma$, i.e., we do not assume a specific parametric family beyond these moments. $\sigma^2_k$ is the genetic variance component for the $k$th genomic partition and $\sigma^2_e$ is the environmental variance component. We assume $\bbeta_1,...,\bbeta_K, \bepsilon$ are independent. 

The total heritability is defined as
\begin{equation}
    h^2_{\mathrm{SNP}} = \frac{\sum_{k=1}^K \sigma_k^2}{\sum_{k=1}^K \sigma_k^2 + \sigma^2_e}
\end{equation} and the partitioned heritability for the partition $k$ is defined as
\begin{equation}
    h^2_k = \frac{\sigma_k^2}{\sum_{k=1}^K \sigma_k^2 + \sigma^2_e},\quad k \in \{1,...,K\}.
\end{equation}

Let $\bK_k = \bX_k\bX_k\tp / M_k$ be the genetic relationship matrix (GRM) associated with the $k$th genomic partition. Then, 
\begin{equation}
    \E\by\by\tp = \sum_{k=1}^K \sigma^2_k\bK_k + \sigma^2_e\bI_N.
\end{equation} 
The method-of-moments estimates of variance components is obtained by minimizing the least squares criterion:
\begin{equation}\label{eqn:uncensored_ranef_ls}
 (\bsigmahat^2_g\tp,\sigmahat^2_e) = \arg \min_{\bsigma^2_g,\sigma^2_e}   ||\by\by\tp - (\sum_k \sigma^2_k \bK_k + \sigma^2_e\bI)||_F^2,
\end{equation} 
where $\bsigma_g^2 = (\sigma^2_1,...,\sigma^2_K)\tp$ and $\norm{\cdot}_F$ denotes the Frobenius norm. This is equivalent to solving the following normal equations
\begin{equation}\label{eqn:uncensored_ranef_normaleqn}
     \bmat{\bT & \bb \\ \bb\tp & N}\bmat{\bsigma^2_g \\ \sigma^2_e} = \bmat{\bc \\ \by\tp\by},
\end{equation} where 
$T_{k\ell} = \tr\bK_k\bK_\ell$, $b_k = \tr\bK_k = N$, and $c_k = \by\tp\bK_k\by$ for $k, \ell = 1,2,...,K.$

Now, in the context of survival traits, we assume $y_i = \log t_i$, where $t_i$ denotes the observed time-to-event outcome. Heritability estimates on the log scale are first-order approximations to those on the original scale (Section~\ref{sec:h2_obs_scale} of Supplementary Material). Since $y_i$ are subject to right censoring, what we observe are right-censored traits $u_i= \min(y_i,r_i)$ with their censoring indicators $\delta_i=\I_{\{y_i \leq r_i\}}$, where $r_i$ are censoring variables on a log scale, which is assumed to be independent of one another and $y_i$. This motivates us to introduce synthetic variables by \cite{Zheng87PseudoVariableCensor} who proposed a transformation of the form
$$
    \ystar_{i} = \delta_i\varphi_1(u_i) + (1-\delta_i)\varphi_2(u_i)
$$ such that $\E(\ystar_i) = \E(y_i).$ 
Here we employ the following first moment synthetic variable:
\begin{equation}
    \ystar_{i1} = \ystar_{i1}(G) = u_i + \int_{-\infty}^{u_i} \frac{G(t)}{1 - G(t)} dt,
\end{equation} where $G(t)$ is the cumulative distribution function (CDF) of $r_i$. This form of the synthetic variable was first introduced by \cite{Leurgans87RandomCensoring}. We further generalize it to the second moment synthetic variable
\begin{equation}
    \ystar_{i2} = \ystar_{i2}(G) = u_i^2 + \int_{-\infty}^{u_i} \frac{2tG(t)}{1 - G(t)}dt,
\end{equation} which satisfies $\E(\ystar_{i2}) = \E( y_i^2)$. Define the synthetic variable matrix $\bYstar$ with entries
\begin{equation}
    \bYstar_{ij} = \bYstar_{ij}(G) = \begin{cases}
        \ystar_{i2} & i = j,\\
        \ystar_{i1}\ystar_{j1} & i \neq j.
    \end{cases}
\end{equation} Then, 
\begin{equation}
    \E(\bYstar) = \E\by\by\tp
\end{equation} since $\E(\ystar_{i2}) = \E(y_i^2)$ for all $i$, and $\E(\ystar_{i1}\ystar_{j1}) = \E(y_iy_j)$ for $i \neq j.$ In Section~\ref{sec:synvar_derivation} of Supplementary Material, we show derivations of first-, second- and mixed-moment synthetic variables.

Returning to our original least-squares criterion \eqref{eqn:uncensored_ranef_ls}, since $\E\by\by\tp = \E\bYstar$, we can write
\begin{equation}\label{eqn:censored_ranef_ls}
    \E\bYstar = \sum_{k=1}^K \sigma^2_k\bK_k + \sigma^2_e\bI_N,
\end{equation}
leading to the least-squares criterion:
$$
    (\bsigmahat^2_g\tp,\sigmahat^2_e) = \arg\min_{\bsigma_g^2,\sigma_e^2}||\bYstar(\Ghat) - (\sum_k \sigma^2_k \bK_k + \sigma^2_e\bI)||_F^2,
$$
where $\Ghat$ is the Kaplan-Meier (KM) estimator for $G$. The normal equation becomes
\begin{equation}\label{eqn:censored_ranef_normaleqn}
    \bmat{\bT & \bb \\ \bb\tp & N}\bmat{\bsigma^2_g \\ \sigma^2_e} = \bmat{\bc \\ \tr\bYstar(\Ghat)},
\end{equation} which is the same as \eqref{eqn:uncensored_ranef_normaleqn} except for $\tr\bYstar$ on the right-hand side of the equation. We opt not to impose nonnegative constraints on the variance components, since the unconstrained estimator is unbiased in the ideal case where the nuisance function $G$ is known. When $G$ is replaced by $\Ghat$, unbiasedness is no longer guaranteed; nevertheless, simulations show minimal bias under low to moderately high censoring and diminishing bias as sample size increases. The package \texttt{CVC.jl} provides both constrained and unconstrained estimates, and all analyses in this paper use the unconstrained version.

\subsection{Censored multi-component linear mixed-effects model}\label{sec:censored_multi_component_linear_mixed_effects}
Assume the following linear mixed-effects model
\begin{equation}\label{eqn:mixef_model}
    \by = \bW\balpha + \sum_{k = 1}^K \bX_k\bbeta_k + \bepsilon,
\end{equation} where $\bW$ is an $N \times C$ covariate matrix and $\balpha$ is a $C$-vector of fixed effects. Note that \eqref{eqn:mixef_model} now incorporates the fixed effect $\balpha$, hence the name \textit{mixed}-effects model, as opposed to the \textit{random}-effects model in \eqref{eqn:ranef_model}. Let $\bV = \bI - \bP_W$ be the orthogonal projector onto the null space of $\bW$. Then
$$
    \bV\by = \sum_{k = 1}^K \bV\bX_k\bbeta_k + \bV\bepsilon
$$
and
\begin{equation}
    \E \bV\by\by\tp\bV\tp = \sum_{k = 1}^K \sigma^2_k \bV\bK_k\bV\tp + \sigma^2_e\bV\bV\tp.
\end{equation} Since $\E(\bV\bYstar\bV\tp) = \E(\bV\by\by\tp\bV\tp)$, the least squares criterion becomes
$$
    (\bsigmahat^2_g\tp,\sigmahat^2_e) = \arg\min_{\bsigma_g^2,\sigma_e^2}\left\Vert\bV\bYstar(\Ghat)\bV\tp - \left(\sum_{k = 1}^K \sigma^2_k \bV\bK_k\bV\tp + \sigma^2_e\bV\bV\tp\right)\right\Vert^2_F,
$$ which is equivalent to solving the following normal equations
\begin{equation}\label{eqn:censored_mixef_normaleqn}
    \bmat{\bT & \bb \\ \bb\tp & N - r}\bmat{\bsigma^2_g \\ \sigma_e^2} = \bmat{\bc \\ \tr\bYstar(\Ghat)\bV}
\end{equation} with $T_{k\ell} = \tr\bK_k\bV\bK_\ell\bV$, $b_k = \tr\bV\bK_k$, and $c_k = \tr\bYstar(\Ghat)\bV\bK_k\bV$ for $k,\ell = 1,2,...,K.$

We note that if $y_i$ and $r_i$ are conditionally independent given covariates, then the synthetic variables should be constructed using the conditional censoring distribution to ensure that $\E(\bYstar) = \E(\by\by\tp)$. In this paper, we focus on the simpler setting of marginal independence. We return to this assumption and its potential limitations in the Discussion.

\subsection{Randomized trace estimator}\label{sec:rand_tr_estimator}
The computational bottleneck in solving the normal equation \eqref{eqn:censored_ranef_normaleqn} is evaluating the terms $T_{k\ell}$ for $k,\ell \in \{1,...,K\}$, whose computational complexity is $O(N^2M)$. It is quadratic in $N$ and becomes intractable for biobank-scale $N$. To improve the computational efficiency, we employ a randomized trace estimator
\[
    \widehat{\tr\bK_k\bK_\ell} = \frac{1}{B}\sum_{b = 1}^B \bz_b\tp\bK_k\bK_\ell\bz_b,\quad z_1,...,z_B \overset{\text{i.i.d.}}{\sim} \Ncal(\bzero, \bI_N),
\] where $B$ is the number of independent random Gaussian vectors $z_b$. Taking the expectation of the estimator, one can readily see that it is an unbiased estimator of the trace. The computational complexity reduces to $O(NMB)$, which is now linear in $N$. It has been shown that $B = 10$ is sufficient to obtain a sufficiently accurate estimate of the trace while only minimally impacting the accuracy of heritability estimates \citep{wu_scalable_2018}. Our sensitivity analysis shows that increasing $B$ beyond $10$ has negligible impact on total heritability estimates and their corresponding standard errors (Section~\ref{sec:impact_B_J}), suggesting that for $B \geq 10$, the additional variability induced by randomized trace estimation is small relative to the overall variability of the heritability estimator.

\subsection{Jackknife standard error estimation of heritability estimates}\label{sec:jackknife_se}
To obtain the standard errors of the heritability estimates, we employ column-wise block jackknife. We partition an $N \times M_k$ standardized genotype matrix $\bX_k$ associated with the variance component $\sigma_k^2$ into $J$ non-overlapping blocks
$$
    \bX_k = \bmat{\bX_k\iter{1} & \bX_k\iter{2} & \cdots & \bX_k\iter{J}}.
$$ Let $\bX_k\iter{-j}$ be the $k$th genotype matrix $\bX_k$ with $\bX_k\iter{j}$ block removed, and $\bK_k\iter{-j}$ be the GRM constructed from $\bX_k\iter{-j}.$ 
The normal equation for the $j$th block jackknife estimate is
$$
    \bmat{\bT\iter{j} & \bb\iter{j} \\ \bb\iter{j}{\tp} & N - C}\bmat{\bsigma_g{^2}\iter{j} \\ \sigma_e{^2}\iter{j}} = \bmat{\bc\iter{j} \\ \tr\bYstar(\Ghat)\bV}
$$ where $T_{k\ell}\iter{j} = \tr\bK_k\iter{-j}\bV\bK_\ell\iter{-j}\bV$, $b_k\iter{j} = \tr\bV\bK_k\iter{-j}$ and $c_k\iter{j} = \tr\bYstar(\Ghat)\bV\bK_k\iter{-j}\bV$ for $k, \ell \in \{1,...,K\}$ and $j \in \{1,...,J\}.$ If we define the $j$th block jackknife estimate of $h^2$ as $\hhat^{2(j)} = \frac{\bone\tp\bsigmahat_g{^2}\iter{j}}{\bone\tp\bsigmahat_g{^2}\iter{j} + \sigmahat_e{^2}\iter{j}}$, then the block jackknife standard error estimate of $\hhat^2$ is given by
\begin{equation}
    \sigmahat_{\text{jack}} = \sqrt{\frac{J-1}{J}\sum_{j=1}^J(\hhat^2{\iter{j}} - \muhat_{\text{jack}})^2},\quad \muhat_{\text{jack}} = \frac{1}{J} \sum_{j=1}^J\hhat^2{\iter{j}}.
\end{equation} Although this procedure differs from the classical observation-wise jackknife~\citep{kunsch_jackknife_1989}, it is analogous in spirit to the SNP-level block jackknife used in LD score regression~\citep{schizophrenia_working_group_of_the_psychiatric_genomics_consortium_ld_2015, finucane_heritability_2018}, where variability is assessed by deleting blocks of SNP-indexed contributions under local dependence induced by linkage disequilibrium. In our formulation, these SNP-indexed contributions enter the estimating equation through the columns of $\bX_k$, and deleting column blocks corresponds to omitting groups of such contributions. This argument is heuristic, and a formal theoretical justification is beyond the scope of this paper. Empirically, our simulation studies (Sections~\ref{sec:h2_calibration} and \ref{sec:comparison}) indicate that the resulting standard error estimates are well calibrated across a range of scenarios.

For all our data analyses, we set $J = 100$, as this has been shown to be sufficient for obtaining unbiased standard error estimates~\citep{pazokitoroudi_efficient_2020}. Our sensitivity analysis indicates that the bias of jackknife standard error estimates remains stable for moderate values (e.g., $J \geq 10$), with diminishing gains beyond this range (Section~\ref{sec:impact_B_J}).

\section{Model misspecification induced by underlying genetic architectures}\label{sec:mis_spec}
Before examining robustness under realistic deviations from model assumptions, we briefly summarize results under the correctly specified scenario. When all SNPs within a partition are causal and share a common variance component, the CVC model accurately estimates total and partitioned heritability, as well as their standard errors, across a range of censoring rates and simulation settings (including sample size, number of SNPs, and number of genomic partitions; see Section~\ref{sec:correct_spec} of Supplementary Material). We therefore focus on misspecified settings, which more closely reflect plausible genetic architectures in practice.

To assess the robustness of the CVC model, we evaluate its accuracy in estimating total and partitioned heritability $h^2_{\text{SNP}}$ under genetic architectures in which only a subset of SNPs are causal and effect size variances may depend on minor allele frequency (MAF) and linkage disequilibrium (LD) score~\citep{haplotype_reference_consortium_comparison_2018}. We will refer to this dependence as MAF-LD coupling. The LD score of a SNP is defined as the sum of squared correlations between that SNP and all other SNPs.

We simulated the phenotypes using the covariate matrix $\bW$ and standardized genotype matrix $\bX$ consisting of $N = 276,169$ unrelated participants of self-reported British white ancestry from the UKB study. We excluded subjects who were outliers for the genotype heterozygosity or missingness, showed mismatch between genetic and self-reported sex, and had putative sex chromosome aneuploidy. The covariate matrix $\bW$ included an intercept and 23 covariates, including sex (self-reported), birth year, assessment center and top 20 genetic principal components (PCs). The genotype data included $M = 592,454$ SNPs meeting the following criteria: the UKB Axiom array SNPs, autosomal SNPs with $\leq 1\%$ missingness, passing Hardy–Weinberg test at the significance level of $10^{-7}$, and residing outside MHC region (Chr6:28-33 Mb). Section~\ref{sec:ukb_field_id} of Supplementary Material contains the UKB Field IDs used to apply the filtering criteria. The simulated data were created using \href{https://github.com/dohyunkim116/CVCData.jl}{\texttt{CVCData.jl}}, a package written in \texttt{Julia} programming language for reproducibility.

We simulated uncensored survival traits with heritability $h^2 \in \{0.2, 0.5, 0.8\}$ under different genetic architectures:
\begin{equation}\label{eqn:mis_spec_model}
\begin{aligned}
    \sigma_j^2 &= s_jw_j^b[f_j(1-f_j)]^a,\\
    \bbeta\tp = (\beta_1,\beta_2,...,\beta_M)\tp &\sim N(\bzero,\diag(\sigma_1^2,\sigma_2^2,...,\sigma^2_M)),\\
    \by\mid \balpha, \bbeta &\sim N(\bW\balpha + \bX\bbeta, \sigma_e^2\bI_N),
\end{aligned}
\end{equation}  where $a \in \{0, 0.75\}, b \in \{0, 1\}$ and $\sigma_e^2 = \left(\frac{1}{h^2} - 1\right)\sum_{j=1}^{M} \sigma_j^2$. Here, $s_j \in \{0, 1\}$ denotes the causal variant status, $w_j$ is the LDAK score, and $f_j$ is the MAF of SNP $j$. The LDAK (linkage-disequilibrium adjusted kinships) scores are modified LD scores that account for local LD structure by assigning higher weights to SNPs in regions of low LD~\citep{speed_improved_2012}. The parameters $a$ and $b$ control the extent to which the variances of genetic effects depend on LD and MAF, respectively. For example, when both $a$ and $b$ are zero with $s_j = 1$, the variances of all genetic effects are nonzero and identical across SNPs, reducing to the correctly specified model.

For each model, we varied the causal variant rates (CVRs) which were set to either 1\% or 100\%. Causal SNPs were selected based on MAFs from one of three intervals: $[0,0.5]$, $[0.01, 0.05]$ and $[0.05, 0.5]$. Specifically, for CVR = 1\%, we randomly selected 1\% of all SNPs to be causal, restricting selection to SNPs whose MAFs were within either $[0,0.5]$, $[0.01,0.05]$ or $[0.05,0.5]$. For CVR = 100\%, all SNPs with MAF in $[0,0.5]$ were set as causal. The fixed effects $\alpha_c$ were sampled uniformly from the interval $(0,1).$ 

We partitioned the genotype data into 24 MAF-LD bins, where four LD bins were based on LDAK score quartiles, and six MAF bins were based on the intervals created by the MAF knots: $0.01, 0.02, 0.03, 0.04$ and $0.05$. This partitioning assumed each partition shared the same variance component magnitude, which differed from the generative model in \eqref{eqn:mis_spec_model}, thereby inducing model misspecification. We additionally examine whether increasing the granularity of MAF-LD partitioning mitigates this misspecification and improves estimation accuracy. The number of simulation replicates was fixed at 100 for all settings.

\subsection{Accuracy of total heritability estimates}\label{sec:h2_accuracy_mis_spec}
\begin{figure}[H]
    \centering\includegraphics[width=\textwidth]{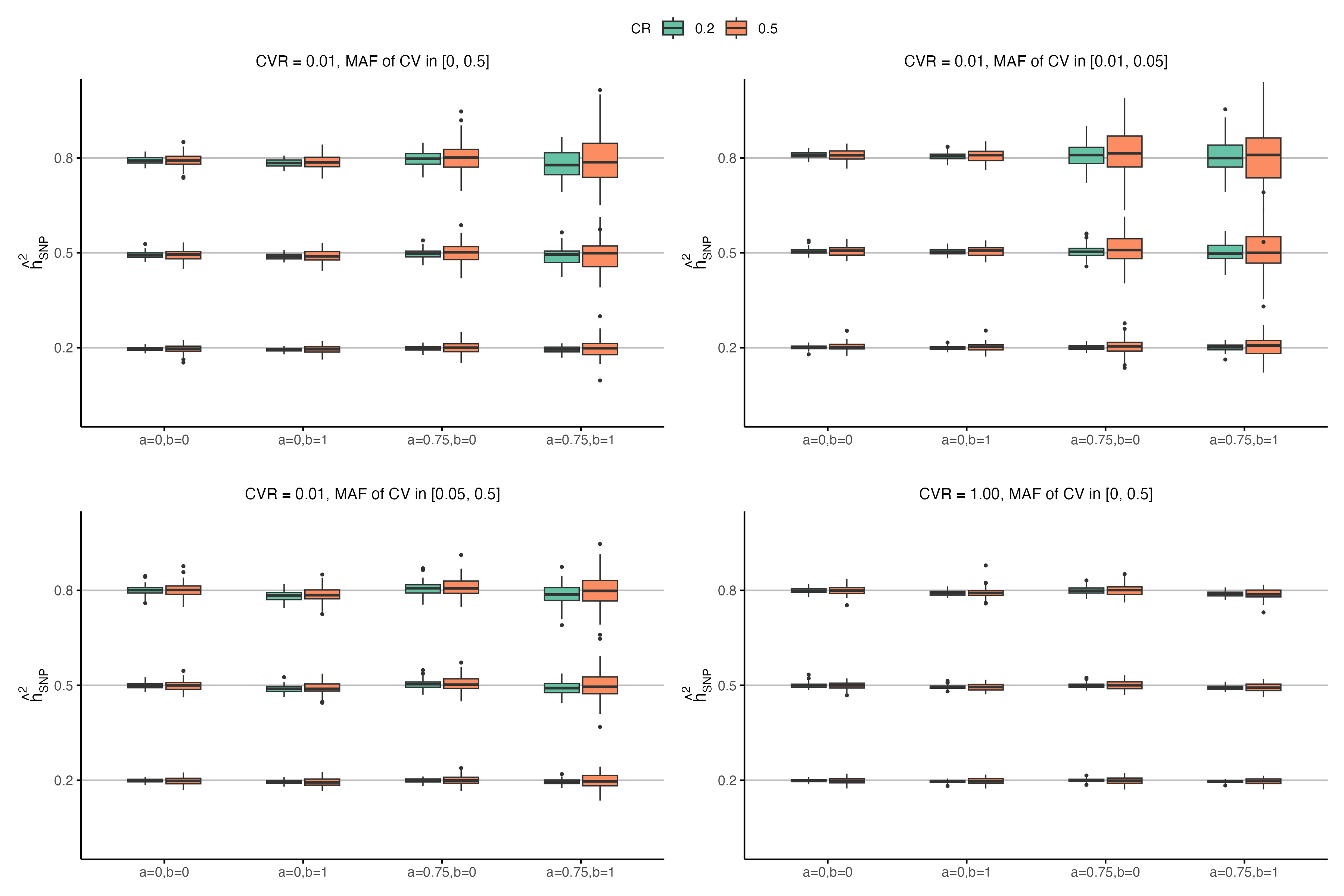}
    \caption{Accuracy of total heritability estimation under 48 different genetic architectures stratified by $\Delta \in \{0.2, 0.5\}$ over 100 runs ($N = 276,169$, $M = 592,454$, $C = 24$, $J = 100$, $B = 10$). The number of genomic partitions $K$ was set to $24$ created from 4 LD bins based on quartiles of LDAK scores and 6 MAF intervals based on 5 MAF knots, 0.01, 0.02, 0.03, 0.04 and 0.05. The horizontal lines indicate $h_{\text{SNP}}^2$. CVR: causal variant rate; MAF: minor allele frequency; CV: causal variant.}
    \label{fig:mis_spec_accuracy}
\end{figure}

Tables~\ref{tab:cvc_mis_spec_accuracy_h2_0.2}--\ref{tab:cvc_mis_spec_accuracy_h2_0.8}, together with Figures~\ref{fig:mis_spec_accuracy} and \ref{fig:mis_spec_accuracy_supp}, summarize total heritability estimation results across 48 different underlying genetic architectures. Across all genetic architectures, the total heritability estimates are relatively unbiased. When $h^2_{\text{SNP}} = 0.2$, the relative bias ranges from $-2.946\%$ to $0.522\%$ for $\Delta = 0.2$, and $-3.081\%$ to $2.052\%$ for $\Delta = 0.5$. When $h^2_{\text{SNP}} = 0.5$, the relative bias ranges from $-2.341\%$ to $0.927\%$ for $\Delta = 0.2$, and $-2.124\%$ to $2.128\%$ for $\Delta = 0.5$. When $h^2_{\text{SNP}} = 0.8$, the relative bias ranges from $-2.345\%$ to $1.206\%$ for $\Delta = 0.2$, and $-1.822\%$ to $2.598\%$ for $\Delta = 0.5$. Relative biases tend to be larger when CVR is lower, MAF-LD coupling is stronger, and censoring rates are higher. 

The jackknife standard error estimates are also nearly unbiased at low and moderate censoring rates. Across $\Delta \in \{0.2,0.5\}$, the bias of the jackknife standard error estimates ranges from $-0.006$ to $0.001$ for $h^2_{\text{SNP}} = 0.2$, $-0.018$ to $0.004$ for $h^2_{\text{SNP}} = 0.5$, and $-0.035$ to $0.009$ for $h^2_{\text{SNP}} = 0.8$. Standard errors of total heritability tend to be larger for scenarios with lower CVR, stronger MAF-LD coupling, and higher censoring rates.

For moderately high censoring rate $\Delta = 0.7$, the relative bias of total heritability estimates and bias of jackknife stand error estimates do not show substantial change from those observed at low and moderate censoring rates, while at high censoring rate $\Delta = 0.8$, these values are approximately an order of magnitude larger than those observed at $\Delta = 0.5$. Across $h^2_{\text{SNP}} \in \{0.2, 0.5, 0.8\}$, the relative bias of total heritability estimates ranges from $-5.083\%$ to $3.335\%$ for $\Delta = 0.7$ and $-4.106\%$ to $19.793\%$ for $\Delta = 0.8$. The bias of jackknife standard error estimates ranges from $-0.039$ to $-0.003$ for $\Delta = 0.7$ and $-0.164$ to $-0.006$ for $\Delta = 0.8$.

In summary, under low to moderate censoring rates, the CVC model is robust to model misspecification and provides accurate estimates of total heritability and their standard errors. However, with high censoring rates, the estimation accuracy declines across all genetic architectures.

\subsection{Accuracy of partitioned heritability estimates}\label{sec:h2_partitioned_accuracy_mis_spec}
We assess the accuracy of the CVC model in estimating partitioned heritability under model misspecification for $h^2_{\text{SNP}} = 0.5$ (Figures~\ref{fig:h2_partitioning_supp_h2_0.5_cr_low_to_moderate} and \ref{fig:h2_partitioning_supp_h2_0.5_cr_high}). We focus on genetic architectures with CVR $=1\%$ to evaluate the model’s sensitivity for detecting nonzero heritability in causal bins and its specificity for identifying zero heritability in noncausal bins. We define causal bins as LD–MAF bins whose MAF intervals contain the MAFs of causal SNPs, with all remaining LD–MAF bins classified as noncausal. For instance, when CVR $= 1\%$ and the causal MAF interval is $[0.01, 0.05]$, only $1\%$ of SNPs---those with MAFs in $[0.01, 0.05]$---have nonzero variance components, whereas all other SNPs are noncausal and contribute zero heritability.

Figures~\ref{fig:h2_partitioning_supp_h2_0.5_cr_low_to_moderate} ($\Delta = 0.2$ and $0.5$) and \ref{fig:h2_partitioning_supp_h2_0.5_cr_high} ($\Delta = 0.7$ and $0.8$) show that the CVC model can provide visual inspection of a trait’s genetic architecture by yielding nonzero point estimates of partitioned heritability in causal bins, while producing null or near-null estimates in noncausal bins. We observe that the resolution of the inferred genetic architecture becomes noisier as the censoring rate increases.

Table~\ref{tab:cvc_mis_spec_partitioning_h2_0.5} presents overall heritability contributions from causal and noncausal bins across eight genetic architectures for censoring rates $\Delta = 0.2$, $0.5$, $0.7$, and $0.8$. Within each censoring rate, as variance components become more strongly coupled with MAF and LD (through parameters $a$ and $b$), relative bias in causal bins tends to increase, as expected, since stronger coupling induces greater model misspecification. Nonetheless, the CVC model estimates heritability in both causal and noncausal bins with high accuracy for $\Delta = 0.2$ and $0.5$, with a maximum absolute relative bias of $2.798\%$. At higher censoring rates, estimation accuracy decreases, with absolute relative bias reaching up to $3.309\%$ for $\Delta = 0.7$ and $4.683\%$ for $\Delta = 0.8$. 

In particular, for causal bins, relative bias in heritability estimates ranges from $-2.679\%$ to $2.327\%$ for $\Delta = 0.2$, $-2.798\%$ to $2.161\%$ for $\Delta = 0.5$, $-3.309\%$ to $1.468\%$ for $\Delta = 0.7$, and $-2.054\%$ to $4.683\%$ for $\Delta = 0.8$. For noncausal bins, bias ranges from $-0.009$ to $0.002$ for $\Delta = 0.2$, $-0.007$ to $0.004$ for $\Delta = 0.5$, $0.000$ to $0.017$ for $\Delta = 0.7$, and $-0.009$ to $0.063$ for $\Delta = 0.8$.

Overall, these results demonstrate that the CVC model can reveal a rough landscape of a trait’s complex genetic architecture by accurately estimating heritability contributions from both causal and noncausal bins across a wide range of genetic architectures and censoring rates.

\subsection{Calibration of heritability inference}\label{sec:h2_calibration}
We examine calibration of heritability inference with jackknife standard error estimates. To summarize calibration quality across different genetic architectures, we used the Integrated Calibration Index (ICI), which is defined as a weighted average of the absolute deviation of empirical coverage from nominal coverage~\citep{austin_integrated_2019}. Since coverage was evaluated at non-uniform spaced nominal levels---with increments of $0.1$ for the interval between $0.1$ and $0.9$, along with the levels $0.95$ and $0.99$---we used trapezoidal weights to approximate the ICI. 

Figure~\ref{fig:calibration_summary_normal}a shows a heatmap of ICI across $16$ different genetic architectures for each $h^2_{\text{SNP}} \in \{0.2, 0.5, 0.8\}$ and $\Delta \in \{0.2, 0.5, 0.7, 0.8\}$. Figures~\ref{fig:mis_spec_calibration_plots_cvr_0.01_maflb_0_mafub_0.5}--\ref{fig:mis_spec_calibration_plots_cvr_1_maflb_0_mafub_0.5} show individual calibration curves for each genetic architecture. For the majority of settings ($168$ out of $192$), we observe negative signed ICI (i.e., the average difference between empirical and nominal coverage), indicating overconfident standard error estimates. 

Although calibration of heritability inference generally deteriorates as overall MAF-LD coupling increases, the extent of miscalibration is substantially more severe when the variance component depends on MAF ($a = 0.75$). When there is no dependence on MAF ($a = 0$), even with additional dependence through the LDAK score (increasing $b$ from $0$ to $1$), we do not observe substantial change in ICI. In contrast, introducing dependence between MAF and the variance component (increasing $a$ from $0$ to $0.75$) leads to a marked increase in ICI even when $b = 0$; this effect is most pronounced when $h^2_{\text{SNP}} = 0.8$ and $\Delta = 0.2$. 

Calibration deteriorates the most when the variance component depends on both MAF and LD, with increasingly overconfident standard errors observed as $h^2_{\text{SNP}}$ increases when $\Delta = 0.2$. Indeed, the worst calibration ($\text{ICI} = 0.307$) is observed for $h^2_{\text{SNP}} = 0.8$, $\text{CVR} = 0.01$, $a = 0.75$, $b = 1$, and MAF of CV is in $[0.01, 0.05]$, whereas the best calibration ($\text{ICI} = 0.015$) is observed when $b = 0$ and MAF of CV is in $[0, 0.5]$ with all other parameters equal to the worst case (Figure~\ref{fig:calibration_summary_normal}b). 

Interestingly, when $\text{CVR} = 0.01$ and MAF of CV is in $[0.05, 0.50]$, calibration degrades only mildly as MAF-LD coupling increases. Similarly, when $\text{CVR} = 1.00$ and MAF of CV is in $[0, 0.5]$, the model remains well calibrated across all settings. These results indicate the CVC model maintains good calibration when most causal variants are common, or when all SNPs are casual with MAFs uniformly distributed across the spectrum (rare, low-frequent and common).

\subsection{Impact of finer MAF-LD partitioning}\label{sec:finer_bins_misspec}
In simulations under different genetic architectures, we fixed the number of partitions at $K = 24$, corresponding to four LDAK bins and six MAF bins. Because coarse partitioning may induce model misspecification when the true genetic architecture has continuous MAF–LD coupling, we examine whether increasing the granularity of MAF-LD partitioning improves estimation performance for the genetic architecture showing the poorest calibration of heritability inference (Figure~\ref{fig:calibration_summary_normal}b).

To this end, we consider two finer partitioning schemes with $K = 48$ and $K = 96$. For $K = 48$, we double the number of LDAK bins to eight bins based on LDAK octiles while retaining the same six MAF bins. For $K = 96$, we further double the number of MAF bins by splitting each of the six MAF bins at its median, resulting in 12 MAF bins combined with the eight LDAK bins.

Increasing the number of LD partitions alone, or jointly increasing both LD and MAF partitions, results in only modest changes in estimation performance. In particular, neither the mean squared error (MSE) nor the ICI is substantially reduced when increasing $K$ from 24 to 48 or 96 (MSE: $0.042$, $0.036$, and $0.040$; ICI: $0.313$, $0.245$, and $0.259$, respectively). These changes correspond to only small relative improvements, so we retain $K=24$ as a reasonable default and recommend sensitivity checks when architecture is uncertain.

\section{Comparisons to other methods}\label{sec:comparison}
\begin{figure}[H]
    \centering\includegraphics[width=0.9\textwidth]{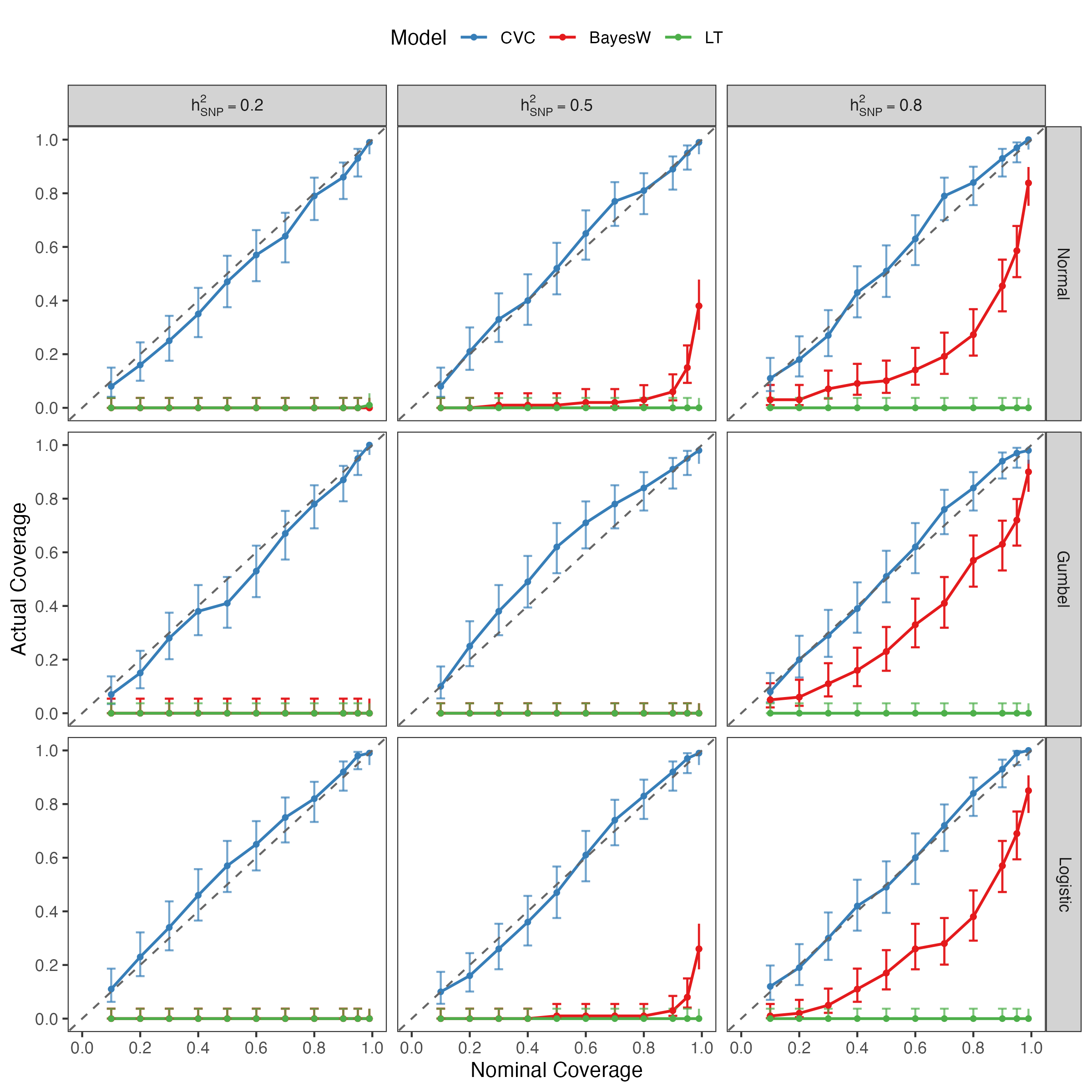}
    \caption{Calibration plots for estimating total heritability under correct model specifications over 100 runs for $h^2_{\text{SNP}} \in \{0.2, 0.5, 0.8\}$ and $\Delta = 0.2$ ($N = 20,000$,  $M = 20,000$, $C = 10$, $K = 5$, $J = 100$, $B = 10$). For BayesW, we used 5 chains with chain length of 3,000, burn-in of 1,000, and thinning of 5. Posterior mean was used as the point estimate. For censoring rates 0.5 and 0.8, BayesW produced errors across all replicates and chains. The error bars of actual coverages indicate 95\% confidence interval.}
    \label{fig:correct_spec_h2_calibration}
\end{figure}

We compare the performance of the CVC, BayesW and LT models by evaluating bias, MSE, mean absolute error (MAE) and coverage of true heritability under three different error distributions: normal, Gumbel and logistic (Tables~\ref{tab:correct_spec_h2snp_comp},~\ref{tab:correct_spec_h2k_comp}, Figures~\ref{fig:correct_spec_h2_calibration},~\ref{fig:correct_spec_h2_calibration_supp}--\ref{fig:correct_spec_h2k_calibration_logistic}). Coverage is defined as the proportion of confidence intervals (for CVC or LT) or credible intervals (for BayesW) containing the true heritability across data replicates. For total heritability, we compare the CVC model with both the BayesW and LT models; for partitioned heritability, we compare the CVC model with the BayesW model only. We do not include HERRA in the comparisons because it is not designed to accommodate multiple variance components, and no publicly available software implementation exists.

Simulation data were generated under the correctly specified model described in Section~\ref{sec:correct_spec}, except that we additionally consider the Gumbel and logistic error distributions. The log-scale survival traits were generated as
\[
    \by \sim \Dcal\left(\bW\balpha + \sum_k \bX_k\bbeta_k, \sigma_e^2\bI\right),
\] where $\Dcal \in \{\mathrm{normal}, \mathrm{Gumbel}, \mathrm{logistic}\}$. When $\Dcal = \mathrm{Gumbel}$, the BayesW model is correctly specified, given its assumption that the conditional distribution of survival traits in their original units (conditioning on genetic effects) is a Weibull distribution. Since the BayesW model produced errors for all data replicates under misspecified model, we were unable to compare the methods in this setting.

Simulation settings included $h^2_{\text{SNP}} \in \{0.2, 0.5, 0.8\}$, $\Delta \in \{0.2, 0.5, 0.8\}$, $N = 20{,}000$, $M = 20{,}000$, $C = 10$ and $K = 5$ with 100 replicates per setting. Across all error distributions, replicates and chains, the BayesW model failed to fit datasets with $\Delta = 0.5$ and $0.8$.

For the BayesW model, we generated MCMC samples from five independent chains, each of length $3{,}000$ with a burn-in of $1{,}000$ and thinning of $5$, for each data replicate. The number of quadrature points was set to $25$. The mean of the posterior samples was used as the point estimate for heritability, and credible intervals were computed from the corresponding posterior quantiles. In the normal error scenario, all chains failed in the $93$rd replicate when $h^2_{\text{SNP}} = 0.8$. In Gumbel error scenario, $33$ data replicates failed in at least one chain for $h^2_{\text{SNP}} = 0.2$; thus, results for this setting are based on the remaining $67$ replicates. See Section~\ref{sec:bayesw} of Supplementary Material for BayesW software configurations and errors messages.

For the LT model, we fit the CVC model without the synthetic-variable transformation, treating the censoring indicator $\delta_i$ as a continuous quantitative trait (case: $\delta_i = 1$, control: $\delta_i = 0$). The observed-scale heritability estimate, $\hhat^2_{\text{binary}}$, was then converted to the liability scale $\hhat^2_\ell$, following \citep{hou_accurate_2019}:
$$
    \hhat^2_\ell = \frac{d^2_\text{pop}(1 - d_\text{pop})^2}{[f(L)]^2 d_\text{GWAS}(1 - d_\text{GWAS})}\hhat^2_{\text{binary}},
$$ where $f(z)$ is the standard normal density and $L = \Phi\inv(1 - d_{\text{pop}})$ is the liability threshold, with $\Phi(z)$ denoting the standard normal CDF. We assumed the population prevalence $d_\text{pop} = 1 - \Delta$, and because case-control status was derived from time-to-event data, we set the sample prevalence $d_\text{GWAS}$ equal to $d_\text{pop}$. Under these assumptions,
$$
    \hhat^2_{\ell} = \frac{\Delta(1 - \Delta)}{[f\left(\Phi\inv(\Delta)\right)]^2}\hhat^2_{\text{binary}}.
$$ 

Tables~\ref{tab:correct_spec_h2snp_comp} and \ref{tab:correct_spec_h2snp_comp_supp} compare the relative bias, MSE, and MAE of total heritability estimates across all models. In all simulation settings, the CVC model consistently achieves superior accuracy, showing smaller relative biases than both of the BayesW and LT models. Across all settings, the LT model performs poorly, with relative bias ranging from $-95.39\%$ to $81.61\%$, whereas the CVC model's relative bias ranges from $-22.49\%$ to $29.52\%$. Across all error distributions, the CVC model's relative bias is substantially smaller than the BayesW model's. While the CVC model's relative bias ranges from $-0.39\%$ to $1.66\%$, the BayesW model's relative bias ranges from $-1.47\%$ to $116.52\%$. The CVC model also consistently yields lower MSE and MAE than the BayesW model across all settings. These results indicate that the CVC model is robust to different underlying survival trait distributions, whereas the BayesW and LT models are not, leading to substantial inaccuracies in total heritability estimates.

For uncertainty quantification of total heritability, Figures~\ref{fig:correct_spec_h2_calibration} and~\ref{fig:correct_spec_h2_calibration_supp} show that the CVC model is well calibrated across all nominal levels (0.10-0.99) and simulation settings, with no statistically significant differences between nominal and actual coverage at the 95\% level in most cases. This suggests that the CVC model’s block jackknife standard errors provide accurate confidence intervals when the model is correctly specified. By contrast, the BayesW and LT models are severely overconfident and show substantial miscalibration.

Table~\ref{tab:correct_spec_h2k_comp} presents the partitioned heritability estimation results, comparing the CVC and BayesW models. Across all $h^2_{\text{SNP}} \in \{0.2, 0.5, 0.8\}$, bins and error distributions, the CVC model has markedly lower relative biases than the BayesW model; the relative bias ranges from $-3.01\%$ to $48.87\%$ for the CVC model, and from $-29.13\%$ to $1880.41\%$ for the BayesW model. In majority of the settings, the CVC model also attains lower MSE and MAE, although the BayesW model shows improvements in these values when $h^2_{\text{SNP}} = 0.8$. Overall, the CVC model provides more accurate partitioned heritability estimates than the BayesW model.

For uncertainty quantification of partitioned heritability, Figures~\ref{fig:correct_spec_h2k_calibration_normal}--~\ref{fig:correct_spec_h2k_calibration_logistic} show that across all settings and bins---except for a few cases---the CVC model's actual coverage does not significantly differ from nominal coverage at the 95\% level, whereas the BayesW model is overconfident in most scenarios. These findings underscore the CVC model's superior uncertainty quantification for partitioned heritability compared to the BayesW model.

\section{Computational benchmark}
We report the computational efficiency of fitting the CVC model (Table~\ref{tab:cvc_computational_benchmark}, Figure~\ref{fig:computational_benchmark}). Table \ref{tab:cvc_computational_benchmark} summarizes the running time and computational resources used to fit the CVC model across different simulation settings. Benchmarks were performed using up to 16 CPU cores and 96 GB of RAM, with up to 1 TB of temporary disk space for memory-mapped working arrays. Derivations of the working arrays for one-pass scanning of genotype data are provided in Supplementary Material (Section~\ref{sec:computational_note}). An efficient implementation of the proposed method is available in our \texttt{CVC.jl} package. The CVC model required less than 9 hours to analyze genotype data with 1 million subjects, 1 million SNPs, and 100 genomic partitions, demonstrating excellent scalability. 

We also compared the CVC model's running time to that of the BayesW model when fitting datasets with $100{,}000$ SNPs, $10$ covariates and $10$ partitions across varying sample sizes. The CVC model can be fitted several orders of magnitude faster than the BayesW model (Figure~\ref{fig:computational_benchmark}). For the BayesW model, the runtime for a chain of length $3{,}000$ was extrapolated from a $20$-chain run due to the high computational cost of fitting.

\section{Heritability of UKB survival traits}\label{sec:ukb_real_analysis}
We quantify the heritability of age-at-diagnosis traits from the UKB study using the partitioned genotype data with $K = 24$ described in Section~\ref{sec:mis_spec} (excluding SNPs with MAF below $0.1\%$). Analyzed traits were ``Started wearing glasses or contact lenses", ``First live birth", ``Hypertension" and ``Started smoking in current and former smokers." The definition for ``Hypertension" follows \cite{dey_efficient_2022}, while the other phenotypes were defined according to Section~\ref{sec:pheno_def} of Supplementary Material. For the female-only trait ``First live birth," analyses were restricted to self-reported females. We selected traits with low to moderately high censoring rates, as our simulation studies indicated that the CVC model yields accurate heritability estimates with acceptable precision in this range.

Table~\ref{tab:ukb_age_at_dx} presents heritability estimates for the selected traits obtained by using the CVC and LT models. All models were adjusted for sex, birth year, assessment center and the top 20 genetic PCs; the sex covariate was excluded for sex-specific traits. Heritability estimates from the BayesW model could not be obtained, as the model produced errors for all selected traits (see Section~\ref{sec:bayesw} in Supplementary Material for details).

For age at ``Started wearing glasses or contact lens," the CVC model estimated the total heritability at $0.101$ (95\% CI: $0.093$--$0.109$), significantly lowered than the LT model estimate of $0.249$ ($0.233$--$0.265$). This trait is considered a proxy for spherical equivalent (SPHE) myopia and shows evidence of genetic correlation with SPHE~\citep{patasova_assessment_2018, kiefer_genome-wide_2013, patasova_genome-wide_2022, the_consortium_for_refractive_error_and_myopia_meta-analysis_2020, le_polygenic_2024}. Treating the age of first spectacle wear (AOSW) as a continuous trait, \cite{patasova_assessment_2018} estimated total heritability at $0.10$ (or $0.13$ using a different software) in a UKB cohort of European descent, excluding censored individuals. These estimates align with our results although their estimates reflect the heritability among affected individuals rather than the full population.

For age at ``First live birth," the CVC model estimated total heritability at $0.192$ ($0.078$--$0.306$) compared to $0.143$ ($0.127$--$0.159$) for the LT model. Prior UKB analyses report heritability estimates of age at first birth among affected individuals ranging from $0.09$ ($0.04$--$0.14$) to $0.22$ ($0.19$--$0.25$) depending on birth cohort~\citep{mills_identification_2021}, and $0.077$ for age at first pregnancy and childbirth~\citep{feng_findings_2020}. \cite{tropf_human_2015} reported $0.15$ ($0.07$--$0.23$) for age at first child birth in a combined Netherlands--UK cohort, excluding censored subjects. These values lie within the CVC model's confidence interval.

For age at “Hypertension,” the CVC model estimate was $-0.018$ ($-0.048$–$-0.012$) and the LT model estimate was $-0.008$ ($-0.016$–$0.000$). \cite{feng_findings_2020} reported estimates among affected individuals in UKB at $0.055$ (self-reported diagnosis) and $0.049$ (hospital records), and $0.116$ in a Finnish cohort. These findings suggest that the total heritability of age at first hypertension occurrence is low in European-descent populations, although family studies~\citep{ruiz_giolo_genetic_2009, Niiranenj1949} show evidence of genetic factors affecting hypertension risk. This discrepancy may arise because the heritability of disease risk and the heritability of age at diagnosis capture related but distinct genetic components.

For age at ``Started smoking," the CVC model estimate was $0.301$ ($0.201$--$0.401$) compared to $0.225$ ($0.211$--$0.239$) for the LT model. Using a cohort of European ancestry, \cite{jang_rare_2022} estimated heritability of smoking initiation age among affected individuals at $0.226$ ($-0.006$--$0.458$), while \cite{evans_genetic_2021} reported $0.05$ ($0.03$--$0.07$) in UKB. Discrepancies with our estimates may in part reflect differences in the inclusion of censored subjects or in phenotype definitions (e.g., our definition included both current and former smokers regardless of regular smoking status).

\begin{figure}[H]
    \centering\includegraphics[trim=0 0 0 0, clip, width=\linewidth]{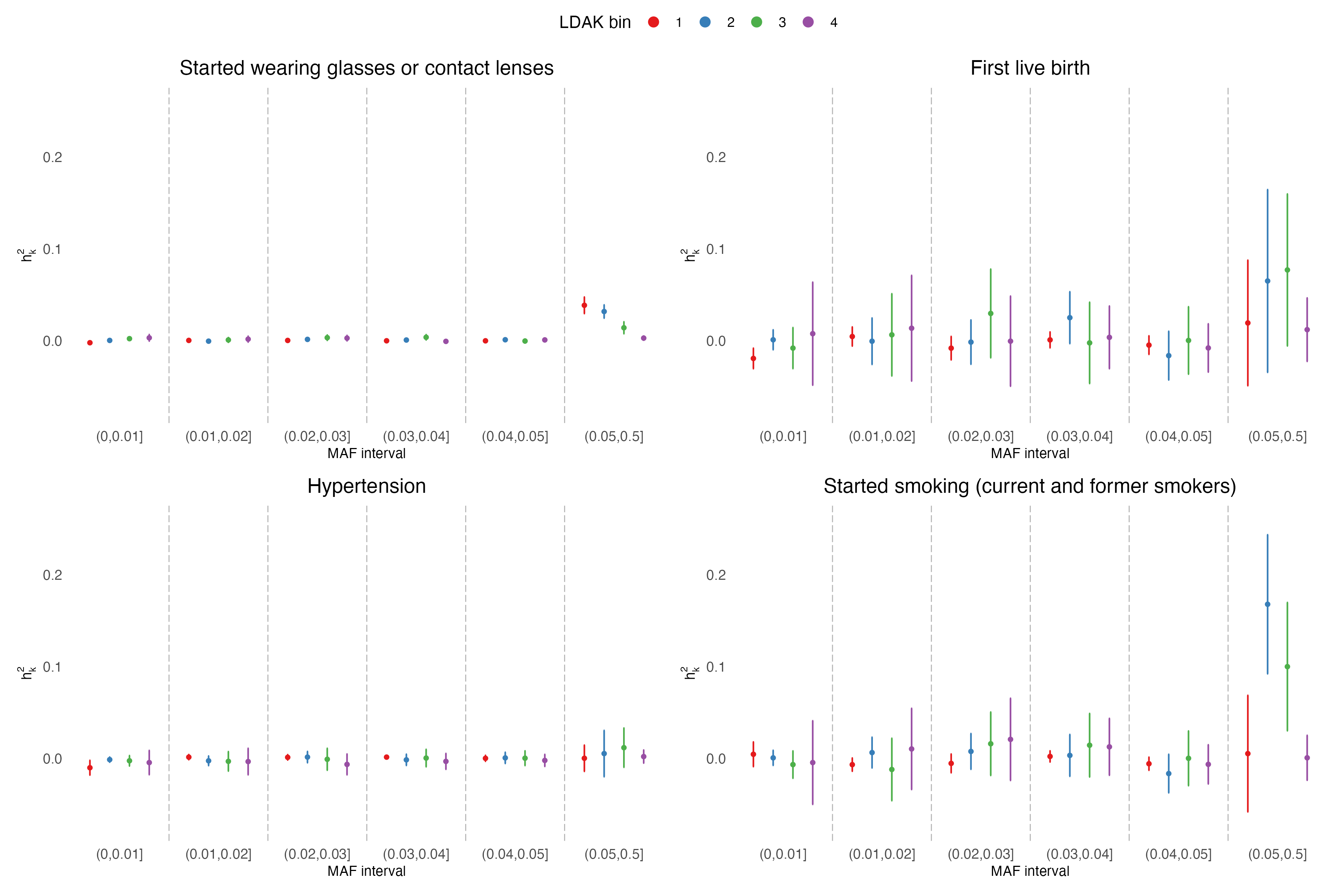}
    \caption{Estimated partitioned heritability of UKB survival traits ($K = 24$, $J = 100$, $B = 10$). The circles represent partitioned heritability estimates. The error bars are formed by two times the standard error estimates of partitioned heritability estimates. MAF intervals indicate the MAF bins formed by the MAF knots 0.01, 0.02, 0.03, 0.04 and 0.05. The four LDAK score bins are based on quartiles of LDAK scores of the SNPs. Each SNP belongs to one of 24 partitions formed by 6 MAF bins and 4 LDAK score bins. MAF: minor allele frequency; LDAK: linkage-disequilibrium adjusted kinships.}
    \label{fig:ukb_phecode_heritability_partitioning}
\end{figure}

Figure~\ref{fig:ukb_phecode_heritability_partitioning} shows partitioned heritability across LDAK-MAF bins. For all traits, most heritability was attributable to common variants (MAF $0.05$--$0.5$), but the distribution of heritability across LDAK score bins varied by trait, indicating distinct underlying genetic architectures. For ``Started wearing glasses or contact lens," the first LDAK score quartile contributed the largest proportion of heritability, whereas for ``Started smoking," the second quartile contributed the most. These results demonstrate the utility of the CVC model for dissecting the genetic architecture of survival traits.

Finally, sensitivity analyses for the CVC model using finer MAF–LD partitioning schemes (Section~\ref{sec:finer_bins_misspec}) yield total heritability estimates and standard errors that are nearly identical to those obtained with $K = 24$ (Table~\ref{tab:ukb_age_at_dx_finer_bins}), indicating robustness to moderate increases in partition granularity.

\section{Discussion}
We have proposed a novel method to robustly quantify both total and partitioned heritability of right-censored survival traits. The CVC model addresses censoring through the use of synthetic variables, which replace partially observed survival outcomes with surrogate quantities whose moments match those of the uncensored trait, thereby enabling unbiased variance component estimation. The method is computationally efficient and highly scalable, as demonstrated by the analysis of genotype data with one million subjects and one million SNPs in under nine hours. Simulation studies demonstrate that the CVC model accurately estimates total heritability across diverse genetic architectures and censoring rates, and outperforms the BayesW model in terms of accuracy, uncertainty quantification, and computational efficiency.

The synthetic variable used in our model is conceptually related to the pseudo-values introduced by \citet{andersen_generalised_2003}. Both approaches address censoring by replacing partially observed outcomes with surrogate quantities based on the KM estimator under an independent censoring assumption. The surrogates nevertheless differ in construction: pseudo-values rely on a leave-one-out jackknife estimator of the survival function, whereas synthetic variables are a closed-form transformation involving the estimated censoring distribution. The pseudo-value framework is broadly applicable to a wide range of estimands, while the synthetic-variable approach offers computational advantages that are important for large-scale variance component modeling. Investigating whether pseudo-observations can be adapted to such settings is an interesting direction for future work.

When the underlying data-generating process follows an accelerated failure time model with a given censoring rate, the LT model fails to provide accurate heritability and standard error estimates. This is expected, as the LT model uses only disease status and ignores the timing of the event. Consequently, direct comparison between the heritability of disease status and that of a survival trait such as age at onset may be inappropriate: the former measures the proportion of the variance in lifetime risk attributable to genetic factors, whereas the latter measures the proportion of variance in a censored continuous trait attributable to genetic factors. Indeed, \cite{pitkaniemi_genetic_2007} reported that, while substantial familial variation exists in susceptibility to diabetic nephropathy (DN), variation in time to DN is less influenced by shared familial factors. These findings suggest that the two trait types are related phenotypes but not identical, with both shared and distinct genetic architectures \citep{feng_findings_2020}. 

In our real-data analysis of age at diagnosis of smoking initiation, our heritability estimates using the entire cohort deviated somewhat from previous estimates obtained after excluding censored subjects. This is not surprising, as removing subjects still at risk changes the definition of heritability to apply only to the affected subset. Similarly, ignoring censoring and treating truncated values of censored traits as uncensored can lead to substantial bias, with the magnitude of bias increasing with the censoring rate \citep{taraszka_abstract_2024, epstein_tobit_2003}. In contrast, the CVC model analyzes the full cohort, accounting for right censoring in a principled manner and avoiding ad hoc preprocessing.

Several limitations of the proposed method suggest directions for future research. First, the accuracy and precision of estimates decline as censoring rates increase, with substantial deterioration for traits with very high censoring rate ($\Delta > 0.8$). Nevertheless, many traits of scientific interest, such as age at menopause and time to menarche \citep{ojavee_genomic_2021}, have relatively low censoring rates, for which accurate and precise estimates can be obtained. The method’s computational efficiency also enables analysis of biobank-scale datasets (e.g., $500{,}000$ subjects or more), which can improve precision even for highly censored traits. Since estimation is based on least squares, future work could explore weighted least squares using working variances, which may further improve precision.

Second, our method assumes marginal independence between survival and censoring times, which is stronger than the assumption of conditional independence given observed covariates. Additional simulation results under conditional independence indicate that violations of marginal independence can lead to non-negligible bias in total heritability estimation (Section~\ref{sec:conditional_independence}). As expected, bias increases with the magnitude of dependence between survival and censoring times, with negative dependence generally inducing smaller bias than positive dependence. For a fixed level of dependence, the absolute bias increases with the censoring rate (Table~\ref{tab:conditional_independence}). In principle, this limitation may be mitigated by constructing synthetic variables using the conditional censoring distribution given covariates, for example by estimating $G(t \mid \cdot)$ via a Cox proportional hazards model or more sophisticated machine learning tools~\citep{wolock_framework_2024, westling_inference_2024}. We leave a systematic investigation of this extension to future work.

Third, we have not yet established formal consistency of the proposed variance component estimator, and hence of the resulting heritability estimator. This limitation is shared by many widely used heritability estimation methods, which have historically been motivated and validated through empirical studies~\citep{yang_gcta:_2011, schizophrenia_working_group_of_the_psychiatric_genomics_consortium_ld_2015}, with rigorous asymptotic theory developed subsequently~\citep{jiang_high-dimensional_2016, jiang_high-dimensional_2024, wang_estimation_2022, xue_high-dimensional_2025}. Our estimator is defined implicitly through the estimating equation~\eqref{eqn:censored_mixef_normaleqn} and can be viewed as a method-of-moments estimator that extends the classical Haseman–Elston regression~\citep{haseman_investigation_1972}, which is conceptually closely linked to LD score regression (LDSC)~\citep{bulik-sullivan_relationship_2015}. Recent advances have established asymptotic properties for LDSC under high-dimensional settings~\citep{xue_high-dimensional_2025}, suggesting that a formal theoretical analysis of our estimator may be feasible under appropriate regularity conditions. Extending such results to the censored setting would additionally require careful treatment of the KM estimator used in constructing the synthetic variables, including its uniform consistency~\citep{gill_large_1983}, while separately accounting for the additional error introduced by the randomized trace estimator. We leave a comprehensive theoretical treatment of the proposed estimator to future work.

Fourth, the column-wise block jackknife used for variance estimation is motivated by analogy to the SNP-level block jackknife employed in LDSC. While our simulation studies suggest that the resulting standard error estimates are well calibrated across a range of scenarios, a rigorous theoretical analysis of this procedure under realistic linkage disequilibrium dependence structures among SNPs remains an important direction for future work.

Finally, the CVC model requires access to individual-level genotype and phenotype data. As more biobanks with linked genomic data become available, developing methods for summary-level data will be crucial for both scalability and privacy. Several approaches have been proposed for estimating heritability of continuous traits from summary statistics \citep{schizophrenia_working_group_of_the_psychiatric_genomics_consortium_ld_2015, speed_sumher_2019, jeong_scalable_2024}. Building on these ideas to accommodate survival traits would enable large-scale heritability analyses and enrichment studies in datasets with tens of millions of SNPs and multiple functional annotations \citep{finucane_heritability_2018}, thereby providing a more detailed understanding of their genetic architecture.

From a practical standpoint, the choice of the number of partitions and the placement of MAF cutpoints depends on the empirical distributions of LD scores and MAF across SNPs. We recommend beginning with a relatively small number of partitions (e.g., $K = 24$) and increasing this number to assess the stability of the resulting estimates. In practice, finer partitioning reduces the number of SNPs per partition, which may limit the number of feasible jackknife blocks. Our simulation results under correctly specified models indicate that jackknife standard error estimates remain accurate for moderate numbers of blocks (e.g., $J \ge 10$), with diminishing gains beyond this range (Table~\ref{tab:impact_J}). In real-data applications with unknown genetic architecture, sensitivity analyses over both the number of partitions and jackknife blocks may therefore be useful.

Despite the limitations discussed above, the ability to model right-censored survival traits at biobank scale represents an important step toward a more comprehensive understanding of the genetic architecture of time-to-event outcomes. In the biological and health sciences, the heritability of time-to-event traits provides critical insight into the genetic architecture of complex outcomes where event timing is key. Our proposed method offers a scalable approach for inference on the heritability of survival traits from high-dimensional genetic data. Continued methodological development in this area will be essential for dissecting complex genetic architectures, including those influenced by rare variants, epistasis, and gene–environment interactions.

\section*{Software}
The software implementing the CVC model is available at 
\url{https://github.com/dohyunkim116/CVC.jl}, and simulated data under misspecified model scenarios can be generated using 
\url{https://github.com/dohyunkim116/CVCData.jl}.

\section*{Supplementary material}
Supplementary material is available at \url{http://biostatistics.oxfordjournals.org}.

\section*{Data availability}
UK Biobank data are available to approved researchers via application 
(\url{https://www.ukbiobank.ac.uk/enable-your-research/apply-for-access}). 
This research was conducted under application ID~48152.

\section*{Funding}
This work was supported by National Institutes of Health under grants P30 CA-16042 (GL), UL1TR000124-02 (GL), P50CA211015 (GL), T32 HG002536 (DK), R35 GM141798 (HZ), R01 HG006139 (HZ and JZ), and R01 DK142026 (HZ, JZ and GL); National Science Foundation under grants DMS 2054253 (HZ and JZ) and IIS 2205441 (HZ, JZ and GL).

\section*{Acknowledgments}
We thank Seyoon Ko, developer of \texttt{SnpArrays.jl}, for helpful discussions on its use in the computationally efficient implementation of \texttt{CVC.jl}.

\section*{Conflict of interest}
None declared.

\bibliographystyle{myapalike}
\bibliography{references}

\begin{table}[H]
\caption{Comparison of different methods for total heritability estimation ($\Delta = 0.2$, $N$ = 20000, $M$ = 20000, $C$ = 10, $K$ = 5). For BayesW, we used 5 chains with chain length of 3,000, burn in of 1000, and thinning of 5 for each run. Posterior mean was used as the point estimate. Relative bias is computed using the formula $(\hat{h}^2_{\text{SNP}} - h^2_{\text{SNP}}) / h^2_{\text{SNP}} \times 100$. MSE: Mean Squared Error; MAE: Mean Absolute Error. Values displayed as $<0.001$ indicate positive values smaller than the respective threshold.}
\label{tab:correct_spec_h2snp_comp}
\centering
\resizebox{\ifdim\width>\linewidth\linewidth\else\width\fi}{!}{
\begin{tabular}[t]{ccrrrrrrrrr}
\toprule
 &  & \multicolumn{3}{c}{Normal} & \multicolumn{3}{c}{Gumbel} & \multicolumn{3}{c}{Logistic} \\
\cmidrule(l{3pt}r{3pt}){3-5} \cmidrule(l{3pt}r{3pt}){6-8} \cmidrule(l{3pt}r{3pt}){9-11}
$h^2_{\text{SNP}}$ & Model & Rel. Bias (\%) & MSE & MAE & Rel. Bias (\%) & MSE & MAE & Rel. Bias (\%) & MSE & MAE\\
\midrule
\multirow{3}{*}{0.2} & CVC & 0.20 & $<0.001$ & 0.014 & 1.66 & $<0.001$ & 0.017 & 0.83 & $<0.001$ & 0.012\\
 & BayesW & 59.41 & 0.014 & 0.119 & 116.52 & 0.054 & 0.233 & 65.02 & 0.017 & 0.130\\
 & LT & 57.76 & 0.014 & 0.116 & 81.61 & 0.027 & 0.163 & 61.11 & 0.015 & 0.122\\ \midrule[1pt]
\multirow{3}{*}{0.5} & CVC & 0.15 & $<0.001$ & 0.015 & -0.39 & $<0.001$ & 0.015 & -0.24 & $<0.001$ & 0.016\\
 & BayesW & 6.75 & 0.001 & 0.034 & 18.54 & 0.009 & 0.093 & 7.40 & 0.001 & 0.037\\
 & LT & 56.58 & 0.081 & 0.283 & 66.02 & 0.110 & 0.330 & 56.49 & 0.081 & 0.282\\ \midrule[1pt]
\multirow{3}{*}{0.8} & CVC & -0.15 & $<0.001$ & 0.021 & -0.12 & $<0.001$ & 0.022 & -0.06 & $<0.001$ & 0.022\\
 & BayesW & -1.47 & $<0.001$ & 0.012 & 1.02 & $<0.001$ & 0.008 & -1.25 & $<0.001$ & 0.010\\
 & LT & 28.56 & 0.053 & 0.228 & 29.67 & 0.057 & 0.237 & 27.82 & 0.051 & 0.223\\
\bottomrule
\end{tabular}}
\end{table}

\begin{table}[H]
  \caption{Estimated heritability of UKB age-at-diagnosis traits ($K = 24$, $J = 100$, $B = 10$). The code indicates the UKB Data ID or Phecode. The models were  adjusted for sex, birth year, assessment center and 20 genetic PCs. The sex covariate was excluded for female-only traits. CR: censoring rate; JSE: jackknife standard error.}
  \label{tab:ukb_age_at_dx}
  \centering
  \adjustbox{max width=\textwidth,center}{%
  \begin{tabular}{@{}r l r r r c c c@{}}
    \toprule
    \textbf{Code} 
      & \textbf{Description}   
      & \textbf{N}
      & \textbf{M}
      & \textbf{CR}
      & \textbf{Model}
      & $\boldsymbol{\hat{h}_\mathrm{SNP}^2}$ 
      & \textbf{JSE}  \\
    \midrule
    \multirow{2}{*}{2217}       
      & \multirow{2}{*}{Started wearing glasses or contact lenses} 
      & \multirow{2}{*}{276,169} 
      & \multirow{2}{*}{592,454}
      & \multirow{2}{*}{0.13} 
      & CVC & 0.101 & 0.004 \\
      &  
      & & & & LT  & 0.249 & 0.008 \\
      \cmidrule(lr){1-8}
    \multirow{2}{*}{2754}       
      & \multirow{2}{*}{First live birth} 
      & \multirow{2}{*}{147,224} 
      & \multirow{2}{*}{592,454}
      & \multirow{2}{*}{0.32} 
      & CVC & 0.192 & 0.057 \\
      & 
      & & & & LT  & 0.143 & 0.008 \\
      \cmidrule(lr){1-8}
    \multirow{2}{*}{401}        
      & \multirow{2}{*}{Hypertension} 
      & \multirow{2}{*}{239,151} 
      & \multirow{2}{*}{592,454}
      & \multirow{2}{*}{0.67} 
      & CVC & -0.018 & 0.015 \\
      & 
      & & & & LT  & -0.008 & 0.004 \\
      \cmidrule(lr){1-8}
    \multirow{2}{*}{3436, 2867} 
      & \multirow{2}{*}{Started smoking in current and former smokers} 
      & \multirow{2}{*}{276,169} 
      & \multirow{2}{*}{592,454}
      & \multirow{2}{*}{0.68} 
      & CVC & 0.301 & 0.050 \\
      & 
      & & & & LT  & 0.225 & 0.007 \\
    \bottomrule
  \end{tabular}%
  }
\end{table}

%% file: supp_content.tex
\clearpage
\begin{center}
    {\LARGE Supplementary Materials for ``Estimating heritability of survival traits using censored multiple variance component model"}
\end{center}
\vspace{1em}

\setcounter{section}{0}
\setcounter{table}{0}
\setcounter{figure}{0}
\setcounter{equation}{0}
\renewcommand{\thesection}{S.\arabic{section}}
\renewcommand{\thetable}{S.\arabic{table}}
\renewcommand{\thefigure}{S.\arabic{figure}}
\renewcommand{\theequation}{S.\arabic{equation}}
\renewcommand{\theHsection}{supp.\arabic{section}}
\renewcommand{\theHtable}{supp.\arabic{table}}
\renewcommand{\theHfigure}{supp.\arabic{figure}}
\renewcommand{\theHequation}{supp.\arabic{equation}}

\section{Heritability in observed time scale}\label{sec:h2_obs_scale}
We show that the heritability in the observed scale, $h^2_o$, is approximately equal to the heritability in the log-time scale, $h^2$. For the $i$th subject the LMM looks like
$$
y_i = \log T_i = \bw_i\tp\balpha + \sum_{k = 1}^K \bx_{ik}\tp\bbeta_k + \epsilon_i
$$ with $\bbeta_k \sim \Dcal(\bzero,\frac{\sigma_k^2}{M_k}\bI_{M_k})$, $\bepsilon \sim (0, \sigma_e^2\bI)$ and $\bbeta_1,...,\bbeta_K, \bepsilon$ are independent. Let $\mu_i = \bw_i\tp\balpha$ and $g_i = \sum_k \bx_{ik}\tp\bbeta_k$. Then,
$$
\V g_i = \sum_k \sigma_k^2(\bK_k)_{ii} \approx \sum_k\sigma_k^2 =: \sigma_{g}^2.
$$ assuming the estimated genetic similarity of a subject to himself, $(\bK_k)_{ii}$, is close to one. We have defined the total heritability in log-time scale to be
$$
    h^2 = \frac{\sigma_g^2}{\sigma_g^2 + \sigma_e^2}.
$$ The total variance of time-to-event phenotype of the $i$th subject is given by
$$
\V T_i = \V\exp(\mu_i + g_i + \epsilon_i) = \{\exp(\mu_i)\}^2 \V\exp(g_i + \epsilon_i) =: \gamma_i^2\V\exp(g_i + \epsilon_i).
$$ Similar to the $h^2_o$ defined in \cite{ducrocq_two_1999}, we define
$$
    h^2_o = \frac{\V (\gamma_i\exp g_i)}{\V\{\gamma_i\exp(g_i + \epsilon_i)\}} = \frac{\V\exp g_i}{\V\exp(g_i + \epsilon_i)}.
$$ For a random variable $X$ with $\E X = 0$, the first-order Taylor approximation of $\exp X$ about $\E X = 0$ is
$$
    \exp X \approx 1 + X \implies \V \exp X \approx \V X.
$$ Thus, 
$$
h_o^2 \approx \frac{\V g_i}{\V (g_i + \epsilon_i)} = \frac{\sigma_g^2}{\sigma_g^2 + \sigma_e^2} = h^2.
$$ 
Similarly, if we define $g_{ik} = \bx_{ik}\tp\bbeta_k$, then the heritability associated with the $k$th genotype partition in the observed scale can be defined as
$$
h_{ok}^2 = \frac{\V \exp g_{ik}}{ \V \exp(g_i + \epsilon_i)} \approx \frac{\sigma_k^2}{\sigma_g^2 + \sigma_e^2} = h^2_k.
$$

\section{Derivations of first-moment, second-moment and mixed-moment Leurgans synthetic variables}\label{sec:synvar_derivation}

Using Zheng's synthetic variable form \citep{Zheng87PseudoVariableCensor}, we define $\ystar_{i1} = \delta_i\varphi_1(u_i) + (1-\delta_i)\varphi_2(u_i)$. We wish to find $\varphi_i$'s such that $\E \ystar_{i1} = \E y_i.$ The condition that for all $y_i$
\begin{align*}
    y_i = \E(\ystar_{i1}\mid y_i) &= \E_{r_i\sim G}[\delta \varphi_1(u_i) + (1 - \delta)\varphi_2(u_i)\mid y_i]\\
    &= \int_{r_i \leq y_i} \delta \varphi_1(u_i) + (1 - \delta)\varphi_2(u_i) dG(r_i)\\ 
    &\quad + \int_{r_i > y_i} \delta \varphi_1(u_i) + (1 - \delta)\varphi_2(u_i) dG(r_i)\\
    &= \int_{r_i \leq y_i} \varphi_2(r_i)dG(r_i) + \int_{r_i > y_i} \varphi_1(y_i) dG(r_i)\\
    &= \int_{r_i \leq y_i} \varphi_2(r_i)dG(r_i) + \varphi_1(y_i)[1 - G(y_i)]
\end{align*} ensures the unbiasedness $\E\ystar_1 = \E\E(\ystar_1 \mid y_i) = \E y_i.$ Assuming $G$ is continuous with density g and differentiability of $\varphi_1$, the above condition can be written as a differential equation
\[
    \varphi_2(y_i)g(y_i) -\varphi_1(y_i)g(y_i) + \varphi_1'(y_i)[1 - G(y_i)] = 1.
\]
Suppose $\varphi = \varphi_1 = \varphi_2.$ Then,
$$
[1 - G(y_i)]\varphi'(y_i) = 1 \implies \varphi'(y_i) = \frac{1}{1 - G(y_i)}\quad\forall y_i.
$$ If
$$
\varphi(y_i) = y_i + \int_{-\infty}^{y_i} \frac{G(t)}{1 - G(t)}dt,
$$ then
$$
\varphi'(y_i) = \frac{1}{1 - G(y_i)}.
$$ Thus, $\varphi(y_i)$ satisfies the differential equation. Thus,
$$
\ystar_{i1} = \varphi(u_i) = u_i + \int_{-\infty}^{u_i} \frac{G(t)}{1 - G(t)} dt.
$$ We require $F\inv(1) < G\inv(1)$ so that $\frac{1}{1 - G(y_i)}$ is finite where $F$ is the cumulative distribution function of $y_i.$

Next, to derive the second-moment synthetic variable, define $\ystar_{i2} = \delta_i\phi_1(u_i) + (1 - \delta_i) \phi_2(u_i)$. Using the same technique used for deriving the first moment synthetic variable, we can derive
$$
\E(\ystar_{i2}\mid y_i) = \phi_1(y_i)[1 - G(y_i)] + \int_{-\infty}^{y_i}\phi_2(s) dG(s).
$$ We want for all $y_i$,
$$
y_i^2 = \E(\ystar_{i2}\mid y_i) = \phi_1(y_i)[1 - G(y_i)] + \int_{-\infty}^{y_i}\phi_2(s) dG(s).
$$ Assuming $G$ is continuous with density $g$, and $\phi_1$ is differentiable, we obtain a differential equation,
$$
\phi_1'(y_i)[1 - G(y_i)] - \phi_1(y_i)g(y_i) + \phi_2(y_i)g(y_i) = 2y_i.
$$
Suppose $\phi = \phi_1 = \phi_2$. Then,
$$
\phi'(y_i) = \frac{2y_i}{1 - G(y_i)}.
$$ Let $\phi(y_i) =  y_i^2 + \int_{-\infty}^{y_i} \frac{2t G(t)}{1 - G(t)} dt$. Then,
$$
\phi'(y_i) = \frac{2y_i}{1 - G(y_i)}.
$$ We can see $\phi(y_i)$ satisfies the differential equation. Hence,
$$
    \ystar_{i2} = \phi_1(u_i) = \int_{-\infty}^{u_i} \frac{2t}{1 - G(t)} dt = u_i^2 + \int_{-\infty}^{u_i} \frac{2tG(t)}{1 - G(t)}dt,
$$ where we again require $F\inv(1) < G\inv(1)$.

Lastly, we derive mixed-moment Leurgans synthetic variable. The goal is to derive synthetic variable $(y_iy_j)^*$ such that $\E(y_iy_j)^* = \E y_iy_j$. We claim that the mixed-moment synthetic variable is the product of first-moment syntehtic variables:
\begin{equation}\label{mixed_synvar}
\begin{aligned}
    \E y_{i1}^*y_{j1}^* &= \mathbb{E}\E(y_{i1}^*y_{j1}^*|y_i,y_j)\\
    &= \mathbb{E}[\E(y_{i1}^*|y_i)\E(y_{j1}^*|y_j)]\\
    &= \E y_iy_j.
\end{aligned} 
\end{equation} The first equality used the law of iterated expectations. The second equality is true because
\begin{align*}
    &\E(y_i^*y_j^*\mid y_i, y_j)\\
    =&\mathbb{E}_{r_i \sim G, r_j \sim G}[\delta_i\delta_j \varphi_1(u_i)\varphi_1(u_j)\\ 
    &+ \delta_i(1-\delta_j)\varphi_1(u_i)\varphi_2(u_j)\\
    &+ (1-\delta_i)\delta_j\varphi_2(u_i)\varphi_1(u_j)\\
    &+ (1-\delta_i)(1-\delta_j)\varphi_2(u_i)\varphi_2(u_j)\mid y_i, y_j]\\
    =& \int_{y_i}^{\infty}\int_{y_j}^{\infty} \varphi_1(y_i)\varphi_2(y_j)dG(t_j)dG(t_i)\\
    & + \int_{y_i}^{\infty}\int_{-\infty}^{y_j}\varphi_1(y_i)\varphi_2(t_j)dG(t_j)dG(t_i)\\
    & + \int_{-\infty}^{y_i}\int_{y_j}^{\infty} \varphi_1(t_i)\varphi_2(y_j) dG(t_j)dG(t_i)\\
    & + \int_{-\infty}^{y_i}\int_{-\infty}^{y_j} \varphi_1(t_i)\varphi_2(t_j) dG(t_j)dG(t_i)
\end{align*} using the assumption $r_i \indep r_j.$ We know the conditional expectation of $\ystar_{i1}$ given $y_i$ is of the form
$$
\E(\ystar_{i1} \mid y_i) = [1 - G(y_i)]\varphi_1(y_i) + \int_{-\infty}^{y_i} \varphi_2(t)dG(t).
$$ Put $A_i = [1 - G(y_i)]\varphi_1(y_i)$ and $B_i = \int_{-\infty}^{y_i} \varphi_2(t)dG(t).$ Then,
\begin{align*}
    \E(y_i^* y_j^* \mid y_i,y_j) &= A_iA_j + A_iB_j + B_iA_j + B_iB_j\\
    &= (A_i + B_i)(A_j + B_j)\\
    &= \E(y_i^*\mid y_i) \E(y_j^*\mid y_j).
\end{align*} The third equality in \eqref{mixed_synvar} is true by construction requiring the unbiasedness of the first-moment synthetic variables, i.e., $\E(\ystar_{i1}\mid y_i) = y_i$ for all $y_i.$

The cumulative distribution function $G$ of censoring variables $r_i$ is required to construct the synthetic variables. We estimate $G$ using the Kaplan--Meier estimator and plug it into the synthetic variable formula.

\section{Computational note}\label{sec:computational_note}
Let $\bystar_1{\tp} = (\ystar_{11},\ystar_{21},...,\ystar_{N1})$ and $\bystar_2{\tp} = (\ystar_{12},\ystar_{22},...,\ystar_{N2}).$ We can rewrite the synthetic variable matrix $\bYstar$ as 
$$
\bYstar = \diag(\bystar_2) - \diag(\bystar_1\bystar_1{\tp}) + \bystar_1\bystar_1{\tp} = \bD + \bystar_1\bystar_1{\tp},
$$ where $\bD = \diag(\bystar_2) - \diag(\bystar_1\bystar_1{\tp}).$

Recall $\bHtilde$ is the matrix whose columns are the left singular vectors of the thin singular value decomposition of $\bW$ where the size of $\bHtilde$ is $N \times \min\{N, C\}$. Recall also $\bH = \bHtilde[:,1:r]$ where $r$ is the rank of $\bW.$ Let $\bP = \bH\bH\tp$ be the orthogonal projector onto $\Rcal(\bW)$. Suppose $\bz_1,...,\bz_B \overset{\text{i.i.d.}}\sim N_N(\bzero,\bI_N).$ These $\bz_b$ will be used to obtain the unbiased randomized trace estimator of a matrix to improve computational efficiency of computing the trace from $O(N^2M)$ to $O(NM).$

Below, we summarize the terms appearing in the normal equation and define the working arrays to efficiently solve the equation. These are the terms we need to compute:
\begin{enumerate}
    \item $\tr\bYstar\bV,$
    \item $b_k = \tr\bV\bK_k$ for $k = 1,2,...,K,$
    \item $T_{k\ell} = \tr\bK_k\bV\bK_\ell\bV$ for $1 \leq k < \ell \leq K,$
    \item $c_i = \tr\bYstar\bV\bK_i\bV$,
\end{enumerate} where $\bV = \bI - \bP.$\\
\begin{enumerate}[wide]
    \item $\tr \bYstar \bV = \bone\tp\bystar_2 - \tr\bH\bH\tp\bD + \bystar_1{\tp}\bH\bH\tp\bystar_1$.    
    \item $b_k = N - \frac{1}{M_k}\bLtilde_{(k,0)}$
    where
    \begin{align*}
        \bL_{(k,j)} &:= \tr \bX_k\iter{j}{\tp} \bH\bH\tp \bX_k\iter{j} = \langle \bA_{(k,j)}, \bA_{(k,j)}\rangle,\quad \bA_{(k,j)} := \bX_k\iter{j}{\tp}\bH\\
        \bLtilde_{(k,0)} &:= \sum_{j=1}^J \bL_{(k,j)}.
    \end{align*} 
    Let $\bLtilde_{(k,j)} := \bLtilde_{(k,0)} - \bL_{(k,j)}$. Then,
    \begin{align*}
        b_k\iter{-j} = \tr \bV\bK_k\iter{-j} &= N - \frac{1}{M_k\iter{-j}}\bLtilde_{(k,j)},\quad M_k\iter{-j} := M_k - M_{kj},
    \end{align*} where $M_{kj}$ is the number of columns in $\bX_{k}\iter{j}.$
    \item $T_{k\ell} = \frac{1}{BM_kM_\ell} \sum_b(\bU_{(k,0,b)} - \bUtilde_{(k,0,b)})\tp (\bU_{(\ell,0,b)} - \bH\bH\tp \bU_{(\ell,0,b)})$ where
    \begin{align*}
        &\bU_{(k,0,b)} = \sum_{j=1}^J \bZ_{(k,j,b)},\quad \bZ_{(k,j,b)} = \bX_k\iter{j}\bX_k\iter{j}{\tp}\bz_b\\
        &\bUtilde_{(k,0,b)} = \sum_{j=1}^J \bZtilde_{(k,j,b)},\quad \bZtilde_{(k,j,b)} = \bX_k\iter{j}\bX_k\iter{j}{\tp}\bztilde_b, \quad \bztilde_b = \bH\bH\tp\bz_b.
    \end{align*} 
    Let 
    \begin{align*}
        \bU_{(k,j,b)} &= \bU_{(k,0,b)} - \bZ_{(k,j,b)}\\
        \bUtilde_{(k,j,b)} &= \bUtilde_{(k,0,b)} - \bZtilde_{(k,j,b)}.
    \end{align*} Then,
    $$
    T_{k\ell}\iter{-j} = \hat{\tr}\bV\bK_k\iter{-j}\bV\bK_\ell\iter{-j} = \frac{1}{BM_k\iter{-j}M_\ell\iter{-j}}\sum_b (\bU_{(k,j,b)} - \bUtilde_{(k,j,b)})\tp(\bU_{(\ell,j,b)} - \bH\bH\tp\bU_{(\ell,j,b)}).
    $$ Now, letting 
    \begin{align*}
        &\bU_{(k,0)} = \sum_{j = 1}^J \bZ_{(k,j)},\quad \bZ_{(k,j)} = \bX_k\iter{j}\bX_k\iter{j}{\tp}\bZ,\quad \bZ = \bmat{\bz_1 & \cdots & \bz_B}\\
        &\bUtilde_{(k,0)} = \sum_{j = 1}^J \bZtilde_{(k,j)},\quad \bZtilde_{(k,j)} = \bX_k\iter{j}\bX_k\iter{j}{\tp}\bZtilde,\quad \bZtilde = \bmat{\bztilde_1 & \cdots & \bztilde_B}\\
        &\bU_{(k,j)} = \bU_{(k,0)} - \bZ_{(k,j)}\\
        &\bUtilde_{(k,j)} = \bUtilde_{(k,0)} - \bZtilde_{(k,j)},
    \end{align*} we have
    \begin{align*}
        T_{k\ell} &= \frac{1}{BM_kM_\ell}\langle\bU_{(k,0)} - \bUtilde_{(k,0)}, \bU_{(\ell,0)} - \bH\bH\tp\bU_{(\ell,0)}\rangle\\
        T_{k\ell}\iter{-j} &= \frac{1}{BM_k\iter{-j}M_\ell\iter{-j}}\langle\bU_{(k,j)} - \bUtilde_{(k,j)}, \bU_{(\ell,j)} - \bH\bH\tp\bU_{(\ell,j)}\rangle
    \end{align*}

    \item $c_i = \tr\bYstar\bV\bK_i\bV = \tr\bK_k\bYstar - 2\tr\bK_k \bP\bYstar + \tr \bP\bK_k\bP\bYstar$:
    \begin{enumerate}
        \item $\tr\bK_k\bP\bYstar = \frac{1}{M_k} [\bRtilde_{(k,0)} + \bQtilde\tp_{(k,0)}(\bH\bH\tp\bystar_1)]$
        where
        \begin{align*}
            \bR_{(k,j)} &= \langle\bA_{(k,j)}, \bAtilde_{(k,j)}\rangle,\quad \bAtilde_{(k,j)} = \bX_k\iter{j}{\tp}\bHbar,\quad \bHbar = \bD\bH,\\
            \bRtilde_{(k,0)} &= \sum_{j=1}^J \bR_{(k,j)}\\
            \bQ_{(k,j)} &= \bX_k\iter{j}\bX_k\iter{j}{\tp}\bystar_1\\
            \bQtilde_{(k,0)} &= \sum_{j=1}^J \bQ_{(k,j)}.
        \end{align*} 
        Let 
        $$
        \bRtilde_{(k,j)} = \bRtilde_{(k,0)} - \bR_{(k,j)},\quad \bQtilde_{(k,j)} = \bQtilde_{(k,0)} - \bQ_{(k,j)}.
        $$ Then,
        $$
            \tr\bK_k\iter{-j}\bP\bYstar = \frac{1}{M_k\iter{-j}}[\bRtilde_{(k,j)} + \bQtilde\tp_{(k,j)}(\bH\bH\tp\bystar_1)].
        $$
        \item $\tr\bK_k\bYstar = \frac{1}{M_k}[\bystar_1{\tp}\bQtilde_{(k,0)} + \bFtilde_{(k,0)}]$ where
        $$
        \bFtilde_{(k,0)} = \sum_{j=1}^J \bF_{(k,j)},\quad \bF_{(k,j)} = \langle\bX_k\iter{j}, \bD\bX_k\iter{j}\rangle.
        $$ Let $\bFtilde_{k,j} = \bFtilde_{(k,0)} - \bF_{(k,j)}$. Then,
        $$
            \tr\bK_k\iter{-j}\bYstar = \frac{1}{M_k}[\bystar_1{\tp}\bQtilde_{(k,j)} + \bFtilde_{(k,j)}].
        $$

        \item $\tr\bP\bK_k\bP\bYstar = \frac{1}{M_k}\tr\bP\bX_k\bX_k\tp\bP\bD + \frac{1}{M_k}\bystar_1{\tp}\bP\bX_k\bX_k\tp\bP\bystar_1$:
        \begin{enumerate}
            \item $\tr\bP\bX_k\bX_k\tp\bP\bD = \frac{1}{B}\sum_b \bUtilde_{(k,0,b)}\tp\tilde{\bztilde}_b,\quad \tilde{\bztilde}_b = \bH\bHbar\tp\bz_b$, and
            $$
            \hat{\tr}\bP\bX_k\iter{-j}\bX_k\iter{-j}{\tp}\bP\bD = \frac{1}{B}\sum_b \bUtilde\tp_{(k,j,b)}\tilde{\bztilde}_b.
            $$
            \item $\bystar_1{\tp}\bP\bX_k\bX_k\tp\bP\bystar_1 = \bVtilde_{(k,0)}$ where
            $$
            \bV_{(k,j)} = ||\bX_k\iter{j}{\tp}(\bH\bH\tp\bystar_1)||_2^2, \quad \bVtilde_{(k,0)} = \sum_{j=1}^J \bV_{(k,j)}.
            $$ Let $\bVtilde_{(k,j)} = \bVtilde_{(k,0)}- \bV_{(k,j)}$. Then,
            $$
            \bystar_1{\tp}\bP\tp\bX_k\iter{-j}\bX_k\iter{-j}{\tp}\bP\bystar_1 = \bVtilde_{(k,j)}.
            $$
        \end{enumerate}
        Therefore,
        \begin{align*}
            \tr\bP\bK_k\bP\bYstar &= \frac{1}{M_k}\left[\frac{1}{B}\sum_b\bUtilde_{(0,b,k)}\tp \tilde{\bztilde}_b + \bVtilde_{(k,0)}\right],\\
            \tr\bP\bK_k\iter{-j}\bP\bYstar &= \frac{1}{M_k\iter{-j}}\left[\frac{1}{B}\sum_b\bUtilde_{(j,b,k)}\tp \tilde{\bztilde}_b + \bVtilde_{(k,j)}\right].
        \end{align*}
    \end{enumerate}
\end{enumerate}

\section{Correctly specified model}\label{sec:correct_spec}
We assess the performance of the CVC model in estimating the total heritability $h_{\text{SNP}}^2$ and the partitioned heritability $h_k^2$ for the $k$th partition when the model is correctly specified. We used the following generative model to simulate uncensored survival traits on the log scale:
\begin{align*}
    \bbeta_k &\sim N\left(\bzero, \frac{\sigma_k^2}{M_k}\bI_{M_k}\right),\\
    \by &\sim N\left(\bW\balpha + \sum_k \bX_k\bbeta_k, \sigma_e^2\bI\right),
\end{align*} where $\bW$ is a randomly generated covariate matrix of size $N \times C$ (including an intercept column of ones), $\bX_k$ is the $k$th partitioned genotype matrix of size $N \times M_k$, and $\sigma_e^2$ is the environmental variance component. The genomic variance components $\sigma_k^2$ were sampled uniformly from the interval $(0,1).$  The environmental variance component $\sigma_e^2$ was set to $\left(\frac{1}{h^2} - 1\right)\sum_k \sigma_k^2$ with the total heritability $h_{\text{SNP}}^2 = 0.5$. The fixed effects $\alpha_1$,...,$\alpha_C$ were uniformly sampled from the interval $(0,1)$. 

We obtained $\bX_k$ by partitioning standardized genotype matrix $\bX$ into $K$ partitions. Each SNP in the $k$th partition shared the same associated variance component $\sigma_k^2 / M_k$. The standardized genotype matrix $\bX$ is the standardized form of the genotype matrix $\bG = \bmat{\bg_1 & \cdots & \bg_M}$. The $(i,j)$-th entry of $\bX$ is defined as
\[
    x_{ij} = \frac{1}{N}(g_{ij} - \gbar_j)^2,\quad \gbar_j = \frac{1}{N}\sum_{i = 1}^N g_{ij}. 
\] The genotype matrix $\bG = (g_{ij})$ was generated such that
\[
     g_{1j},...,g_{Nj} \overset{\text{i.i.d.}}{\sim}\text{Bin}(2, f_j)
\] with a first-order autoregressive correlation structure among the SNPs, parametrized by $\rho = 0.1$~\citep{jiang_set_2021}. The MAF $f_j$ was generated as $f_j = 0.01 + 0.49 \times \text{Beta}(0.2, 0.8).$ 

To generate right-censored outcomes $u_i = \min(y_i, r_i)$ for $i = 1,...,N$ with a desired censoring rate $\Delta \in \{0.2, 0.5, 0.8\}$, we simulated censoring variables $r_i = r_i(\mu_c)$ from
$$
        c_1,...,c_N \overset{\text{i.i.d.}}{\sim}  \mu_c + \sigma_e N(0,1),
$$ where $\mu_c$ is the solution to the equation
$$
            \frac{\sum_{i = 1}^N \I_{\{y_i > r_i(\mu_c)\}}}{N}  - \Delta = 0.
$$
We set the number of covariates to $C = 10$, the number of jackknife blocks to $J = 100$, and the number of random Gaussian vectors to $B = 10$. The number of Monte Carlo replicates was fixed at 100 for all simulations.

\subsection{Impact of sample size}
\begin{figure}[H]
    \centering\includegraphics[width=0.85\textwidth]{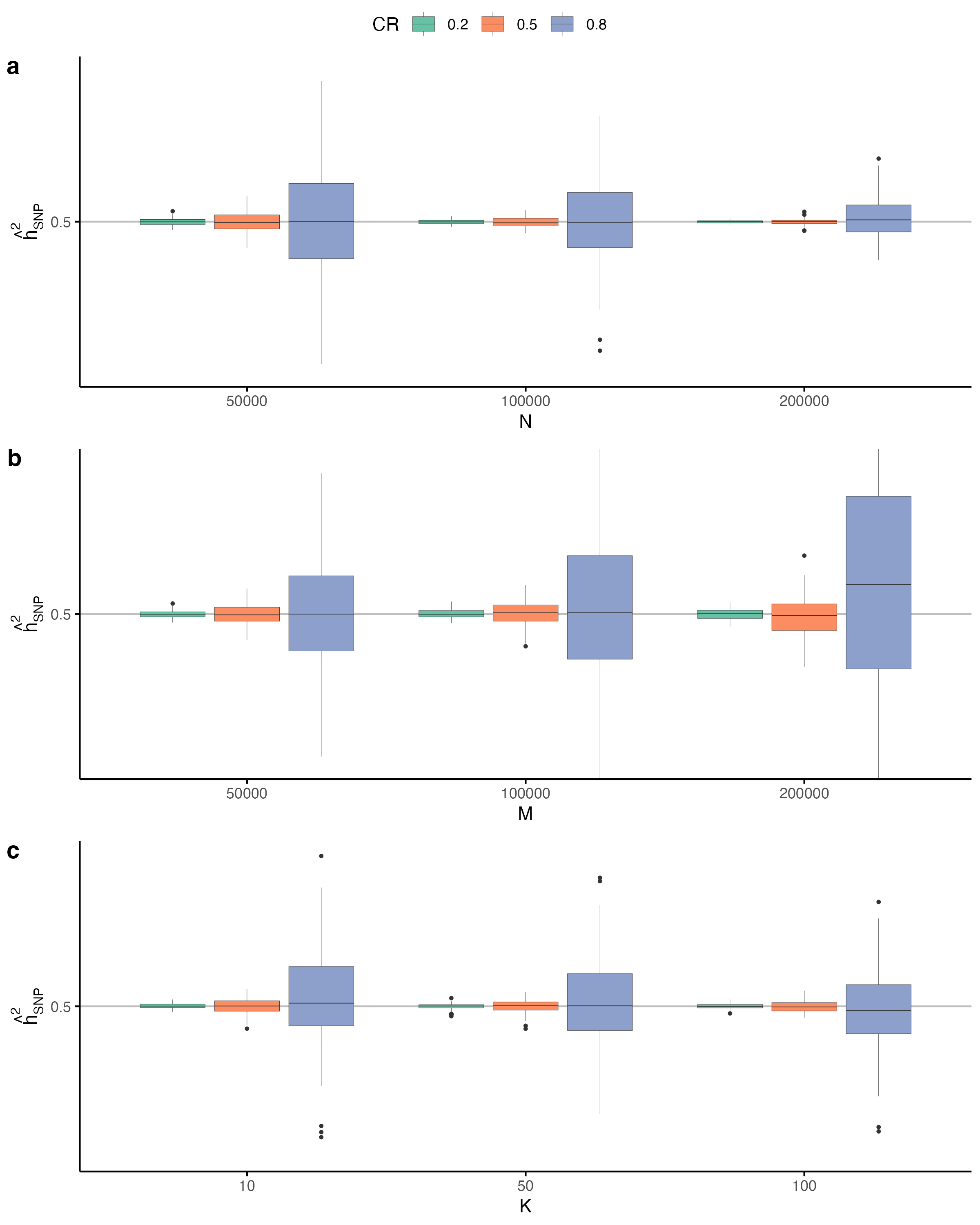}
    \caption{Impact of sample size, number of SNPs, and number of partitions on accuracy of total heritability estimation under correct model specifications over 100 runs ($h^2_{\text{SNP}} = 0.5$, $C = 10$, $J = 100$, $B = 10$). \textbf{a} The impact of sample size $N$ when $M = 50,000$ and $K = 10$. \textbf{b} The impact of number of SNPs when $N = 50,000$ and $K = 10$. \textbf{c} The impact of number of partitions $K$ when $N = M = 100,000$.}
    \label{fig:all_impact_correct_spec}
\end{figure}

We evaluate the impact of sample size on the CVC model's performance on estimating total heritability $h^2_{\text{SNP}} = 0.5$ (Figure~\ref{fig:all_impact_correct_spec}a). Simulations were conducted with $N \in \{50000, 100000, 200000\}$, $M = 50,000$ and $K = 10$. Table~\ref{tab:N_impact_h2_0.5_M_50000_K_10} summarizes results.

The relative bias of total heritability estimates does not necessarily decrease as the sample size increases. While the magnitude of relative bias decreases as sample size increases for $\Delta = 0.2$, it increases for $\Delta = 0.2$ as sample size increases. as As $N$ increases from $50,000$ to $200,000$, the absolute relative bias decreases from $0.100\%$ to $0.052\%$ for $\Delta 0.2$, whereas the absolute relative bias increases from $0.451\%$ to $1.940\%$ for $\Delta = 0.8$. For $\Delta = 0.5$, the absolute relative bias is highest at $N = 100,000$, while it is lowest at $N = 50,000$.

For each censoring rate, the standard errors decrease as the sample size increases from $50,000$ to $200,000$; the standard error decreases from $0.012$ to $0.005$ for $\Delta = 0.2$, $0.034$ to $0.011$ for $\Delta = 0.5$, and $0.211$ to $0.068$ for $\Delta = 0.8$.

The jackknife standard error estimates show little bias and are close to the Monte Carlo standard errors. The biases tend to decrease as the sample size increases except for $\Delta = 0.2$, albeit marginal increase ($0.001$ to $0.002$ from $N = 50,000$ to $N = 200,000$).

In summary, total heritability estimates and their standard error estimates are nearly unbiased across sample sizes, with larger sample sizes generally reducing the bias of standard error estimates.

\subsection{Impact of the number of SNPs}
We evaluate the impact of number of SNPs on the CVC model's performance by varying $M$ while fixing the sample size (Figure~\ref{fig:all_impact_correct_spec}b). The simulation parameters were $N = 50,000$, $M \in \{50000, 100000, 200000\}$ and $K = 10$. Table~\ref{tab:M_impact_h2_0.5_N_50000_K_10} presents the simulation results.

Across all censoring rates, the absolute relative bias tend to increase as $M$ increases from $50,000$ to $200,000$: from $0.100\%$ to $0.103\%$ for $\Delta = 0.2$, $0.035\%$ to $1.682\%$ and $0.451\%$ to $19.850\%$.

Similarly, the standard errors of the total heritability estimates increase as the number of SNPs increases: from $0.012$ to $0.018$ for $\Delta = 0.2$, $0.034$ to $0.066$ for $\Delta = 0.5$, and $0.211$ to $0.475$ for $\Delta = 0.8$.

The absolute bias of the jackknife standard error also shows increasing pattern as the number of SNPs increases: $0.001$ to $0.002$ for $\Delta = 0.2$, $0.003$ to $0.008$ for $\Delta = 0.5$, and $0.003$ to $0.024$ for $\Delta = 0.8$.

In summary, increasing the number of SNPs reduces both the accuracy and precision of heritability estimates. A possible explanation is that as the number of SNPs grows, the per-SNP variance component becomes vanishingly small, resulting in the sampling of many near-zero genetic effect sizes. We may be observing a phenomenon analogous to that in classic linear regression, where smaller fixed-effect estimates are associated with larger standard errors.

\subsection{Impact of the number of partitions}
We assess how increasing the number of partitions influences the model's performance (Figures~\ref{fig:all_impact_correct_spec}c). The simulations were conducted with $N = 100,000$, $M = 100,000$ and $K \in \{10, 50, 100\}$. Table~\ref{tab:K_impact_h2_0.5_N_100000_M_100000} presents results. 

The relative biases of the total heritability estimates show no clear trend as the number of partitions increases from $10$ to $100$ across all censoring rates. As $K$ increases from $10$ to $100$, the magnitude of relative bias drops from $0.269\%$ to $0.031\%$ for $\Delta = 0.2$, and from $5.278\%$ to $2.685\%$ for $\Delta = 0.8$, while it increases from $0.043\%$ to $0.201\%$ at $\Delta = 0.5$.

The standard errors of the total heritability estimates are nearly invariant to changes in the number of partitions. It ranges from $0.009$ to $0.011$ for $\Delta = 0.2$, $0.019$ to $0.027$ for $\Delta = 0.5$, and $0.143$ to $0.178$ for $\Delta = 0.8$.

The bias of the jackknife standard error estimates is also largely invariant to changes in the number of partitions. It ranges from $-0.002$ to $0.000$ for $\Delta = 0.2$, $-0.003$ to $0.002$ for $\Delta = 0.5$, and $-0.012$ to $-0.004$ for $\Delta = 0.8$.

Overall, the number of partitions has relatively little effect on the standard errors of the total heritability or on the biases of the jackknife standard error estimates, while the magnitude of relative bias tends to decrease with an increasing number of partitions.  

\subsection{Impact of number of random Gaussian vectors and jackknife blocks}\label{sec:impact_B_J}
In all analyses we set the number of independent random Gaussian vectors to $B = 10$ and the number of jackknife blocks to $J = 100$, following previous work~\citep{wu_scalable_2018, pazokitoroudi_efficient_2020}. Simulation results suggest that these choices are sufficient to obtain accurate point estimates of heritability and their standard error estimates across a range of scenarios. 

Nevertheless, to assess sensitivity to these tuning parameters, we conducted additional simulations under correctly specified model settings (Tables~\ref{tab:impact_B} and~\ref{tab:impact_J}). The experiments were conducted with $N = 10,000$, $M = 10,000$, $C = 10$, and $K = 10$. We fixed $J = 100$ for sensitivity analysis of $B$, and fixed $B = 10$ for the sensitivity analysis of $J$.

As $B$ increases ($B \in \{1, 5, 10, 50, 100\}$), the Monte Carlo standard error of the total heritability estimates and their standard error estimates decrease only modestly, whereas the relative bias remains largely unchanged across values of $B$ within each censoring rate ($\Delta \in \{0.2, 0.5, 0.8\}$). Importantly, the stability of both point estimates and their Monte Carlo standard errors for $B \geq 10$ indicates that the additional variability induced by the randomized trace estimator is negligible relative to the overall sampling variability of the heritability estimator in our setting. Similarly, as $J$ increases ($J \in \{2, 10, 100, 500\}$), the variability of the jackknife standard error estimates decreases, while their bias shows little variation for $J \geq 10$.

We note that the weak empirical dependence of heritability accuracy on $B$ observed in our simulations is more favorable than conservative worst-case theory would suggest. Because our estimator depends on randomized trace approximations, one might expect $B$ to have a stronger impact on downstream accuracy. In particular, existing results for randomized trace estimators suggest that $B = O(\log(1/\delta)/\epsilon^2)$ is sufficient to ensure that the ratio between the randomized trace estimator and the true trace lies within an $\epsilon$-neighborhood of $1$ with probability at least $1 - \delta$~\citep{meyer_hutch++:_2020}. For a fixed failure probability upper bound $\delta$, this implies a requirement of $B = O(1/\epsilon^2)$, which may be substantial when high accuracy is desired. We conjecture that the weak sensitivity we observe reflects the non-worst-case spectral properties of standardized genotype matrices (and functions thereof), which may yield much smaller trace-estimation error than general worst-case bounds predict~\citep{patel_scientific_2025}.

\subsection{Accuracy of partitioned heritability estimates}
Finally, we assess the CVC model's performance on estimating partitioned heritability. The simulation settings were $N = 100,000$, $M = 100,000$ and $K = 10$. Table~\ref{tab:correct_spec_partitioned_h2_0.5} presents the results.

For low to moderate censoring rates ($\Delta =  0.2, 0.5$), the CVC model accurately estimates partitioned heritability across all bins. However, at a high censoring rate ($\Delta = 0.8$), the accuracy deteriorates: the bias ranges from -0.0013 to 0.0006 for $\Delta = 0.2$, -0.0009 to 0.0013 for $\Delta = 0.5$, and -0.0264 to 0.0207 for $\Delta = 0.8$. 

Similarly, the jackknife standard error estimates are more accurate for low and moderate censoring rates compared to the high censoring rate across all bins: the bias ranges from -0.0003 to 0.0002 for $\Delta = 0.2$, -0.0027 to -0.0010 for $\Delta = 0.5$, and -0.0166 to 0.0031 for $\Delta = 0.8$. 

These results show that the CVC model can accurately estimate the partitioned heritability for traits with low to moderate censoring rates, but that accuracy degrades at higher censoring rates.

\section{Impact of conditional independence assumption on total heritability estimates}\label{sec:conditional_independence}
We investigate how the violation of marginal independence assumption arising from conditional independence assumption impacts the relative bias and standard error estimates of total heritability. We assume the following model that specifies conditional independence between $\by$ and $\br$ given $\bW$:
\begin{align*}
    \by &\sim \Ncal\left(\bW\balpha + \sum_k \bX_k\bbeta_k, \sigma_e^2\bI\right),\quad\bbeta_k \sim N\left(\bzero, \frac{\sigma_k^2}{M_k}\bI_{M_k}\right),\\
    \br &\sim \Ncal(\bone\mu + \xi\bW\balpha, s^2\bI),\quad
    s = \sqrt{\frac{1}{N - 1}\sum_{i = 1}^N (y_i - \ybar)^2},
\end{align*} where $\bW$ is a randomly generated covariate matrix of size $N \times C$ (including an intercept column of ones) and $\bX_k$ is the $k$th partitioned genotype matrix of size $N \times M_k$. The genetic variance components $\sigma_k^2$, the environmental variance component $\sigma_e^2$, and the fixed effects $\alpha_1$,...,$\alpha_C$ were generated following the same specifications in \ref{sec:correct_spec}. A shift parameter $\mu$ was chosen to achieve a target censoring rate, and values of $\xi \in \{-1, -0.5, 0, 0.5, 1\}$ were used to induce varying degrees of marginal correlation between $y_i$ and $r_i$. We considered censoring rates $\Delta \in \{0.2, 0.5, 0.8\}$ with $h^2_{\mathrm{SNP}} = 0.5$, using 100 replicates per setting.

The results are summarized in Table~\ref{tab:conditional_independence}. The average empirical marginal correlation $\rho$ between $y_i$ and $r_i$ ranged from $-0.17$ to $0.17$ across all settings. As the magnitude of this correlation increased, the relative bias of total heritability estimates also increased, with negative correlation generally inducing smaller bias than positive correlation. For a fixed correlation level, the absolute bias increased with the censoring rate. Relative bias ranged from $-4.76\%$ to $11.13\%$ for $\Delta=0.2$, $-10.69\%$ to $114.43\%$ for $\Delta=0.5$, and $-177.58\%$ to $73.48\%$ for $\Delta=0.8$. At low censoring ($\Delta=0.2$), total heritability estimates remained reasonably accurate, whereas at higher censoring rates even modest dependence led to substantial bias. 

\section{Criteria and UKB Field ID}\label{sec:ukb_field_id}
The following \href{https://biobank.ctsu.ox.ac.uk/ukb/search.cgi}{UKB data fields} were used to apply the criteria and curate the cohort used in our data analyses. The Axiom array SNPs were chosen using the UKB resource 1955.
\begin{table}[H]
\centering
\begin{tabular}{ll}
\hline
\textbf{Criteria} & \textbf{UKB Field ID} \\ \hline
$>$3rd degree relatives                               & 22021 \\
Self-reported British white ancestry                & 22006 \\
Outliers for genotype heterozygosity or missingness & 22027 \\
Genetic sex                                         & 22001 \\
Self-reported sex                                   & 31    \\
Putative sex chromosome aneuploidy                  & 22019 \\
Birth year                                          & 34    \\
Assessment center                                   & 54    \\
Top 20 PCs                                          & 22009 \\ \hline
\end{tabular}
\caption{Criteria and UK Biobank Field IDs}
\label{tab:ukb-fields}
\end{table}

\section{Survival trait phenotyping definition}\label{sec:pheno_def}
PheCode-based survival traits were used following the PheCode and ICD9/1CD10 mapping file provided in \cite{dey_efficient_2022}. We also followed their phenotyping algorithm to create each survival trait:
\begin{itemize}
    \item An event is censored ($\delta = 0$) if a subject does not have a PheCode.
    \item An event is uncensored ($\delta = 1$) if a subject has a PheCode.
    \item Censoring time is defined as $t - \text{date of birth}$ where $t$ is the earliest date among the last non-imaging visit to any of hte UKB assessment centers, the last time any ICD code associated with the PheCode was recorded, the time of death (if death was recorded).
    \item Uncensored time is defined as $s - \text{date of birth}$ where $s$ is the earliest recorded date among any of PheCode-specific ICD codes.
    \item The date of birth was rounded to the nearest full month.
\end{itemize} 
In addition to PheCode-based survival traits, we curate UKB Field ID based survival traits as shown in Table \ref{tab:ukb-field-traits}. These traits were selected for their lower censoring rates compared to PheCode-based traits. The phenotyping algorithm for each trait looks like
\begin{itemize}
    \item An event is censored if a subject does not have any record among all visits (also known as instances in UKB terminology) corresponding to the Field ID.
    \item An event is uncensored if a subject has a record of visit corresponding to the Field ID.
    \item Censoring time is defined as $\text{the last assessment center visit} - \text{date of birth}$
    \item Uncensored time is defined as the earliest assessment center visit corresponding to the Field ID.
\end{itemize}

\begin{table}[H]
\centering
\begin{tabular}{ll}
\hline
\textbf{Trait Description} & \textbf{UKB Field ID} \\ \hline
Age started wearing glasses or contact lenses & 2217 \\
Age at first live birth & 2754 \\ 
Age started smoking in current and former smokers & 3436 \& 2867 \\
Age high blood pressure diagnosed & 2966 \\
Age stopped smoking & 2897 \\
\hline
\end{tabular}
\caption{UKB Field ID based traits}
\label{tab:ukb-field-traits}
\end{table}

\section{BayesW software configuration and error message summary}\label{sec:bayesw}
We compiled the software using the Intel compiler (version 2020.4), following the instructions provided in \href{https://github.com/medical-genomics-group/hydra}{this} GitHub repository. The compilation utilized the \texttt{boost} module (version 1.73.0) and  the \texttt{eigen} module (version 3.3.9). Below, we provide the configuration script used to set the environmental variables for MPI:
\begin{verbatim}
#!/bin/bash      
MPI_RANKS=$1
export OMP_NUM_THREADS=$(( NSLOTS / MPI_RANKS ))
export OMP_PLACES=cores
export OMP_PROC_BIND=close
export OMP_WAIT_POLICY=active
export I_MPI_FABRICS=shm:ofi
export UCX_TLS=tcp
export I_MPI_HYDRA_BOOTSTRAP=ssh
export I_MPI_DEBUG=5
export I_MPI_PIN=1
export I_MPI_PIN_DOMAIN=omp
export KMP_MALLOC_POOL_INCR=16M
export KMP_INIT_AT_FORK=FALSE
\end{verbatim}

For all analyses, we opted to use BED files directly rather than the sparse files that can be generated by the BayesW software. For each group, we set the slab variances to 0.0001, 0.001, 0.01 and 0.1. The additional options used are listed below:
\begin{verbatim}
     --shuf-mark 1
     --sync-rate 10
     --bed-sync
\end{verbatim}

For the correctly specified model scenario with errors following a normal distribution, we set \texttt{MPI\_RANKS = 4} with \texttt{NSLOTS = 8}. An error code 1003 was produced in all chains across all data replicates with censoring rates 0.5 and 0.8. The same error occurred in all chains in the 93rd data replicate with censoring rate of 0.2. Inspection of the source code indicated that the error code 1003 was generated within the \texttt{arms} function in  the \texttt{BayesW\_arms.cpp} file as shown in the following code snippet:
\begin{verbatim}
if((xinit[0] <= xl) || (xinit[ninit-1] >= xr)){
    /* initial points do not satisfy bounds */
    return 1003;
}
\end{verbatim} 
For scenarios in which errors followed a Gumbel distribution, we set \texttt{MPI\_RANKS = 1} with \texttt{NSLOTS = 4}. In the setting with heritability of 0.2 and a censoring rate of 0.2, error code 1003 was produced in all chains for 14 data replicates, while error code 2000 occurred in at least one chain for 19 data replicates. Altogether, 33 data replicates produced either error code 1003 or 2000 with no overlap between the two error types. For this scenario, we included only the 67 data replicates without any errors in any chain in our analyses. For censoring rates of 0.5 and 0.8, error code 1003 was produced in all chains across all data replicates. We were unable to determine the nature or cause of error code 2000.

For scenarios in which errors followed a logistic distribution, we also set \texttt{MPI\_RANKS = 1} with \texttt{NSLOTS = 4}. An error code 1003 was produced in all chains across all data replicates with censoring rates 0.5 and 0.8.

For misspecified model scenario, we set \texttt{MPI\_RANKS = 1} with \texttt{NSLOTS = 2}. Error code 1003 was produced for all datasets across all censoring rates and genetic architectures. 

For the analysis of survival traits from the UKB, we set \texttt{MPI\_RANKS = 1} with \texttt{NSLOTS = 8}, allocating a total of 188GB of RAM distributed evenly across eight slots. The BayesW software produced errors for all four age-at-diagnosis traits. In particular, the following error message was generated when analyzing the ``first live birth" trait:
\begin{WrappedVerbatim}
hydra: /u/local/apps/boost/1_73_0/gcc-4.8.5/include/boost/random/gamma_distribution.hpp:118:
boost::random::gamma_distribution<RealType>::gamma_distribution(const RealType &, const RealType &) [with RealType = double]: 
Assertion `_beta > result_type(0)' failed.
\end{WrappedVerbatim}
All other traits produced the error message 
\begin{verbatim}
    #FATAL#: malloc failed on line 730 of data.cpp
\end{verbatim} 
which may be due to exceeding the allocated 188GB of RAM. We also attempted to run the software using the sparse file format with the same amount of memory, but errors were produced for all four traits.

\section{Supplementary Figures}

\begin{figure}[H]
    \centering\includegraphics[width=\textwidth]{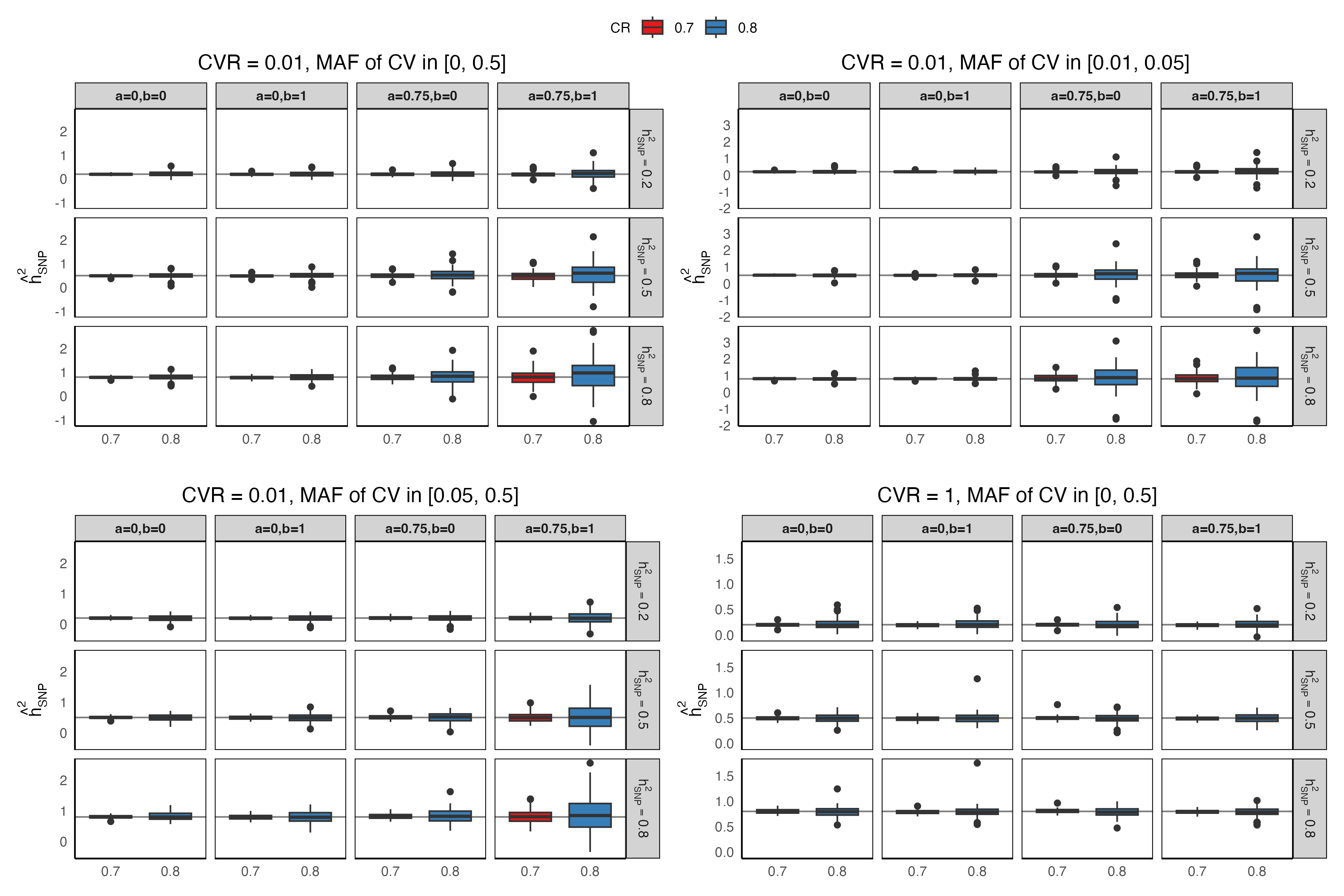}
    \caption{Accuracy of genome-wide SNP heritability estimation under 48 different genetic architectures stratified by CR $\in \{0.2, 0.5, 0.7, 0.8\}$ over 100 runs ($N = 276,169$, $M = 592,454$, $C = 24$, $J = 100$, $B = 10$). The number of genomic partitions $K$ was set to $24$ created from 4 LD bins based on quartiles of LDAK scores and 6 MAF intervals based on 5 MAF knots, 0.01, 0.02, 0.03, 0.04 and 0.05. The horizontal lines indicate true $h_{\text{SNP}}^2$.}
    \label{fig:mis_spec_accuracy_supp}
\end{figure}

\begin{figure}[H]
    \centering\includegraphics[trim=0 0 0 0, clip, width=0.9\linewidth]{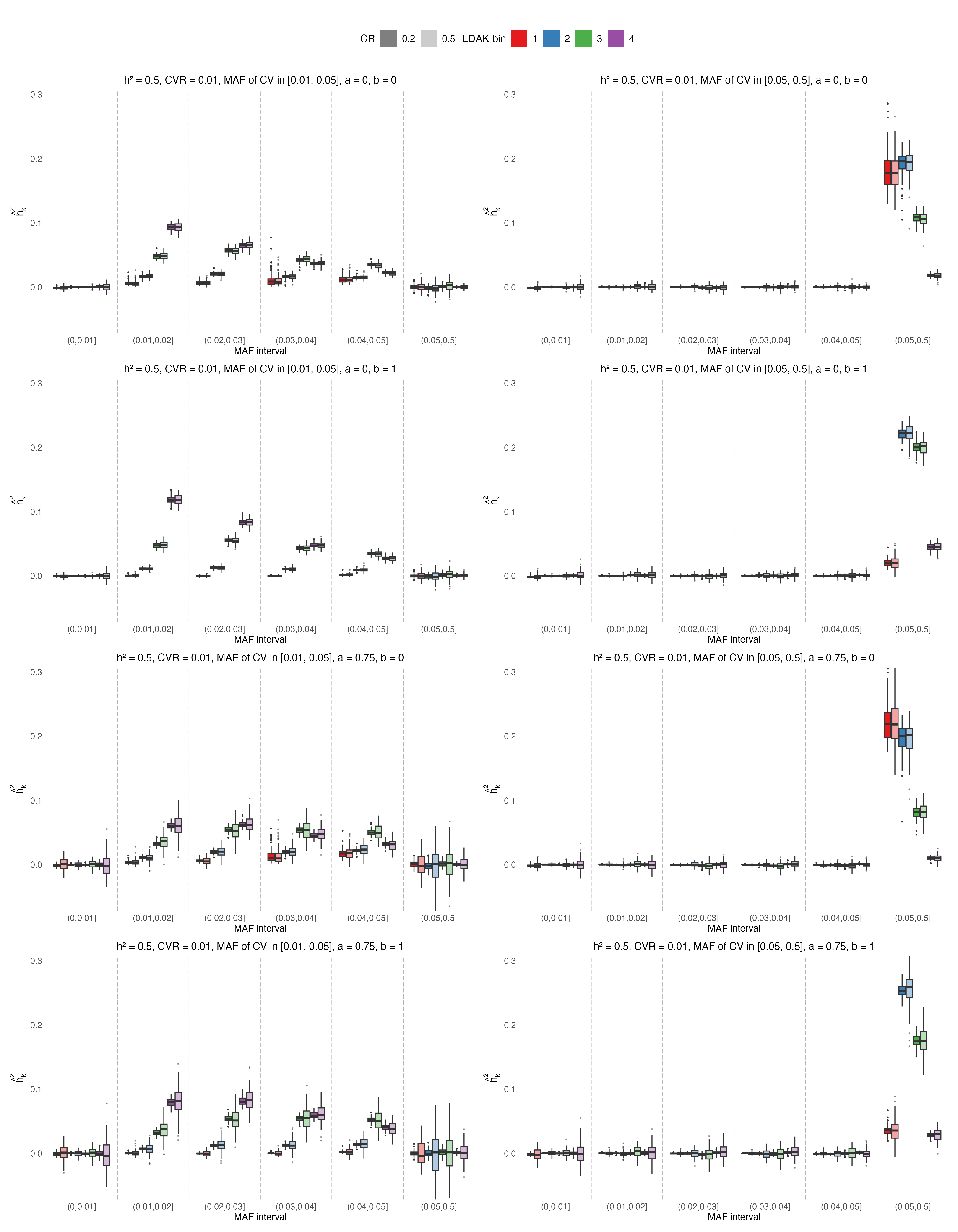}
    \caption{Heritability partitioning into LDAK-MAF bins under different genetic architectures with $\text{CVR} = 1\%$ and $h^2_{\text{SNP}} = 0.5$ for censoring rates $0.2$ and $0.5$ ($N = 276,169$,  $M = 592,454$, $C = 24$, $K = 24$, $J = 100$, $B = 10$). Only causal variants within a given MAF interval should contribute to heritability (for example, ``CV in $[0.05,0.5]$" indicates that heritability should be attributable only to the causal SNPs within the MAF interval [0.05,0.5]). CVR: casual variant rate; MAF: minor allele frequency; LDAK: linkage-disequilibrium adjusted kinships.}
    \label{fig:h2_partitioning_supp_h2_0.5_cr_low_to_moderate}
\end{figure}

\begin{figure}[H]
    \centering\includegraphics[trim=0 0 0 0, clip, width=0.9\linewidth]{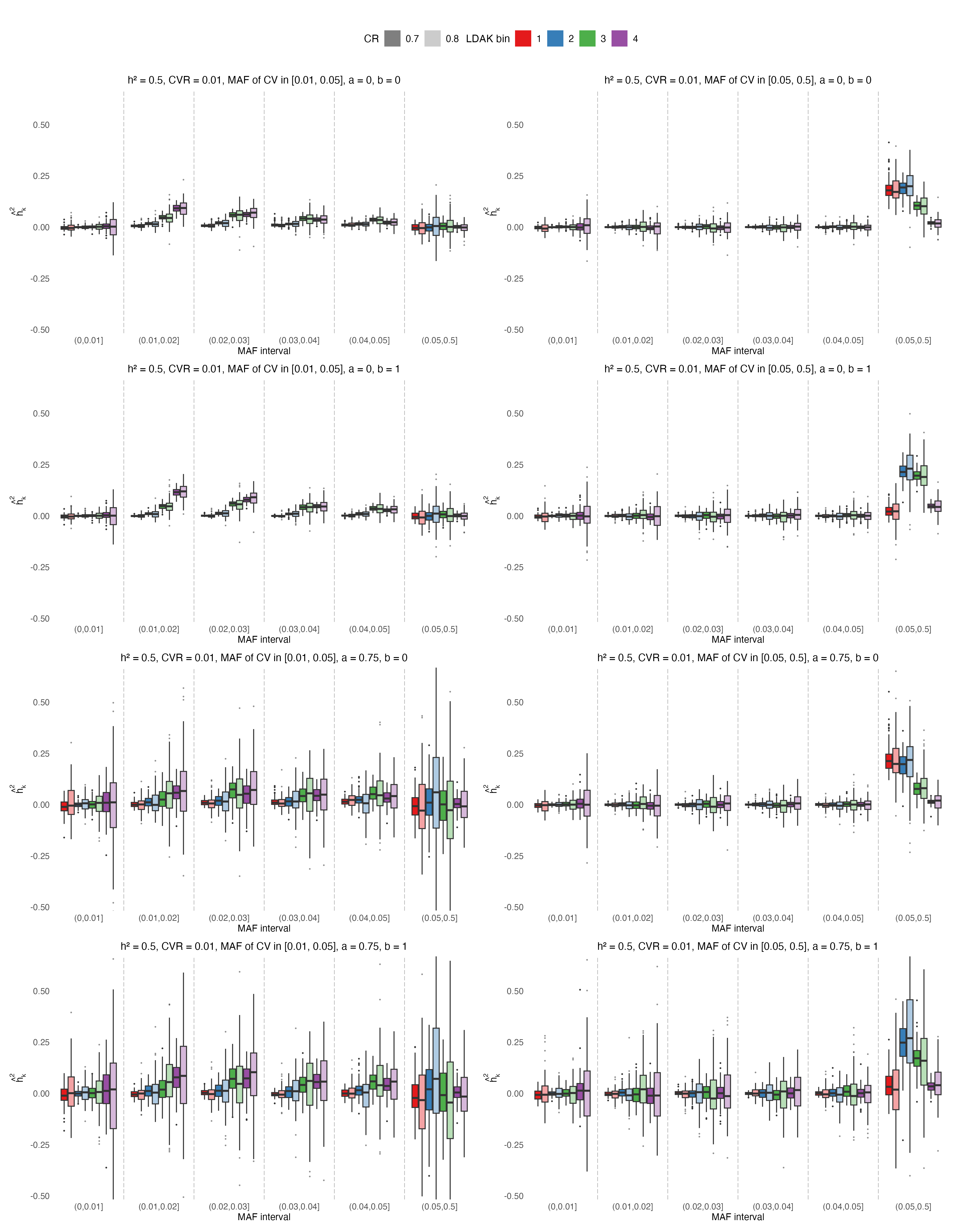}
    \caption{Heritability partitioning into LDAK-MAF bins under different genetic architectures with $\text{CVR} = 1\%$ and $h^2_{\text{SNP}} = 0.5$ for censoring rates $0.7$ and $0.8$ ($N = 276,169$,  $M = 592,454$, $C = 24$, $K = 24$, $J = 100$, $B = 10$). Only causal variants within a given MAF interval should contribute to heritability (for example, ``CV in $[0.05,0.5]$" indicates that heritability should be attributable only to the causal SNPs within the MAF interval [0.05,0.5]). CVR: casual variant rate; MAF: minor allele frequency; LDAK: linkage-disequilibrium adjusted kinships.}
    \label{fig:h2_partitioning_supp_h2_0.5_cr_high}
\end{figure}

\begin{figure}[H]
    \centering\includegraphics[width=\textwidth]{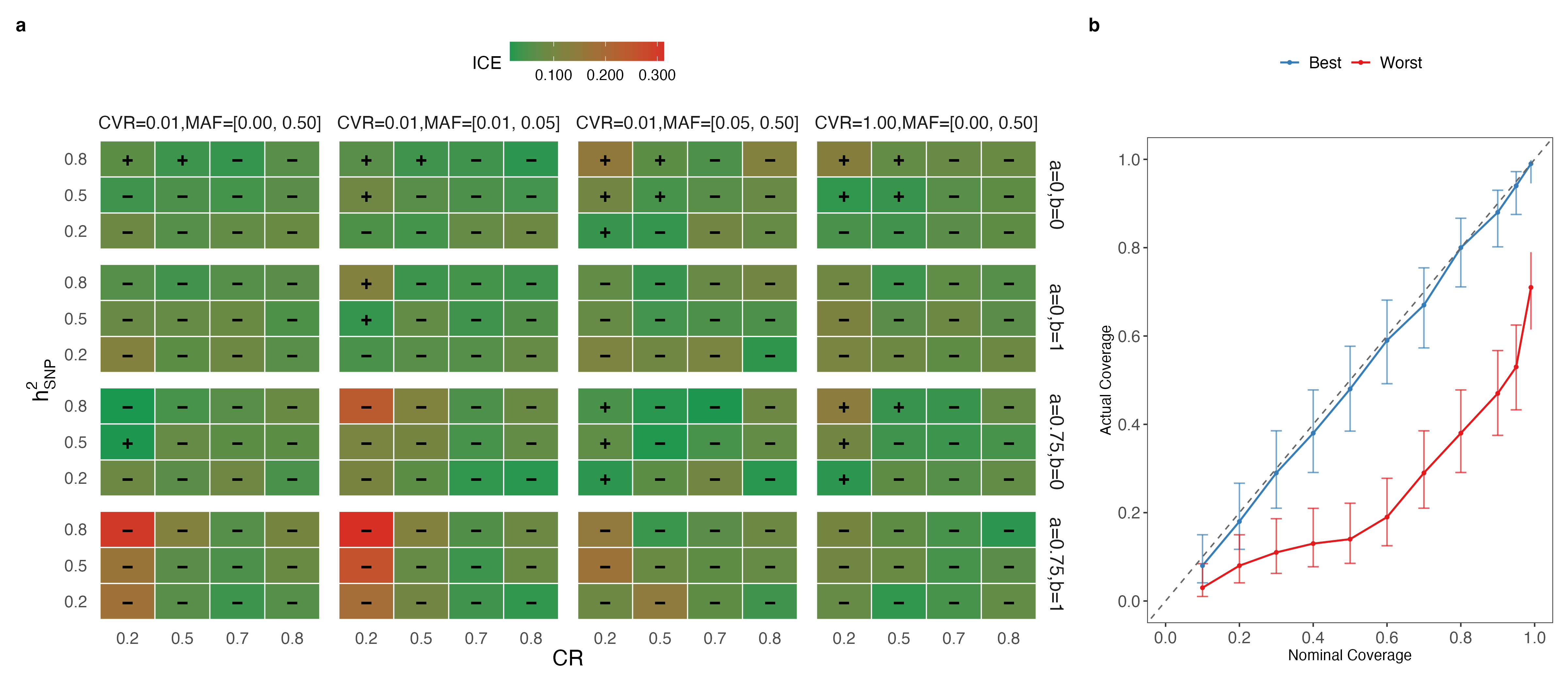}
    \caption{Summary of calibration of heritability inference across 48 genetic architectures, stratified by $\Delta \in \{0.2, 0.5, 0.7, 0.8\}$, based on 100 simulation replicates ($N = 276,169$, $M = 592,454$, $C = 24$, $J = 100$, $B = 10$). The number of genomic partitions was fixed at $K = 24$, constructed from 4 LD bins defined by quartiles of LDAK scores and 6 MAF bins defined by five MAF knots at 0.01, 0.02, 0.03, 0.04 and 0.05. \textbf{a} Integrated Calibration Index (ICI) heatmap summarizing calibration accuracy; $+$ indicates underconfident standard error estimates and $-$ indicates overconfident standard error estimates on average. \textbf{b} Calibration curves for genetic architectures with the best ($b = 0$, MAF of CV in $[0,0.5]$) and worst ($b = 1$, MAF of CV in $[0.01, 0.05]$) ICI. For both cases, the settings were $h^2_{\text{SNP}} = 0.8$, $\Delta = 0.2$, $\text{CVR} = 0.01$, and $a = 0.75$. The dashed diagonal line denotes perfect calibration.}
    \label{fig:calibration_summary_normal}
\end{figure}

\begin{figure}[H]
    \centering\includegraphics[width=\textwidth]{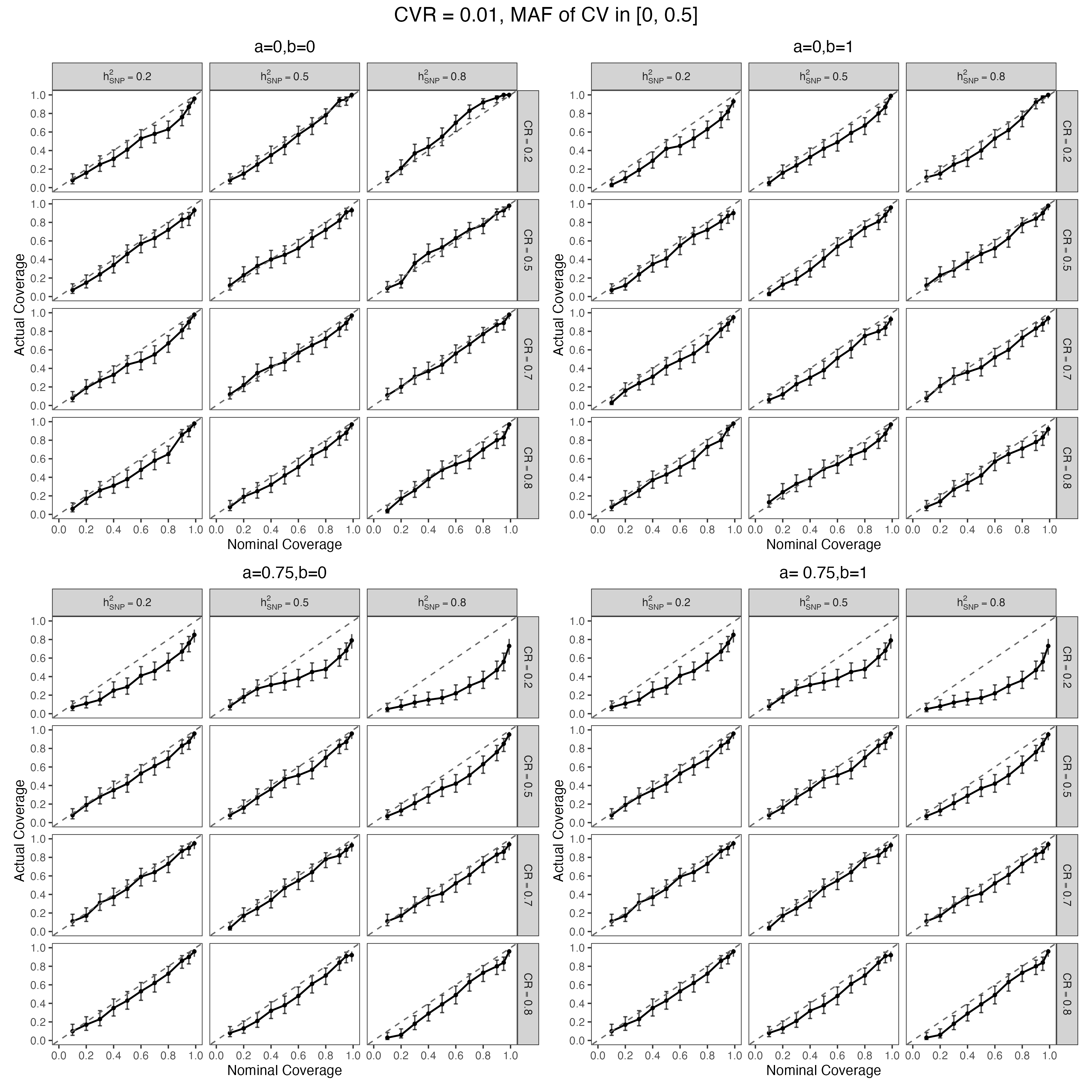}
    \caption{Calibration curves stratified by $h^2_{\text{SNP}} \in \{0.2, 0.5, 0.8\}$ and CR $\in \{0.2, 0.5, 0.7, 0.8\}$ over 100 runs ($N = 276,169$, $M = 592,454$, $C = 24$, $J = 100$, $B = 10$) for the setting $\text{CVR} = 0.01$ and MAF of CVs are from $[0, 0.5]$. The number of genomic partitions $K$ was set to $24$ created from 4 LD bins based on quartiles of LDAK scores and 6 MAF intervals based on 5 MAF knots, 0.01, 0.02, 0.03, 0.04 and 0.05. The dashed diagonal line represents perfect calibration.}
    \label{fig:mis_spec_calibration_plots_cvr_0.01_maflb_0_mafub_0.5}
\end{figure}

\begin{figure}[H]
    \centering\includegraphics[width=\textwidth]{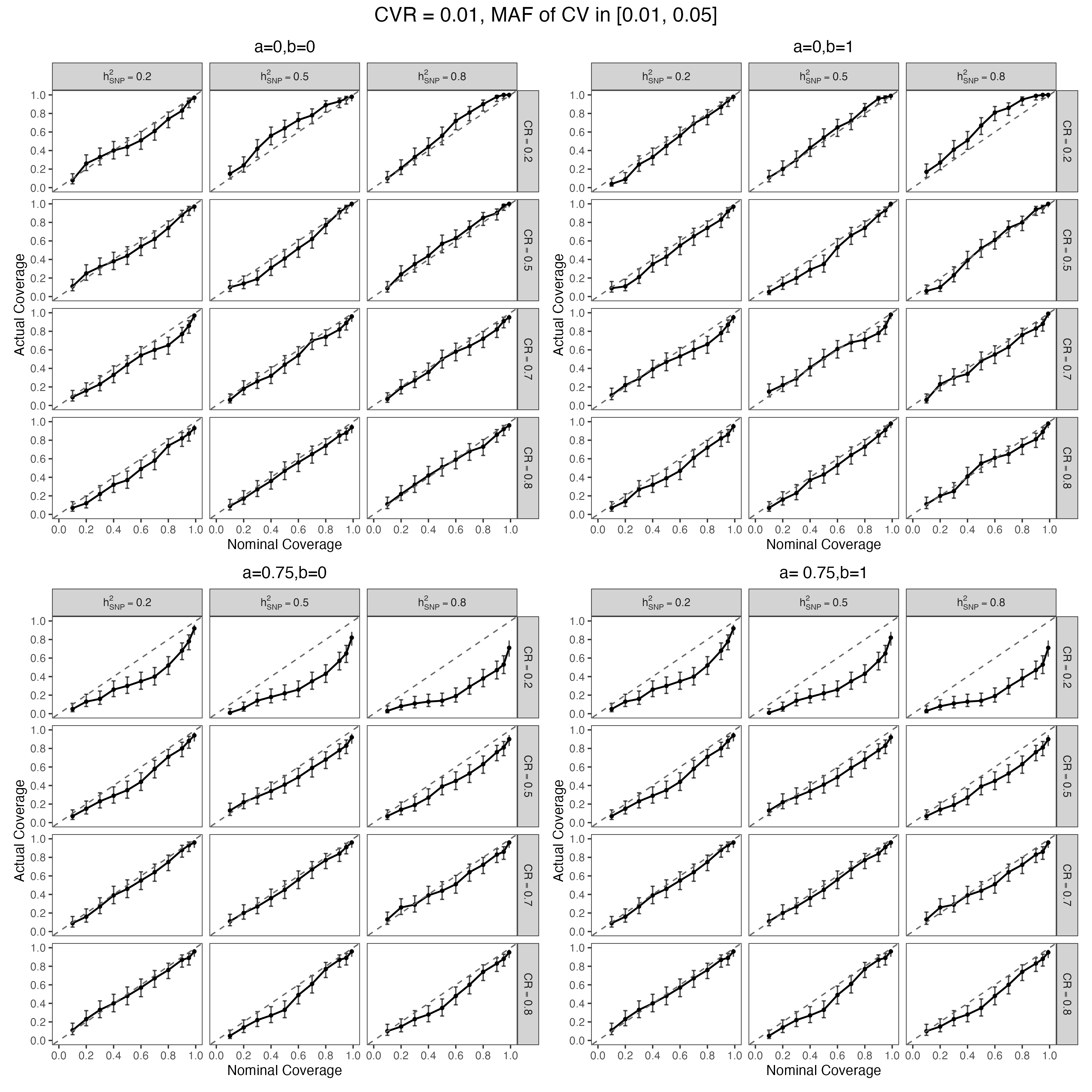}
    \caption{Calibration curves stratified by $h^2_{\text{SNP}} \in \{0.2, 0.5, 0.8\}$ and CR $\in \{0.2, 0.5, 0.7, 0.8\}$ over 100 runs ($N = 276,169$, $M = 592,454$, $C = 24$, $J = 100$, $B = 10$) for the setting $\text{CVR} = 0.01$ and MAF of CVs are from $[0.01, 0.05]$. The number of genomic partitions $K$ was set to $24$ created from 4 LD bins based on quartiles of LDAK scores and 6 MAF intervals based on 5 MAF knots, 0.01, 0.02, 0.03, 0.04 and 0.05. The dashed diagonal line represents perfect calibration.}
    \label{fig:mis_spec_calibration_plots_cvr_0.01_maflb_0.01_mafub_0.05}
\end{figure}

\begin{figure}[H]
    \centering\includegraphics[width=\textwidth]{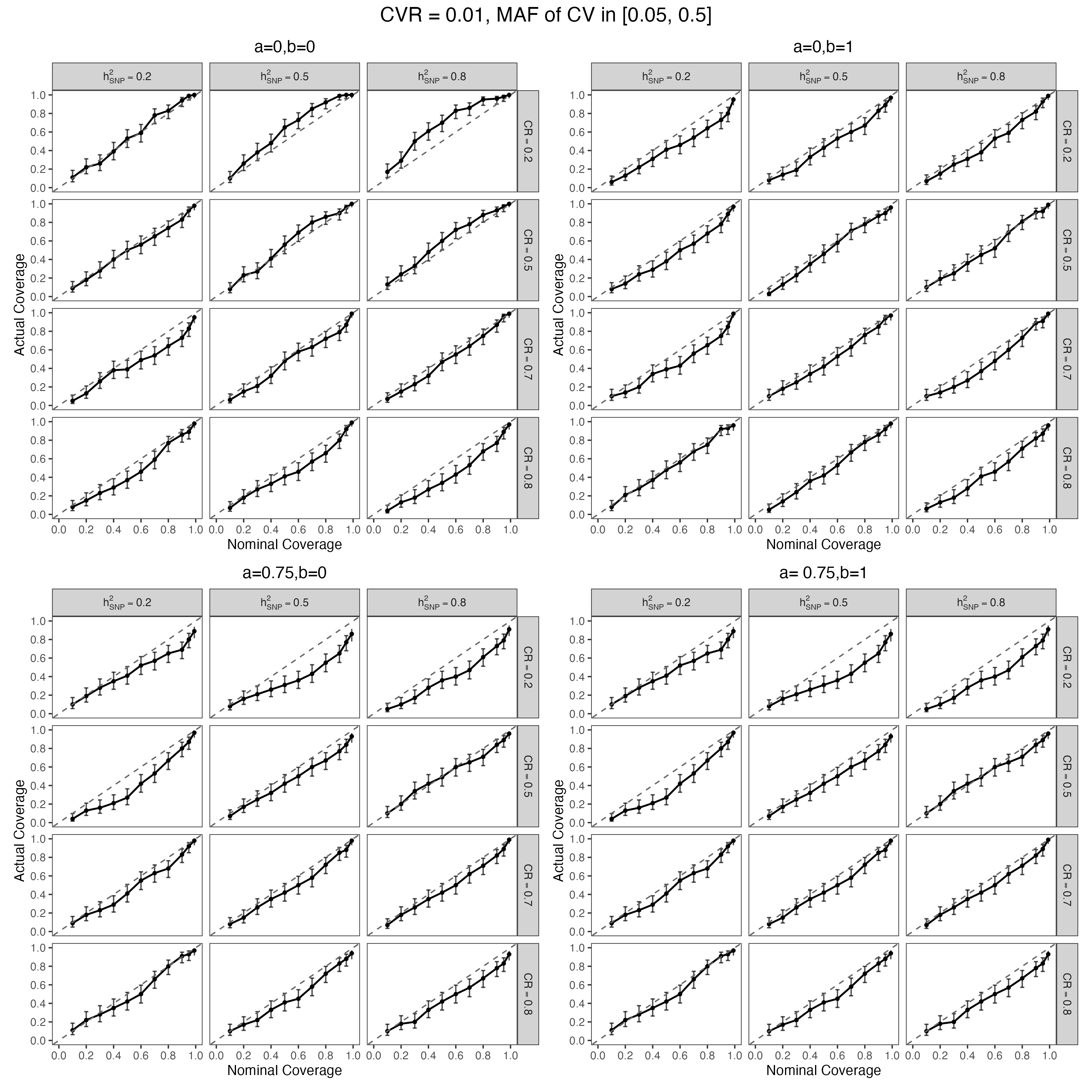}
    \caption{Calibration curves stratified by $h^2_{\text{SNP}} \in \{0.2, 0.5, 0.8\}$ and CR $\in \{0.2, 0.5, 0.7, 0.8\}$ over 100 runs ($N = 276,169$, $M = 592,454$, $C = 24$, $J = 100$, $B = 10$) for the setting $\text{CVR} = 0.01$ and MAF of CVs are from $[0.05, 0.5]$. The number of genomic partitions $K$ was set to $24$ created from 4 LD bins based on quartiles of LDAK scores and 6 MAF intervals based on 5 MAF knots, 0.01, 0.02, 0.03, 0.04 and 0.05. The dashed diagonal line represents perfect calibration.}
    \label{fig:mis_spec_calibration_plots_cvr_0.01_maflb_0.05_mafub_0.5}
\end{figure}

\begin{figure}[H]
    \centering\includegraphics[width=\textwidth]{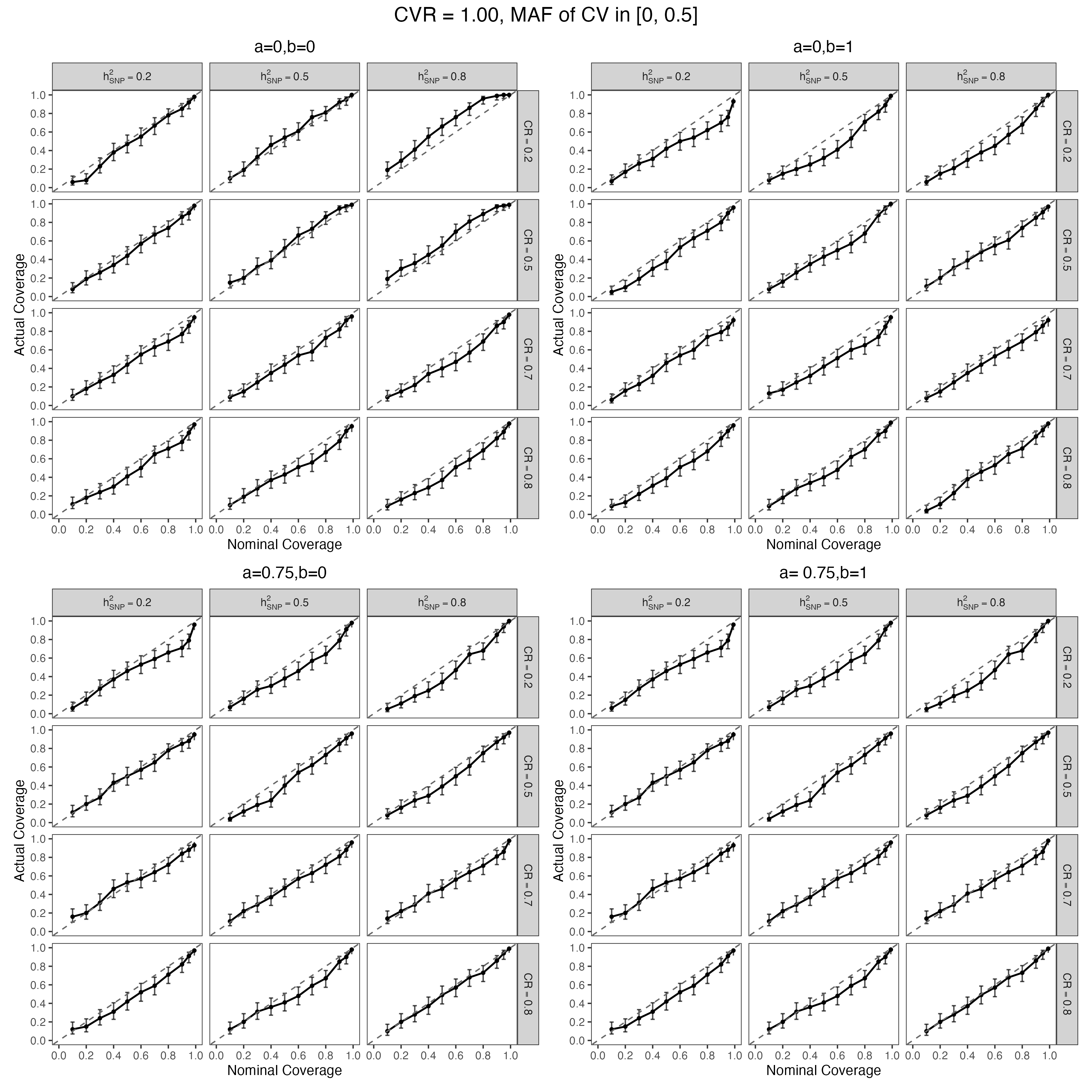}
    \caption{Calibration curves stratified by $h^2_{\text{SNP}} \in \{0.2, 0.5, 0.8\}$ and CR $\in \{0.2, 0.5, 0.7, 0.8\}$ over 100 runs ($N = 276,169$, $M = 592,454$, $C = 24$, $J = 100$, $B = 10$) for the setting $\text{CVR} = 1$ and MAF of CVs are from $[0, 0.5]$. The number of genomic partitions $K$ was set to $24$ created from 4 LD bins based on quartiles of LDAK scores and 6 MAF intervals based on 5 MAF knots, 0.01, 0.02, 0.03, 0.04 and 0.05. The dashed diagonal line represents perfect calibration.}
    \label{fig:mis_spec_calibration_plots_cvr_1_maflb_0_mafub_0.5}
\end{figure}

\begin{figure}[H]
    \centering\includegraphics[width=0.9\textwidth]{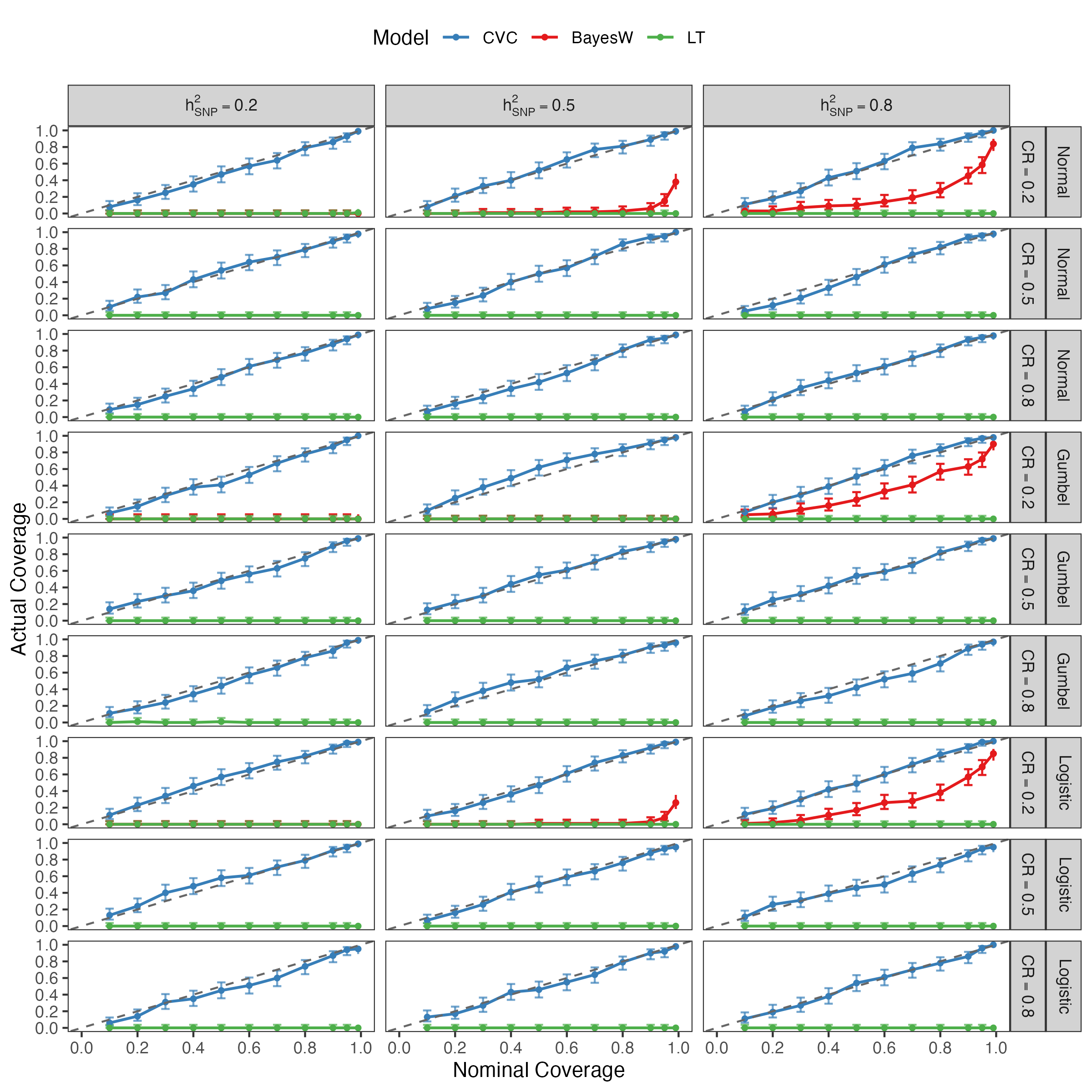}
    \caption{Calibration plots for estimating total heritability under correct model specifications over 100 runs for $h^2_{\text{SNP}} \in \{0.2, 0.5, 0.8\}$ and $\Delta \in \{0.2, 0.5, 0.8\}$ ($N = 20,000$,  $M = 20,000$, $C = 10$, $K = 5$, $J = 100$, $B = 10$). For BayesW, we used 5 chains with chain length of 3,000, burn-in of 1,000, and thinning of 5. Posterior mean was used as the point estimate. For censoring rates 0.5 and 0.8, BayesW produced errors across all replicates and chains. The error bars of actual coverages indicate 95\% confidence interval.}
    \label{fig:correct_spec_h2_calibration_supp}
\end{figure}

\begin{figure}[H]
    \centering\includegraphics[width=\textwidth]{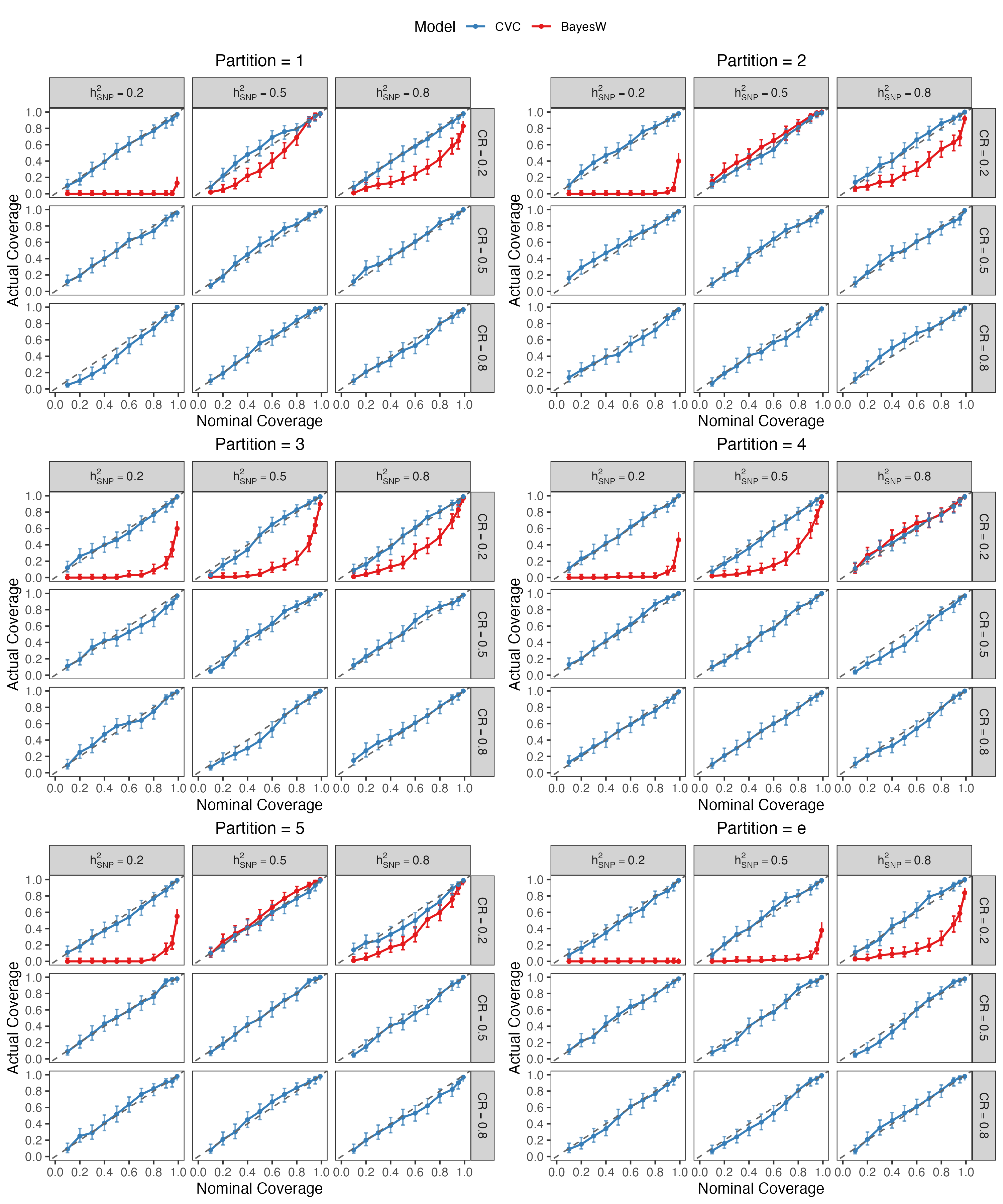}
    \caption{Calibration plots for estimating partitioned heritability under correct model specifications with normal error distribution over 100 runs for $h^2_{\text{SNP}} \in \{0.2, 0.5, 0.8\}$ and $\Delta \in \{0.2, 0.5, 0.8\}$ ($N = 20,000$,  $M = 20,000$, $C = 10$, $K = 5$, $J = 100$, $B = 10$). For BayesW, we used 5 chains with chain length of 3,000, burn in of 1000, and thinning of 5. Posterior mean was used as the point estimate. For censoring rates 0.5 and 0.8, BayesW produced errors across all replicates and chains. The error bars of actual coverages indicate 95\% confidence interval.}
    \label{fig:correct_spec_h2k_calibration_normal}
\end{figure}

\begin{figure}[H]
    \centering\includegraphics[width=\textwidth]{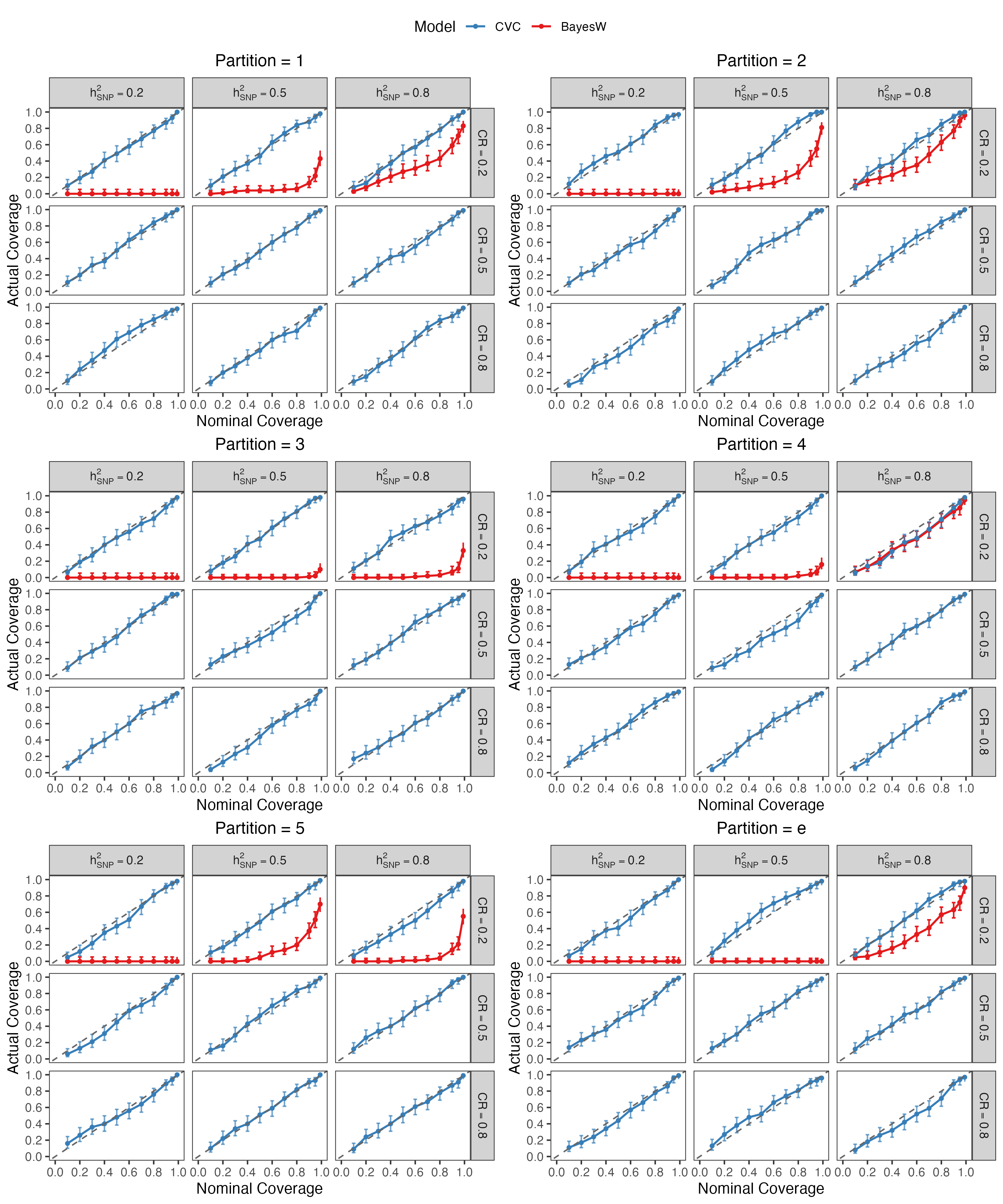}
    \caption{Calibration plots for estimating partitioned heritability under correct model specifications with Gumbel error distribution over 100 runs for $h^2_{\text{SNP}} \in \{0.2, 0.5, 0.8\}$ and $\Delta \in \{0.2, 0.5, 0.8\}$ ($N = 20,000$,  $M = 20,000$, $C = 10$, $K = 5$, $J = 100$, $B = 10$). For BayesW, we used 5 chains with chain length of 3,000, burn in of 1000, and thinning of 5. Posterior mean was used as the point estimate. For censoring rates 0.5 and 0.8, BayesW produced errors across all replicates and chains. The error bars of actual coverages indicate 95\% confidence interval.}
    \label{fig:correct_spec_h2k_calibration_gumbel}
\end{figure}

\begin{figure}[H]
    \centering\includegraphics[width=\textwidth]{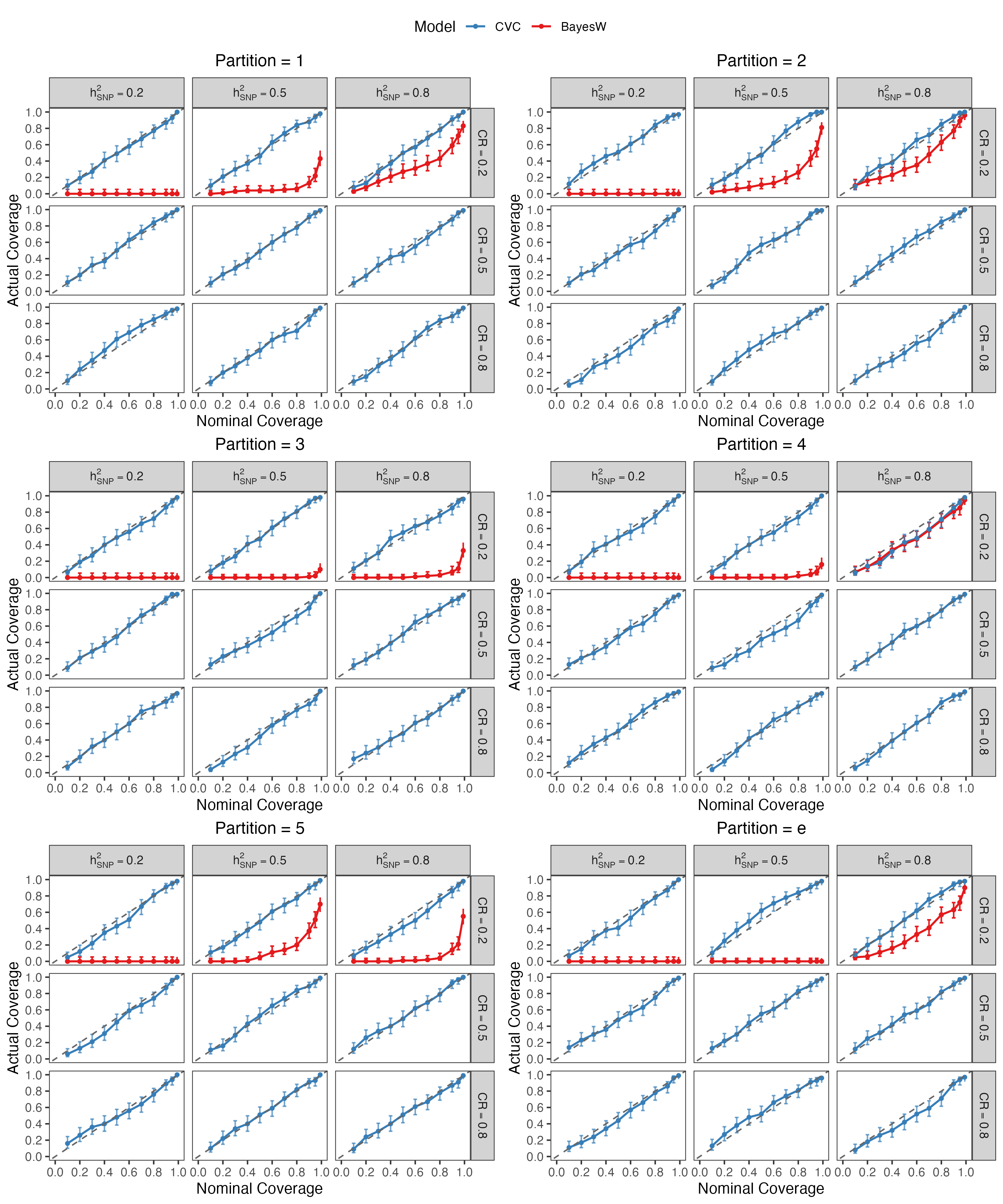}
    \caption{Calibration plots for estimating partitioned heritability under correct model specifications with logistic error distribution over 100 runs for $h^2_{\text{SNP}} \in \{0.2, 0.5, 0.8\}$ and $\Delta \in \{0.2, 0.5, 0.8\}$ ($N = 20,000$,  $M = 20,000$, $C = 10$, $K = 5$, $J = 100$, $B = 10$). For BayesW, we used 5 chains with chain length of 3,000, burn in of 1000, and thinning of 5. Posterior mean was used as the point estimate. For censoring rates 0.5 and 0.8, BayesW produced errors across all replicates and chains. The error bars of actual coverages indicate 95\% confidence interval.}
    \label{fig:correct_spec_h2k_calibration_logistic}
\end{figure}

\begin{figure}[H]
    \centering\includegraphics[width=0.75\textwidth]{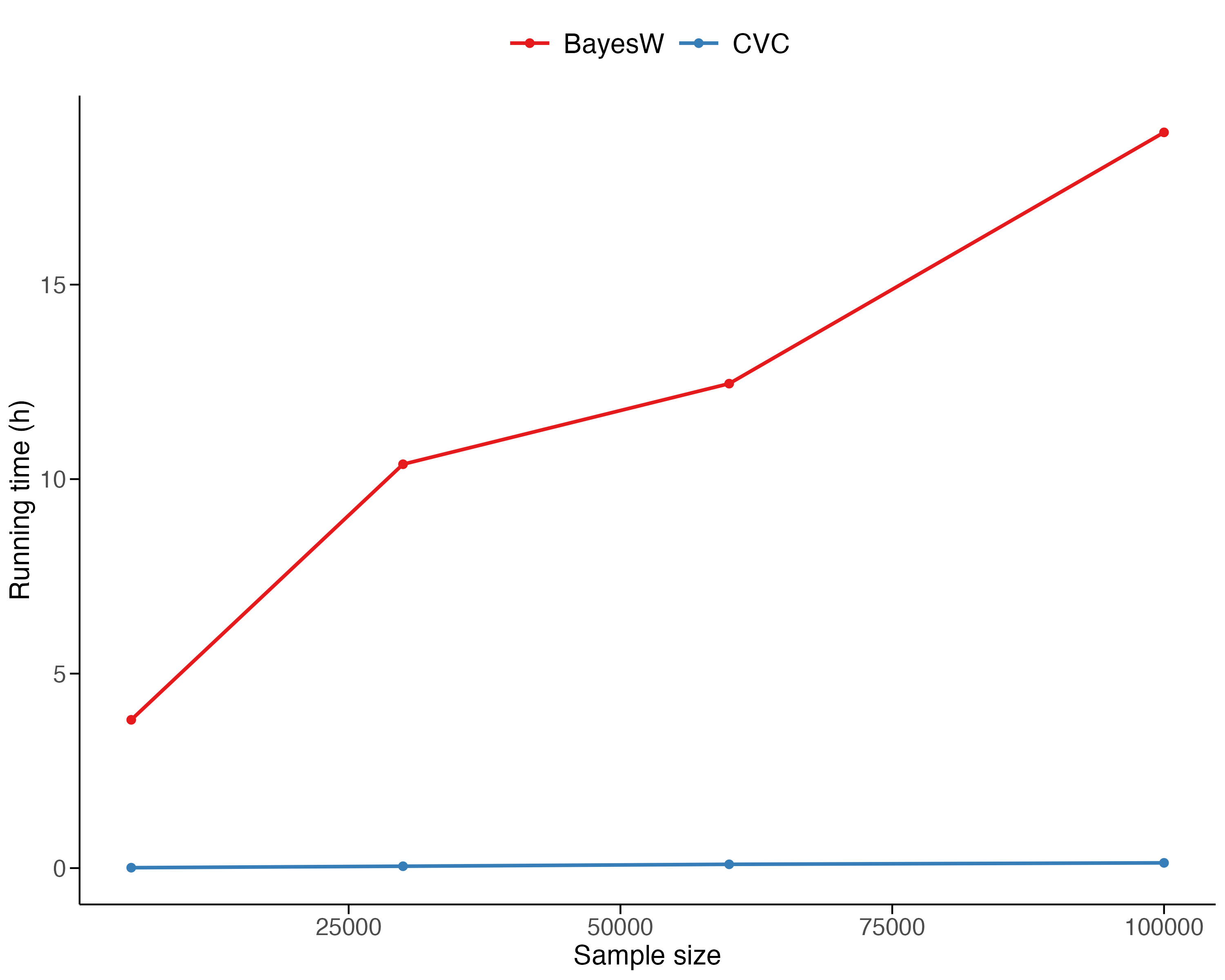}
    \caption{Comparison of median running times for CVC and BayesW across 5 data replicates ($M = 100,000$, $C = 10$, $K = 10$, $J = 100$, $B = 10$, $h^2_{\text{SNP}} = 0.2$, $\Delta = 0.2$). BayesW running times are predicted times to generate 3,000 MCMC samples, extrapolated from 20-chain run. For both models, compute nodes with 8 CPU cores with 16 GB RAM were used. We used 1 MPI rank with 8 threads for BayesW.}
    \label{fig:computational_benchmark}
\end{figure}

\vspace*{-5em}
\section{Supplementary Tables}

\begin{table}[H]
\centering
\caption{\label{tab:N_impact_h2_0.5_M_50000_K_10}Impact of sample size $N$ on total heritability estimation with CVC model ($h^2_{\text{SNP}}$ = 0.5, $M$ = 50,000, $C$ = 10, $K$ = 10, $J$ = 100, $B$ = 10). Relative bias is the bias of estimated heritability with respect to true total heritability computed using the formula $(\hat{h}^2_{\text{SNP}} - h^2_{\text{SNP}}) / h^2_{\text{SNP}} \times 100$. Bias is the difference between the estimated jackknife standard error and the Monte Carlo standard error of total heritability estimates. The values report means from 100 replications. SE: standard error; JSE: Jackknife SE.}
\centering
\resizebox{\ifdim\width>\linewidth\linewidth\else\width\fi}{!}{
\fontsize{11}{13}\selectfont
\begin{tabular}[t]{ccrrrrrr}
\toprule
CR & N & $\hat{h}_{\text{SNP}}^2$ & Relative bias (\%) & SE of $\hat{h}_{\text{SNP}}^2$ & JSE of $\hat{h}_{\text{SNP}}^2$ & Bias & SE of JSE\\
\midrule
\multirow{3}{*}{0.2} & 50000 & 0.500 & -0.100 & 0.012 & 0.013 & 0.001 & 0.001\\
 & 100000 & 0.499 & -0.142 & 0.008 & 0.008 & 0.001 & 0.001\\
 & 200000 & 0.500 & -0.052 & 0.005 & 0.006 & 0.002 & 0.000\\ \cmidrule(l){1-8}
\multirow{3}{*}{0.5} & 50000 & 0.500 & -0.035 & 0.034 & 0.031 & -0.003 & 0.003\\
 & 100000 & 0.499 & -0.208 & 0.018 & 0.019 & 0.001 & 0.002\\
 & 200000 & 0.500 & -0.055 & 0.011 & 0.011 & -0.000 & 0.001\\ \cmidrule(l){1-8}
\multirow{3}{*}{0.8} & 50000 & 0.502 & 0.451 & 0.211 & 0.214 & 0.003 & 0.054\\
 & 100000 & 0.497 & -0.698 & 0.146 & 0.128 & -0.018 & 0.029\\
 & 200000 & 0.510 & 1.940 & 0.068 & 0.068 & 0.001 & 0.015\\
\bottomrule
\end{tabular}}
\end{table}

\begin{table}[H]
\centering
\caption{\label{tab:M_impact_h2_0.5_N_50000_K_10}Impact of number of SNPs $M$ on total heritability estimation with CVC model ($h^2_{\text{SNP}}$ = 0.5, $N$ = 50,000, $C$ = 10, $K$ = 10, $J$ = 100, $B$ = 10). Relative bias is the bias of estimated heritability with respect to true total heritability computed using the formula $(\hat{h}^2_{\text{SNP}} - h^2_{\text{SNP}}) / h^2_{\text{SNP}} \times 100$. Bias is the difference between the estimated jackknife standard error and the Monte Carlo standard error of total heritability estimates. The values report means from 100 replications. SE: standard error; JSE: Jackknife SE.}
\centering
\resizebox{\ifdim\width>\linewidth\linewidth\else\width\fi}{!}{
\fontsize{11}{13}\selectfont
\begin{tabular}[t]{ccrrrrrr}
\toprule
CR & M & $\hat{h}_{\text{SNP}}^2$ & Relative bias (\%) & SE of $\hat{h}_{\text{SNP}}^2$ & JSE of $\hat{h}_{\text{SNP}}^2$ & Bias & SE of JSE\\
\midrule
\multirow{3}{*}{0.2} & 50000 & 0.500 & -0.100 & 0.012 & 0.013 & 0.001 & 0.001\\
 & 100000 & 0.500 & 0.086 & 0.015 & 0.016 & 0.001 & 0.001\\
 & 200000 & 0.501 & 0.103 & 0.018 & 0.020 & 0.002 & 0.002\\ \cmidrule(l){1-8}
\multirow{3}{*}{0.5} & 50000 & 0.500 & -0.035 & 0.034 & 0.031 & -0.003 & 0.003\\
 & 100000 & 0.503 & 0.507 & 0.041 & 0.046 & 0.005 & 0.004\\
 & 200000 & 0.492 & -1.682 & 0.066 & 0.058 & -0.008 & 0.008\\ \cmidrule(l){1-8}
\multirow{3}{*}{0.8} & 50000 & 0.502 & 0.451 & 0.211 & 0.214 & 0.003 & 0.054\\
 & 100000 & 0.526 & 5.154 & 0.335 & 0.347 & 0.011 & 0.088\\
 & 200000 & 0.599 & 19.850 & 0.475 & 0.450 & -0.024 & 0.116\\
\bottomrule
\end{tabular}}
\end{table}

\begin{table}[H]
\centering
\caption{\label{tab:K_impact_h2_0.5_N_100000_M_100000}Impact of number of partitions $K$ on total heritability estimation with CVC model ($h^2_{\text{SNP}}$ = 0.5, $N$ = 100,000, $C$ = 10, $M$ = 100,000, $J$ = 100, $B$ = 10). Relative bias is the bias of estimated heritability with respect to true total heritability computed using the formula $(\hat{h}^2_{\text{SNP}} - h^2_{\text{SNP}}) / h^2_{\text{SNP}} \times 100$. Bias is the difference between the estimated jackknife standard error and the Monte Carlo standard error of total heritability estimates. The values report means from 100 replications. SE: standard error; JSE: Jackknife SE.}
\centering
\resizebox{\ifdim\width>\linewidth\linewidth\else\width\fi}{!}{
\fontsize{11}{13}\selectfont
\begin{tabular}[t]{ccrrrrrr}
\toprule
CR & K & $\hat{h}_{\text{SNP}}^2$ & Relative bias (\%) & SE of $\hat{h}_{\text{SNP}}^2$ & JSE of $\hat{h}_{\text{SNP}}^2$ & Bias & SE of JSE\\
\midrule
\multirow{3}{*}{0.2} & 10 & 0.501 & 0.269 & 0.009 & 0.009 & 0.000 & 0.001\\
 & 50 & 0.499 & -0.274 & 0.011 & 0.009 & -0.002 & 0.001\\
 & 100 & 0.500 & 0.031 & 0.009 & 0.009 & -0.000 & 0.001\\ \cmidrule(l){1-8}
\multirow{3}{*}{0.5} & 10 & 0.500 & 0.043 & 0.027 & 0.024 & -0.003 & 0.003\\
 & 50 & 0.500 & -0.069 & 0.020 & 0.021 & 0.001 & 0.002\\
 & 100 & 0.499 & -0.201 & 0.019 & 0.020 & 0.002 & 0.003\\ \cmidrule(l){1-8}
\multirow{3}{*}{0.8} & 10 & 0.526 & 5.278 & 0.178 & 0.172 & -0.007 & 0.060\\
 & 50 & 0.507 & 1.411 & 0.162 & 0.149 & -0.012 & 0.044\\
 & 100 & 0.487 & -2.685 & 0.143 & 0.140 & -0.004 & 0.039\\
\bottomrule
\end{tabular}}
\end{table}

\begin{table}[H]
\centering
\caption{Impact of number of random Gaussian vectors on total heritability estimates 
($N = 10,000$, $M = 10,000$, $C = 10$, $K = 10$, $J = 100$, $h^2_{\text{SNP}} = 0.5$).}
\label{tab:impact_B}
\begin{tabular}[t]{ccrrrrr}
\toprule
CR & $B$ & $\hat{h}_{\text{SNP}}^2$ & SE & JSE & SE of JSE & Rel. Bias of $\hat{h}_{\text{SNP}}^2$ (\%)\\
\midrule
 & 1 & 0.49 & 0.038 & 0.036 & 0.004 & -2.18\\

 & 5 & 0.49 & 0.026 & 0.028 & 0.003 & -2.77\\

 & 10 & 0.48 & 0.025 & 0.028 & 0.003 & -3.08\\

 & 50 & 0.49 & 0.024 & 0.027 & 0.002 & -2.91\\

\multirow{-5}{*}{\centering\arraybackslash 0.2} & 100 & 0.48 & 0.023 & 0.026 & 0.002 & -3.06\\
\cmidrule{1-7}
 & 1 & 0.50 & 0.069 & 0.074 & 0.010 & -0.96\\

 & 5 & 0.49 & 0.071 & 0.069 & 0.010 & -1.56\\

 & 10 & 0.49 & 0.069 & 0.068 & 0.009 & -2.14\\

 & 50 & 0.49 & 0.068 & 0.068 & 0.009 & -2.07\\

\multirow{-5}{*}{\centering\arraybackslash 0.5} & 100 & 0.49 & 0.068 & 0.068 & 0.009 & -2.06\\
\cmidrule{1-7}
 & 1 & 0.44 & 0.465 & 0.408 & 0.097 & -11.61\\

 & 5 & 0.44 & 0.467 & 0.399 & 0.094 & -12.26\\

 & 10 & 0.44 & 0.462 & 0.398 & 0.091 & -12.16\\

 & 50 & 0.44 & 0.461 & 0.397 & 0.091 & -11.89\\

\multirow{-5}{*}{\centering\arraybackslash 0.8} & 100 & 0.44 & 0.462 & 0.396 & 0.090 & -12.11\\
\bottomrule
\end{tabular}
\end{table}

\begin{table}[H]
\centering
\caption{Impact of number of jackknife blocks on standard error estimates 
($N = 10,000$, $M = 10,000$, $C = 10$, $K = 10$, $B = 10$, $h^2_{\text{SNP}} = 0.5$).}
\label{tab:impact_J}
\begin{tabular}[t]{ccrrrrr}
\toprule
CR & $J$ & $\hat{h}_{\text{SNP}}^2$ & SE & JSE & SE of JSE & Bias of JSE\\
\midrule
 & 2 & 0.49 & 0.025 & 0.013 & 0.010 & -0.01\\

 & 10 & 0.49 & 0.026 & 0.025 & 0.006 & 0.00\\

 & 100 & 0.49 & 0.025 & 0.028 & 0.002 & 0.00\\

\multirow{-4}{*}{\centering\arraybackslash 0.2} & 500 & 0.48 & 0.025 & 0.028 & 0.001 & 0.00\\
\cmidrule{1-7}
 & 2 & 0.49 & 0.069 & 0.026 & 0.020 & -0.04\\

 & 10 & 0.49 & 0.070 & 0.062 & 0.018 & -0.01\\

 & 100 & 0.49 & 0.069 & 0.068 & 0.010 & 0.00\\

\multirow{-4}{*}{\centering\arraybackslash 0.5} & 500 & 0.49 & 0.070 & 0.069 & 0.008 & 0.00\\
\cmidrule{1-7}
 & 2 & 0.44 & 0.465 & 0.170 & 0.133 & -0.29\\

 & 10 & 0.44 & 0.461 & 0.358 & 0.111 & -0.10\\

 & 100 & 0.44 & 0.461 & 0.397 & 0.089 & -0.06\\

\multirow{-4}{*}{\centering\arraybackslash 0.8} & 500 & 0.44 & 0.460 & 0.400 & 0.094 & -0.06\\
\bottomrule
\end{tabular}
\end{table}

\begin{table}[H]
\centering
\caption{\label{tab:correct_spec_partitioned_h2_0.5}Accuracy of partitioned heritability estimation for $h^2_{\text{SNP}} = 0.5$ 
under correct model specification ($N = 100,000$, $M = 100,000$, $C = 10$, $K = 10$, $J = 100$, $B = 10$). 
Relative bias is computed using the formula $(\hat{h}_k^2 - h_k^2) / h_k^2 \times 100$. 
Bias is the difference between the estimated standard error and the Monte Carlo standard error of heritability estimates.
The values report means from 100 replications.
SE: standard error. JSE: jackknife standard error.}
\centering
\resizebox{\ifdim\width>\linewidth\linewidth\else\width\fi}{!}{
\fontsize{11}{13}\selectfont
\begin{tabular}[t]{ccrrrrrrc}
\toprule
CR & Bin & $h^2_k$ & $\hat{h}_k^2$ & Bias of $\hat{h}_k^2$ & SE of $\hat{h}_k^2$ & JSE of $\hat{h}_k^2$ & Bias of JSE & SE of JSE\\
\midrule
\multirow{11}{*}{0.2} & 1 & 0.0219 & 0.0219 & 0.0000 & 0.0025 & 0.0023 & -0.0002 & 0.0002\\
 & 2 & 0.0626 & 0.0627 & 0.0001 & 0.0030 & 0.0029 & -0.0001 & 0.0002\\
 & 3 & 0.0490 & 0.0493 & 0.0002 & 0.0030 & 0.0027 & -0.0003 & 0.0002\\
 & 4 & 0.0664 & 0.0670 & 0.0006 & 0.0030 & 0.0030 & 0.0000 & 0.0002\\
 & 5 & 0.0408 & 0.0408 & 0.0000 & 0.0027 & 0.0026 & -0.0001 & 0.0002\\
 & 6 & 0.0135 & 0.0135 & 0.0000 & 0.0020 & 0.0022 & 0.0001 & 0.0002\\
 & 7 & 0.0705 & 0.0710 & 0.0005 & 0.0028 & 0.0030 & 0.0002 & 0.0002\\
 & 8 & 0.0484 & 0.0485 & 0.0001 & 0.0027 & 0.0027 & 0.0000 & 0.0002\\
 & 9 & 0.0610 & 0.0608 & -0.0003 & 0.0030 & 0.0029 & -0.0001 & 0.0002\\
 & 10 & 0.0660 & 0.0660 & 0.0000 & 0.0029 & 0.0030 & 0.0001 & 0.0002\\
 & e & 0.5000 & 0.4987 & -0.0013 & 0.0086 & 0.0088 & 0.0002 & 0.0007\\ \midrule[0.3pt]
\multirow{11}{*}{0.5} & 1 & 0.0088 & 0.0088 & 0.0000 & 0.0063 & 0.0071 & 0.0008 & 0.0010\\
 & 2 & 0.0567 & 0.0571 & 0.0004 & 0.0081 & 0.0077 & -0.0003 & 0.0009\\
 & 3 & 0.0735 & 0.0731 & -0.0005 & 0.0083 & 0.0080 & -0.0003 & 0.0008\\
 & 4 & 0.0455 & 0.0468 & 0.0013 & 0.0085 & 0.0076 & -0.0010 & 0.0010\\
 & 5 & 0.0866 & 0.0864 & -0.0001 & 0.0102 & 0.0082 & -0.0020 & 0.0009\\
 & 6 & 0.0518 & 0.0522 & 0.0004 & 0.0080 & 0.0076 & -0.0004 & 0.0009\\
 & 7 & 0.0176 & 0.0179 & 0.0002 & 0.0066 & 0.0071 & 0.0005 & 0.0009\\
 & 8 & 0.0445 & 0.0445 & 0.0000 & 0.0065 & 0.0074 & 0.0010 & 0.0008\\
 & 9 & 0.0495 & 0.0486 & -0.0009 & 0.0075 & 0.0076 & 0.0000 & 0.0009\\
 & 10 & 0.0656 & 0.0650 & -0.0006 & 0.0078 & 0.0078 & 0.0000 & 0.0008\\
 & e & 0.5000 & 0.4998 & -0.0002 & 0.0267 & 0.0241 & -0.0027 & 0.0026\\ \midrule[0.3pt]
\multirow{11}{*}{0.8} & 1 & 0.0739 & 0.0795 & 0.0056 & 0.0520 & 0.0550 & 0.0029 & 0.0187\\
 & 2 & 0.0130 & 0.0142 & 0.0011 & 0.0536 & 0.0535 & -0.0001 & 0.0184\\
 & 3 & 0.0347 & 0.0318 & -0.0029 & 0.0508 & 0.0528 & 0.0020 & 0.0178\\
 & 4 & 0.0637 & 0.0615 & -0.0023 & 0.0653 & 0.0545 & -0.0108 & 0.0182\\
 & 5 & 0.0622 & 0.0546 & -0.0076 & 0.0614 & 0.0543 & -0.0071 & 0.0192\\
 & 6 & 0.0880 & 0.1026 & 0.0146 & 0.0522 & 0.0553 & 0.0031 & 0.0186\\
 & 7 & 0.0651 & 0.0545 & -0.0105 & 0.0663 & 0.0549 & -0.0114 & 0.0187\\
 & 8 & 0.0157 & 0.0212 & 0.0054 & 0.0559 & 0.0544 & -0.0015 & 0.0189\\
 & 9 & 0.0537 & 0.0559 & 0.0022 & 0.0563 & 0.0545 & -0.0017 & 0.0205\\
 & 10 & 0.0299 & 0.0506 & 0.0207 & 0.0706 & 0.0539 & -0.0166 & 0.0185\\
 & e & 0.5000 & 0.4736 & -0.0264 & 0.1785 & 0.1716 & -0.0069 & 0.0597\\
\bottomrule
\end{tabular}}
\end{table}

\begin{landscape}\begingroup\fontsize{10}{12}\selectfont
\begin{longtable}[t]{llllllllll}
\caption{\label{tab:cvc_mis_spec_accuracy_h2_0.2}Accuracy of SNP heritability estimation under different genetic architectures. 
The true SNP heritability $h^2_{\text{SNP}}$ was set to 0.2. 
Relative bias is the bias of estimated heritability with respect to true SNP heritability 
$\hat{h}^2_{\text{SNP}}$ 
computed using the formula 
$(\hat{h}^2_{\text{SNP}} - h^2_{\text{SNP}} / h^2_{\text{SNP}} \times 100$). 
Bias is the difference between the estimated jackknife standard error and the Monte Carlo 
standard error of SNP heritability estimates.
The values report means from 100 replications. MAF: minor allele frequency; LD: linkage disequilibrium; 
SE: standard error; JSE: Jackknife SE.}\\
\toprule
\multicolumn{1}{c}{ } & \multicolumn{3}{c}{Genetic architecture} & \multicolumn{6}{c}{Estimates} \\
\cmidrule(l{3pt}r{3pt}){2-4} \cmidrule(l{3pt}r{3pt}){5-10}
CR & CVR & MAF of causal SNPs & MAF and LD coupling & $\hat{h}_{\text{SNP}}^2$ & Relative bias (\%) & SE of $\hat{h}_{\text{SNP}}^2$ & JSE of $\hat{h}_{\text{SNP}}^2$ & Bias & SE of JSE\\
\midrule
\endfirsthead
\caption[]{Accuracy of SNP heritability estimation under different genetic architecture \textit{(continued)}}\\
\toprule
\multicolumn{1}{c}{ } & \multicolumn{3}{c}{Genetic architecture} & \multicolumn{6}{c}{Estimates} \\
\cmidrule(l{3pt}r{3pt}){2-4} \cmidrule(l{3pt}r{3pt}){5-10}
CR & CVR & MAF of causal SNPs & MAF and LD coupling & $\hat{h}_{\text{SNP}}^2$ & Relative bias (\%) & SE of $\hat{h}_{\text{SNP}}^2$ & JSE of $\hat{h}_{\text{SNP}}^2$ & Bias & SE of JSE\\
\midrule
\endhead

\endfoot
\bottomrule
\endlastfoot
0.2 & 0.01 & {}[0, 0.5] & a = 0, b = 0 & 0.196 & -1.942 & 0.006 & 0.005 & -0.001 & 0.000\\
0.2 & 0.01 & {}[0, 0.5] & a = 0, b = 1 & 0.194 & -2.946 & 0.006 & 0.006 & 0.000 & 0.000\\
0.2 & 0.01 & {}[0, 0.5] & a = 0.75, b = 0 & 0.198 & -1.111 & 0.008 & 0.007 & -0.001 & 0.001\\
0.2 & 0.01 & {}[0, 0.5] & a = 0.75, b = 1 & 0.194 & -2.915 & 0.010 & 0.007 & -0.003 & 0.001\\
0.2 & 1.00 & {}[0, 0.5] & a = 0, b = 0 & 0.199 & -0.640 & 0.005 & 0.004 & -0.000 & 0.000\\
0.2 & 1.00 & {}[0, 0.5] & a = 0, b = 1 & 0.196 & -1.872 & 0.004 & 0.004 & -0.000 & 0.000\\
0.2 & 1.00 & {}[0, 0.5] & a = 0.75, b = 0 & 0.200 & -0.037 & 0.005 & 0.005 & 0.000 & 0.001\\
0.2 & 1.00 & {}[0, 0.5] & a = 0.75, b = 1 & 0.196 & -1.970 & 0.004 & 0.004 & -0.000 & 0.000\\
0.2 & 0.01 & {}[0.01, 0.05] & a = 0, b = 0 & 0.201 & 0.400 & 0.006 & 0.005 & -0.001 & 0.000\\
0.2 & 0.01 & {}[0.01, 0.05] & a = 0, b = 1 & 0.199 & -0.321 & 0.006 & 0.005 & -0.000 & 0.000\\
0.2 & 0.01 & {}[0.01, 0.05] & a = 0.75, b = 0 & 0.201 & 0.522 & 0.008 & 0.006 & -0.002 & 0.001\\
0.2 & 0.01 & {}[0.01, 0.05] & a = 0.75, b = 1 & 0.201 & 0.291 & 0.011 & 0.007 & -0.004 & 0.001\\
0.2 & 0.01 & {}[0.05, 0.5] & a = 0, b = 0 & 0.199 & -0.511 & 0.005 & 0.006 & 0.001 & 0.001\\
0.2 & 0.01 & {}[0.05, 0.5] & a = 0, b = 1 & 0.195 & -2.551 & 0.006 & 0.006 & -0.000 & 0.001\\
0.2 & 0.01 & {}[0.05, 0.5] & a = 0.75, b = 0 & 0.200 & -0.082 & 0.006 & 0.007 & 0.001 & 0.001\\
0.2 & 0.01 & {}[0.05, 0.5] & a = 0.75, b = 1 & 0.195 & -2.286 & 0.008 & 0.006 & -0.002 & 0.001\\
\midrule[0.3pt]
0.5 & 0.01 & {}[0, 0.5] & a = 0, b = 0 & 0.196 & -2.201 & 0.014 & 0.011 & -0.003 & 0.003\\
0.5 & 0.01 & {}[0, 0.5] & a = 0, b = 1 & 0.194 & -3.081 & 0.013 & 0.011 & -0.002 & 0.001\\
0.5 & 0.01 & {}[0, 0.5] & a = 0.75, b = 0 & 0.200 & -0.174 & 0.018 & 0.014 & -0.003 & 0.001\\
0.5 & 0.01 & {}[0, 0.5] & a = 0.75, b = 1 & 0.196 & -1.823 & 0.026 & 0.021 & -0.005 & 0.002\\
0.5 & 1.00 & {}[0, 0.5] & a = 0, b = 0 & 0.198 & -1.022 & 0.010 & 0.009 & -0.001 & 0.001\\
0.5 & 1.00 & {}[0, 0.5] & a = 0, b = 1 & 0.197 & -1.699 & 0.010 & 0.009 & -0.001 & 0.001\\
0.5 & 1.00 & {}[0, 0.5] & a = 0.75, b = 0 & 0.198 & -0.817 & 0.011 & 0.010 & -0.002 & 0.001\\
0.5 & 1.00 & {}[0, 0.5] & a = 0.75, b = 1 & 0.196 & -1.863 & 0.010 & 0.009 & -0.001 & 0.001\\
0.5 & 0.01 & {}[0.01, 0.05] & a = 0, b = 0 & 0.202 & 1.175 & 0.012 & 0.010 & -0.002 & 0.001\\
0.5 & 0.01 & {}[0.01, 0.05] & a = 0, b = 1 & 0.201 & 0.686 & 0.013 & 0.011 & -0.002 & 0.002\\
0.5 & 0.01 & {}[0.01, 0.05] & a = 0.75, b = 0 & 0.204 & 2.052 & 0.025 & 0.019 & -0.006 & 0.002\\
0.5 & 0.01 & {}[0.01, 0.05] & a = 0.75, b = 1 & 0.204 & 1.796 & 0.030 & 0.023 & -0.006 & 0.003\\
0.5 & 0.01 & {}[0.05, 0.5] & a = 0, b = 0 & 0.198 & -1.107 & 0.012 & 0.011 & -0.001 & 0.001\\
0.5 & 0.01 & {}[0.05, 0.5] & a = 0, b = 1 & 0.194 & -2.861 & 0.013 & 0.012 & -0.001 & 0.001\\
0.5 & 0.01 & {}[0.05, 0.5] & a = 0.75, b = 0 & 0.200 & 0.019 & 0.015 & 0.013 & -0.002 & 0.001\\
0.5 & 0.01 & {}[0.05, 0.5] & a = 0.75, b = 1 & 0.198 & -1.100 & 0.023 & 0.018 & -0.005 & 0.002\\
\midrule[0.3pt]
0.7 & 0.01 & {}[0, 0.5] & a = 0, b = 0 & 0.194 & -3.188 & 0.038 & 0.033 & -0.005 & 0.006\\
0.7 & 0.01 & {}[0, 0.5] & a = 0, b = 1 & 0.191 & -4.653 & 0.044 & 0.035 & -0.010 & 0.005\\
0.7 & 0.01 & {}[0, 0.5] & a = 0.75, b = 0 & 0.195 & -2.361 & 0.059 & 0.048 & -0.010 & 0.007\\
0.7 & 0.01 & {}[0, 0.5] & a = 0.75, b = 1 & 0.191 & -4.358 & 0.093 & 0.079 & -0.014 & 0.014\\
0.7 & 1.00 & {}[0, 0.5] & a = 0, b = 0 & 0.195 & -2.251 & 0.039 & 0.031 & -0.008 & 0.009\\
0.7 & 1.00 & {}[0, 0.5] & a = 0, b = 1 & 0.190 & -5.083 & 0.036 & 0.031 & -0.006 & 0.009\\
0.7 & 1.00 & {}[0, 0.5] & a = 0.75, b = 0 & 0.198 & -0.819 & 0.039 & 0.031 & -0.008 & 0.007\\
0.7 & 1.00 & {}[0, 0.5] & a = 0.75, b = 1 & 0.192 & -4.093 & 0.035 & 0.031 & -0.004 & 0.008\\
0.7 & 0.01 & {}[0.01, 0.05] & a = 0, b = 0 & 0.201 & 0.602 & 0.040 & 0.033 & -0.007 & 0.006\\
0.7 & 0.01 & {}[0.01, 0.05] & a = 0, b = 1 & 0.201 & 0.577 & 0.040 & 0.033 & -0.007 & 0.007\\
0.7 & 0.01 & {}[0.01, 0.05] & a = 0.75, b = 0 & 0.195 & -2.343 & 0.087 & 0.072 & -0.014 & 0.014\\
0.7 & 0.01 & {}[0.01, 0.05] & a = 0.75, b = 1 & 0.196 & -1.881 & 0.110 & 0.091 & -0.020 & 0.018\\
0.7 & 0.01 & {}[0.05, 0.5] & a = 0, b = 0 & 0.200 & -0.164 & 0.044 & 0.033 & -0.011 & 0.005\\
0.7 & 0.01 & {}[0.05, 0.5] & a = 0, b = 1 & 0.194 & -3.197 & 0.047 & 0.038 & -0.009 & 0.007\\
0.7 & 0.01 & {}[0.05, 0.5] & a = 0.75, b = 0 & 0.203 & 1.276 & 0.053 & 0.042 & -0.011 & 0.008\\
0.7 & 0.01 & {}[0.05, 0.5] & a = 0.75, b = 1 & 0.198 & -0.906 & 0.076 & 0.067 & -0.009 & 0.015\\
\midrule[0.3pt]
0.8 & 0.01 & {}[0, 0.5] & a = 0, b = 0 & 0.211 & 5.644 & 0.098 & 0.078 & -0.020 & 0.021\\
0.8 & 0.01 & {}[0, 0.5] & a = 0, b = 1 & 0.203 & 1.382 & 0.098 & 0.080 & -0.018 & 0.017\\
0.8 & 0.01 & {}[0, 0.5] & a = 0.75, b = 0 & 0.208 & 4.127 & 0.138 & 0.118 & -0.020 & 0.029\\
0.8 & 0.01 & {}[0, 0.5] & a = 0.75, b = 1 & 0.240 & 19.793 & 0.231 & 0.190 & -0.041 & 0.040\\
0.8 & 1.00 & {}[0, 0.5] & a = 0, b = 0 & 0.208 & 3.953 & 0.098 & 0.073 & -0.025 & 0.026\\
0.8 & 1.00 & {}[0, 0.5] & a = 0, b = 1 & 0.207 & 3.555 & 0.094 & 0.072 & -0.022 & 0.019\\
0.8 & 1.00 & {}[0, 0.5] & a = 0.75, b = 0 & 0.205 & 2.630 & 0.092 & 0.074 & -0.018 & 0.023\\
0.8 & 1.00 & {}[0, 0.5] & a = 0.75, b = 1 & 0.202 & 0.845 & 0.091 & 0.073 & -0.018 & 0.020\\
0.8 & 0.01 & {}[0.01, 0.05] & a = 0, b = 0 & 0.206 & 3.064 & 0.097 & 0.075 & -0.022 & 0.018\\
0.8 & 0.01 & {}[0.01, 0.05] & a = 0, b = 1 & 0.204 & 2.173 & 0.097 & 0.077 & -0.020 & 0.017\\
0.8 & 0.01 & {}[0.01, 0.05] & a = 0.75, b = 0 & 0.213 & 6.497 & 0.226 & 0.182 & -0.044 & 0.051\\
0.8 & 0.01 & {}[0.01, 0.05] & a = 0.75, b = 1 & 0.226 & 13.171 & 0.285 & 0.227 & -0.058 & 0.062\\
0.8 & 0.01 & {}[0.05, 0.5] & a = 0, b = 0 & 0.194 & -2.915 & 0.094 & 0.077 & -0.017 & 0.017\\
0.8 & 0.01 & {}[0.05, 0.5] & a = 0, b = 1 & 0.192 & -4.106 & 0.094 & 0.088 & -0.006 & 0.020\\
0.8 & 0.01 & {}[0.05, 0.5] & a = 0.75, b = 0 & 0.200 & -0.090 & 0.109 & 0.100 & -0.009 & 0.022\\
0.8 & 0.01 & {}[0.05, 0.5] & a = 0.75, b = 1 & 0.204 & 1.968 & 0.182 & 0.164 & -0.018 & 0.044\\*
\end{longtable}
\endgroup{}
\end{landscape}

\begin{landscape}\begingroup\fontsize{10}{12}\selectfont
\begin{longtable}[t]{llllllllll}
\caption{\label{tab:cvc_mis_spec_accuracy_h2_0.5}Accuracy of SNP heritability estimation under different genetic architectures. 
The true SNP heritability $h^2_{\text{SNP}}$ was set to 0.5. 
Relative bias is the bias of estimated heritability with respect to true SNP heritability 
$\hat{h}^2_{\text{SNP}}$ 
computed using the formula 
$(\hat{h}^2_{\text{SNP}} - h^2_{\text{SNP}}) / h^2_{\text{SNP}} \times 100$. 
Bias is the difference between the estimated jackknife standard error and the Monte Carlo 
standard error of SNP heritability estimates.
The values report means from 100 replications. MAF: minor allele frequency; LD: linkage disequilibrium; 
SE: standard error; JSE: Jackknife SE.}\\
\toprule
\multicolumn{1}{c}{ } & \multicolumn{3}{c}{Genetic architecture} & \multicolumn{6}{c}{Estimates} \\
\cmidrule(l{3pt}r{3pt}){2-4} \cmidrule(l{3pt}r{3pt}){5-10}
CR & CVR & MAF of causal SNPs & MAF and LD coupling & $\hat{h}_{\text{SNP}}^2$ & Relative bias (\%) & SE of $\hat{h}_{\text{SNP}}^2$ & JSE of $\hat{h}_{\text{SNP}}^2$ & Bias & SE of JSE\\
\midrule
\endfirsthead
\caption[]{Accuracy of SNP heritability estimation under different genetic architecture \textit{(continued)}}\\
\toprule
\multicolumn{1}{c}{ } & \multicolumn{3}{c}{Genetic architecture} & \multicolumn{6}{c}{Estimates} \\
\cmidrule(l{3pt}r{3pt}){2-4} \cmidrule(l{3pt}r{3pt}){5-10}
CR & CVR & MAF of causal SNPs & MAF and LD coupling & $\hat{h}_{\text{SNP}}^2$ & Relative bias (\%) & SE of $\hat{h}_{\text{SNP}}^2$ & JSE of $\hat{h}_{\text{SNP}}^2$ & Bias & SE of JSE\\
\midrule
\endhead

\endfoot
\bottomrule
\endlastfoot
0.2 & 0.01 & {}[0, 0.5] & a = 0, b = 0 & 0.493 & -1.432 & 0.010 & 0.012 & 0.002 & 0.001\\
0.2 & 0.01 & {}[0, 0.5] & a = 0, b = 1 & 0.488 & -2.341 & 0.010 & 0.013 & 0.003 & 0.001\\
0.2 & 0.01 & {}[0, 0.5] & a = 0.75, b = 0 & 0.497 & -0.680 & 0.014 & 0.015 & 0.000 & 0.001\\
0.2 & 0.01 & {}[0, 0.5] & a = 0.75, b = 1 & 0.489 & -2.165 & 0.027 & 0.015 & -0.012 & 0.001\\
0.2 & 1.00 & {}[0, 0.5] & a = 0, b = 0 & 0.500 & -0.011 & 0.008 & 0.009 & 0.000 & 0.001\\
0.2 & 1.00 & {}[0, 0.5] & a = 0, b = 1 & 0.494 & -1.151 & 0.006 & 0.007 & 0.001 & 0.001\\
0.2 & 1.00 & {}[0, 0.5] & a = 0.75, b = 0 & 0.499 & -0.166 & 0.008 & 0.011 & 0.003 & 0.001\\
0.2 & 1.00 & {}[0, 0.5] & a = 0.75, b = 1 & 0.492 & -1.535 & 0.006 & 0.008 & 0.002 & 0.001\\
0.2 & 0.01 & {}[0.01, 0.05] & a = 0, b = 0 & 0.505 & 0.927 & 0.009 & 0.011 & 0.002 & 0.001\\
0.2 & 0.01 & {}[0.01, 0.05] & a = 0, b = 1 & 0.504 & 0.761 & 0.010 & 0.012 & 0.002 & 0.001\\
0.2 & 0.01 & {}[0.01, 0.05] & a = 0.75, b = 0 & 0.503 & 0.644 & 0.020 & 0.013 & -0.007 & 0.001\\
0.2 & 0.01 & {}[0.01, 0.05] & a = 0.75, b = 1 & 0.502 & 0.458 & 0.028 & 0.014 & -0.014 & 0.001\\
0.2 & 0.01 & {}[0.05, 0.5] & a = 0, b = 0 & 0.500 & 0.006 & 0.010 & 0.014 & 0.004 & 0.002\\
0.2 & 0.01 & {}[0.05, 0.5] & a = 0, b = 1 & 0.489 & -2.200 & 0.012 & 0.013 & 0.002 & 0.001\\
0.2 & 0.01 & {}[0.05, 0.5] & a = 0.75, b = 0 & 0.503 & 0.568 & 0.013 & 0.015 & 0.002 & 0.002\\
0.2 & 0.01 & {}[0.05, 0.5] & a = 0.75, b = 1 & 0.491 & -1.780 & 0.021 & 0.014 & -0.007 & 0.001\\
\midrule[0.3pt]
0.5 & 0.01 & {}[0, 0.5] & a = 0, b = 0 & 0.492 & -1.511 & 0.018 & 0.016 & -0.002 & 0.002\\
0.5 & 0.01 & {}[0, 0.5] & a = 0, b = 1 & 0.489 & -2.124 & 0.019 & 0.018 & -0.001 & 0.002\\
0.5 & 0.01 & {}[0, 0.5] & a = 0.75, b = 0 & 0.499 & -0.207 & 0.030 & 0.025 & -0.005 & 0.003\\
0.5 & 0.01 & {}[0, 0.5] & a = 0.75, b = 1 & 0.493 & -1.354 & 0.052 & 0.040 & -0.011 & 0.005\\
0.5 & 1.00 & {}[0, 0.5] & a = 0, b = 0 & 0.499 & -0.195 & 0.011 & 0.012 & 0.001 & 0.001\\
0.5 & 1.00 & {}[0, 0.5] & a = 0, b = 1 & 0.495 & -1.057 & 0.011 & 0.011 & 0.000 & 0.001\\
0.5 & 1.00 & {}[0, 0.5] & a = 0.75, b = 0 & 0.500 & 0.037 & 0.015 & 0.014 & -0.000 & 0.002\\
0.5 & 1.00 & {}[0, 0.5] & a = 0.75, b = 1 & 0.493 & -1.407 & 0.013 & 0.012 & -0.000 & 0.001\\
0.5 & 0.01 & {}[0.01, 0.05] & a = 0, b = 0 & 0.505 & 1.026 & 0.015 & 0.015 & 0.000 & 0.002\\
0.5 & 0.01 & {}[0.01, 0.05] & a = 0, b = 1 & 0.505 & 0.999 & 0.016 & 0.016 & -0.000 & 0.002\\
0.5 & 0.01 & {}[0.01, 0.05] & a = 0.75, b = 0 & 0.511 & 2.128 & 0.047 & 0.036 & -0.011 & 0.003\\
0.5 & 0.01 & {}[0.01, 0.05] & a = 0.75, b = 1 & 0.508 & 1.565 & 0.064 & 0.046 & -0.018 & 0.005\\
0.5 & 0.01 & {}[0.05, 0.5] & a = 0, b = 0 & 0.498 & -0.400 & 0.016 & 0.017 & 0.002 & 0.002\\
0.5 & 0.01 & {}[0.05, 0.5] & a = 0, b = 1 & 0.490 & -1.922 & 0.019 & 0.019 & 0.000 & 0.002\\
0.5 & 0.01 & {}[0.05, 0.5] & a = 0.75, b = 0 & 0.504 & 0.761 & 0.023 & 0.022 & -0.001 & 0.002\\
0.5 & 0.01 & {}[0.05, 0.5] & a = 0.75, b = 1 & 0.498 & -0.457 & 0.043 & 0.033 & -0.010 & 0.004\\
\midrule[0.3pt]
0.7 & 0.01 & {}[0, 0.5] & a = 0, b = 0 & 0.490 & -2.041 & 0.046 & 0.040 & -0.006 & 0.008\\
0.7 & 0.01 & {}[0, 0.5] & a = 0, b = 1 & 0.487 & -2.669 & 0.059 & 0.045 & -0.014 & 0.007\\
0.7 & 0.01 & {}[0, 0.5] & a = 0.75, b = 0 & 0.496 & -0.859 & 0.098 & 0.077 & -0.020 & 0.012\\
0.7 & 0.01 & {}[0, 0.5] & a = 0.75, b = 1 & 0.483 & -3.444 & 0.186 & 0.151 & -0.035 & 0.028\\
0.7 & 1.00 & {}[0, 0.5] & a = 0, b = 0 & 0.497 & -0.621 & 0.039 & 0.033 & -0.006 & 0.007\\
0.7 & 1.00 & {}[0, 0.5] & a = 0, b = 1 & 0.486 & -2.745 & 0.040 & 0.031 & -0.009 & 0.005\\
0.7 & 1.00 & {}[0, 0.5] & a = 0.75, b = 0 & 0.502 & 0.339 & 0.045 & 0.033 & -0.012 & 0.010\\
0.7 & 1.00 & {}[0, 0.5] & a = 0.75, b = 1 & 0.489 & -2.156 & 0.037 & 0.033 & -0.004 & 0.008\\
0.7 & 0.01 & {}[0.01, 0.05] & a = 0, b = 0 & 0.505 & 0.974 & 0.044 & 0.039 & -0.006 & 0.008\\
0.7 & 0.01 & {}[0.01, 0.05] & a = 0, b = 1 & 0.504 & 0.721 & 0.049 & 0.041 & -0.007 & 0.009\\
0.7 & 0.01 & {}[0.01, 0.05] & a = 0.75, b = 0 & 0.500 & -0.022 & 0.169 & 0.141 & -0.028 & 0.030\\
0.7 & 0.01 & {}[0.01, 0.05] & a = 0.75, b = 1 & 0.505 & 1.070 & 0.224 & 0.185 & -0.039 & 0.040\\
0.7 & 0.01 & {}[0.05, 0.5] & a = 0, b = 0 & 0.501 & 0.269 & 0.049 & 0.039 & -0.009 & 0.006\\
0.7 & 0.01 & {}[0.05, 0.5] & a = 0, b = 1 & 0.491 & -1.731 & 0.056 & 0.051 & -0.005 & 0.010\\
0.7 & 0.01 & {}[0.05, 0.5] & a = 0.75, b = 0 & 0.508 & 1.545 & 0.074 & 0.063 & -0.011 & 0.013\\
0.7 & 0.01 & {}[0.05, 0.5] & a = 0.75, b = 1 & 0.508 & 1.592 & 0.148 & 0.125 & -0.023 & 0.031\\
\midrule[0.3pt]
0.8 & 0.01 & {}[0, 0.5] & a = 0, b = 0 & 0.494 & -1.225 & 0.116 & 0.089 & -0.027 & 0.021\\
0.8 & 0.01 & {}[0, 0.5] & a = 0, b = 1 & 0.499 & -0.151 & 0.135 & 0.102 & -0.033 & 0.023\\
0.8 & 0.01 & {}[0, 0.5] & a = 0.75, b = 0 & 0.521 & 4.238 & 0.244 & 0.186 & -0.058 & 0.040\\
0.8 & 0.01 & {}[0, 0.5] & a = 0.75, b = 1 & 0.563 & 12.698 & 0.473 & 0.374 & -0.099 & 0.094\\
0.8 & 1.00 & {}[0, 0.5] & a = 0, b = 0 & 0.493 & -1.446 & 0.091 & 0.072 & -0.019 & 0.016\\
0.8 & 1.00 & {}[0, 0.5] & a = 0, b = 1 & 0.499 & -0.253 & 0.111 & 0.074 & -0.037 & 0.023\\
0.8 & 1.00 & {}[0, 0.5] & a = 0.75, b = 0 & 0.495 & -1.015 & 0.094 & 0.075 & -0.019 & 0.020\\
0.8 & 1.00 & {}[0, 0.5] & a = 0.75, b = 1 & 0.494 & -1.266 & 0.091 & 0.077 & -0.014 & 0.019\\
0.8 & 0.01 & {}[0.01, 0.05] & a = 0, b = 0 & 0.499 & -0.238 & 0.112 & 0.088 & -0.023 & 0.025\\
0.8 & 0.01 & {}[0.01, 0.05] & a = 0, b = 1 & 0.505 & 1.052 & 0.113 & 0.095 & -0.018 & 0.021\\
0.8 & 0.01 & {}[0.01, 0.05] & a = 0.75, b = 0 & 0.562 & 12.485 & 0.448 & 0.340 & -0.108 & 0.085\\
0.8 & 0.01 & {}[0.01, 0.05] & a = 0.75, b = 1 & 0.568 & 13.569 & 0.585 & 0.447 & -0.138 & 0.126\\
0.8 & 0.01 & {}[0.05, 0.5] & a = 0, b = 0 & 0.498 & -0.398 & 0.109 & 0.090 & -0.019 & 0.018\\
0.8 & 0.01 & {}[0.05, 0.5] & a = 0, b = 1 & 0.491 & -1.863 & 0.138 & 0.123 & -0.015 & 0.031\\
0.8 & 0.01 & {}[0.05, 0.5] & a = 0.75, b = 0 & 0.513 & 2.578 & 0.161 & 0.146 & -0.015 & 0.028\\
0.8 & 0.01 & {}[0.05, 0.5] & a = 0.75, b = 1 & 0.521 & 4.292 & 0.384 & 0.310 & -0.074 & 0.090\\*
\end{longtable}
\endgroup{}
\end{landscape}

\begin{landscape}\begingroup\fontsize{10}{12}\selectfont
\begin{longtable}[t]{llllllllll}
\caption{\label{tab:cvc_mis_spec_accuracy_h2_0.8}Accuracy of SNP heritability estimation under different genetic architectures. 
The true SNP heritability $h^2_{\text{SNP}}$ was set to 0.8. 
Relative bias is the bias of estimated heritability with respect to true SNP heritability 
$\hat{h}^2_{\text{SNP}}$ 
computed using the formula 
$(\hat{h}^2_{\text{SNP}} - h^2_{\text{SNP}} / h^2_{\text{SNP}} \times 100$). 
Bias is the difference between the estimated jackknife standard error and the Monte Carlo 
standard error of SNP heritability estimates.
The values report means from 100 replications. MAF: minor allele frequency; LD: linkage disequilibrium; 
SE: standard error; JSE: Jackknife SE.}\\
\toprule
\multicolumn{1}{c}{ } & \multicolumn{3}{c}{Genetic architecture} & \multicolumn{6}{c}{Estimates} \\
\cmidrule(l{3pt}r{3pt}){2-4} \cmidrule(l{3pt}r{3pt}){5-10}
CR & CVR & MAF of causal SNPs & MAF and LD coupling & $\hat{h}_{\text{SNP}}^2$ & Relative bias (\%) & SE of $\hat{h}_{\text{SNP}}^2$ & JSE of $\hat{h}_{\text{SNP}}^2$ & Bias & SE of JSE\\
\midrule
\endfirsthead
\caption[]{Accuracy of SNP heritability estimation under different genetic architecture \textit{(continued)}}\\
\toprule
\multicolumn{1}{c}{ } & \multicolumn{3}{c}{Genetic architecture} & \multicolumn{6}{c}{Estimates} \\
\cmidrule(l{3pt}r{3pt}){2-4} \cmidrule(l{3pt}r{3pt}){5-10}
CR & CVR & MAF of causal SNPs & MAF and LD coupling & $\hat{h}_{\text{SNP}}^2$ & Relative bias (\%) & SE of $\hat{h}_{\text{SNP}}^2$ & JSE of $\hat{h}_{\text{SNP}}^2$ & Bias & SE of JSE\\
\midrule
\endhead

\endfoot
\bottomrule
\endlastfoot
0.2 & 0.01 & {}[0, 0.5] & a = 0, b = 0 & 0.792 & -0.956 & 0.012 & 0.018 & 0.006 & 0.001\\
0.2 & 0.01 & {}[0, 0.5] & a = 0, b = 1 & 0.783 & -2.070 & 0.012 & 0.020 & 0.009 & 0.001\\
0.2 & 0.01 & {}[0, 0.5] & a = 0.75, b = 0 & 0.796 & -0.477 & 0.023 & 0.023 & 0.000 & 0.002\\
0.2 & 0.01 & {}[0, 0.5] & a = 0.75, b = 1 & 0.781 & -2.345 & 0.043 & 0.022 & -0.020 & 0.002\\
0.2 & 1.00 & {}[0, 0.5] & a = 0, b = 0 & 0.799 & -0.084 & 0.009 & 0.013 & 0.004 & 0.001\\
0.2 & 1.00 & {}[0, 0.5] & a = 0, b = 1 & 0.791 & -1.067 & 0.008 & 0.010 & 0.002 & 0.001\\
0.2 & 1.00 & {}[0, 0.5] & a = 0.75, b = 0 & 0.800 & 0.044 & 0.012 & 0.017 & 0.006 & 0.002\\
0.2 & 1.00 & {}[0, 0.5] & a = 0.75, b = 1 & 0.789 & -1.390 & 0.009 & 0.012 & 0.004 & 0.001\\
0.2 & 0.01 & {}[0.01, 0.05] & a = 0, b = 0 & 0.809 & 1.177 & 0.010 & 0.017 & 0.007 & 0.002\\
0.2 & 0.01 & {}[0.01, 0.05] & a = 0, b = 1 & 0.805 & 0.609 & 0.011 & 0.018 & 0.006 & 0.002\\
0.2 & 0.01 & {}[0.01, 0.05] & a = 0.75, b = 0 & 0.810 & 1.206 & 0.038 & 0.019 & -0.019 & 0.002\\
0.2 & 0.01 & {}[0.01, 0.05] & a = 0.75, b = 1 & 0.802 & 0.293 & 0.052 & 0.021 & -0.031 & 0.002\\
0.2 & 0.01 & {}[0.05, 0.5] & a = 0, b = 0 & 0.801 & 0.152 & 0.015 & 0.022 & 0.007 & 0.002\\
0.2 & 0.01 & {}[0.05, 0.5] & a = 0, b = 1 & 0.783 & -2.110 & 0.016 & 0.021 & 0.005 & 0.002\\
0.2 & 0.01 & {}[0.05, 0.5] & a = 0.75, b = 0 & 0.806 & 0.762 & 0.020 & 0.023 & 0.003 & 0.003\\
0.2 & 0.01 & {}[0.05, 0.5] & a = 0.75, b = 1 & 0.787 & -1.620 & 0.031 & 0.022 & -0.009 & 0.002\\
\midrule[0.3pt]
0.5 & 0.01 & {}[0, 0.5] & a = 0, b = 0 & 0.791 & -1.083 & 0.021 & 0.022 & 0.001 & 0.002\\
0.5 & 0.01 & {}[0, 0.5] & a = 0, b = 1 & 0.785 & -1.822 & 0.023 & 0.025 & 0.002 & 0.002\\
0.5 & 0.01 & {}[0, 0.5] & a = 0.75, b = 0 & 0.801 & 0.151 & 0.043 & 0.036 & -0.007 & 0.004\\
0.5 & 0.01 & {}[0, 0.5] & a = 0.75, b = 1 & 0.797 & -0.333 & 0.083 & 0.060 & -0.023 & 0.008\\
0.5 & 1.00 & {}[0, 0.5] & a = 0, b = 0 & 0.799 & -0.142 & 0.013 & 0.016 & 0.003 & 0.002\\
0.5 & 1.00 & {}[0, 0.5] & a = 0, b = 1 & 0.792 & -0.949 & 0.015 & 0.014 & -0.001 & 0.001\\
0.5 & 1.00 & {}[0, 0.5] & a = 0.75, b = 0 & 0.801 & 0.083 & 0.018 & 0.020 & 0.001 & 0.002\\
0.5 & 1.00 & {}[0, 0.5] & a = 0.75, b = 1 & 0.789 & -1.351 & 0.015 & 0.016 & 0.001 & 0.001\\
0.5 & 0.01 & {}[0.01, 0.05] & a = 0, b = 0 & 0.809 & 1.128 & 0.017 & 0.021 & 0.004 & 0.002\\
0.5 & 0.01 & {}[0.01, 0.05] & a = 0, b = 1 & 0.807 & 0.919 & 0.019 & 0.022 & 0.003 & 0.002\\
0.5 & 0.01 & {}[0.01, 0.05] & a = 0.75, b = 0 & 0.821 & 2.598 & 0.073 & 0.054 & -0.019 & 0.006\\
0.5 & 0.01 & {}[0.01, 0.05] & a = 0.75, b = 1 & 0.816 & 1.945 & 0.104 & 0.069 & -0.035 & 0.008\\
0.5 & 0.01 & {}[0.05, 0.5] & a = 0, b = 0 & 0.801 & 0.126 & 0.022 & 0.025 & 0.003 & 0.002\\
0.5 & 0.01 & {}[0.05, 0.5] & a = 0, b = 1 & 0.787 & -1.684 & 0.024 & 0.027 & 0.002 & 0.003\\
0.5 & 0.01 & {}[0.05, 0.5] & a = 0.75, b = 0 & 0.809 & 1.131 & 0.030 & 0.031 & 0.001 & 0.003\\
0.5 & 0.01 & {}[0.05, 0.5] & a = 0.75, b = 1 & 0.799 & -0.113 & 0.055 & 0.049 & -0.007 & 0.006\\
\midrule[0.3pt]
0.7 & 0.01 & {}[0, 0.5] & a = 0, b = 0 & 0.786 & -1.767 & 0.050 & 0.046 & -0.004 & 0.007\\
0.7 & 0.01 & {}[0, 0.5] & a = 0, b = 1 & 0.780 & -2.503 & 0.065 & 0.056 & -0.008 & 0.009\\
0.7 & 0.01 & {}[0, 0.5] & a = 0.75, b = 0 & 0.795 & -0.665 & 0.130 & 0.110 & -0.020 & 0.020\\
0.7 & 0.01 & {}[0, 0.5] & a = 0.75, b = 1 & 0.783 & -2.086 & 0.295 & 0.235 & -0.060 & 0.069\\
0.7 & 1.00 & {}[0, 0.5] & a = 0, b = 0 & 0.801 & 0.080 & 0.041 & 0.035 & -0.006 & 0.006\\
0.7 & 1.00 & {}[0, 0.5] & a = 0, b = 1 & 0.787 & -1.586 & 0.040 & 0.033 & -0.007 & 0.006\\
0.7 & 1.00 & {}[0, 0.5] & a = 0.75, b = 0 & 0.807 & 0.909 & 0.040 & 0.036 & -0.003 & 0.006\\
0.7 & 1.00 & {}[0, 0.5] & a = 0.75, b = 1 & 0.791 & -1.167 & 0.039 & 0.035 & -0.004 & 0.006\\
0.7 & 0.01 & {}[0.01, 0.05] & a = 0, b = 0 & 0.813 & 1.566 & 0.054 & 0.047 & -0.007 & 0.009\\
0.7 & 0.01 & {}[0.01, 0.05] & a = 0, b = 1 & 0.810 & 1.295 & 0.057 & 0.051 & -0.007 & 0.009\\
0.7 & 0.01 & {}[0.01, 0.05] & a = 0.75, b = 0 & 0.827 & 3.335 & 0.248 & 0.204 & -0.043 & 0.041\\
0.7 & 0.01 & {}[0.01, 0.05] & a = 0.75, b = 1 & 0.820 & 2.466 & 0.330 & 0.278 & -0.052 & 0.071\\
0.7 & 0.01 & {}[0.05, 0.5] & a = 0, b = 0 & 0.803 & 0.321 & 0.054 & 0.047 & -0.007 & 0.007\\
0.7 & 0.01 & {}[0.05, 0.5] & a = 0, b = 1 & 0.790 & -1.215 & 0.076 & 0.066 & -0.010 & 0.014\\
0.7 & 0.01 & {}[0.05, 0.5] & a = 0.75, b = 0 & 0.815 & 1.856 & 0.088 & 0.083 & -0.006 & 0.016\\
0.7 & 0.01 & {}[0.05, 0.5] & a = 0.75, b = 1 & 0.812 & 1.501 & 0.210 & 0.180 & -0.030 & 0.043\\
\midrule[0.3pt]
0.8 & 0.01 & {}[0, 0.5] & a = 0, b = 0 & 0.789 & -1.408 & 0.129 & 0.099 & -0.030 & 0.023\\
0.8 & 0.01 & {}[0, 0.5] & a = 0, b = 1 & 0.788 & -1.455 & 0.154 & 0.123 & -0.031 & 0.031\\
0.8 & 0.01 & {}[0, 0.5] & a = 0.75, b = 0 & 0.829 & 3.676 & 0.334 & 0.263 & -0.071 & 0.070\\
0.8 & 0.01 & {}[0, 0.5] & a = 0.75, b = 1 & 0.880 & 9.972 & 0.688 & 0.559 & -0.129 & 0.187\\
0.8 & 1.00 & {}[0, 0.5] & a = 0, b = 0 & 0.791 & -1.117 & 0.101 & 0.078 & -0.023 & 0.021\\
0.8 & 1.00 & {}[0, 0.5] & a = 0, b = 1 & 0.795 & -0.681 & 0.129 & 0.076 & -0.054 & 0.023\\
0.8 & 1.00 & {}[0, 0.5] & a = 0.75, b = 0 & 0.791 & -1.100 & 0.094 & 0.077 & -0.017 & 0.018\\
0.8 & 1.00 & {}[0, 0.5] & a = 0.75, b = 1 & 0.784 & -1.983 & 0.094 & 0.081 & -0.013 & 0.020\\
0.8 & 0.01 & {}[0.01, 0.05] & a = 0, b = 0 & 0.801 & 0.126 & 0.115 & 0.103 & -0.012 & 0.025\\
0.8 & 0.01 & {}[0.01, 0.05] & a = 0, b = 1 & 0.807 & 0.896 & 0.130 & 0.112 & -0.018 & 0.027\\
0.8 & 0.01 & {}[0.01, 0.05] & a = 0.75, b = 0 & 0.882 & 10.294 & 0.670 & 0.506 & -0.164 & 0.170\\
0.8 & 0.01 & {}[0.01, 0.05] & a = 0.75, b = 1 & 0.923 & 15.368 & 0.856 & 0.695 & -0.161 & 0.289\\
0.8 & 0.01 & {}[0.05, 0.5] & a = 0, b = 0 & 0.814 & 1.756 & 0.132 & 0.106 & -0.026 & 0.028\\
0.8 & 0.01 & {}[0.05, 0.5] & a = 0, b = 1 & 0.787 & -1.569 & 0.193 & 0.156 & -0.037 & 0.047\\
0.8 & 0.01 & {}[0.05, 0.5] & a = 0.75, b = 0 & 0.832 & 3.977 & 0.231 & 0.190 & -0.041 & 0.042\\
0.8 & 0.01 & {}[0.05, 0.5] & a = 0.75, b = 1 & 0.835 & 4.397 & 0.594 & 0.441 & -0.153 & 0.127\\*
\end{longtable}
\endgroup{}
\end{landscape}

\begin{table}[H]
\centering
\caption{Heritability contribution of causal bins and non-causal bins under different genetic architectures. 
The CVR was set to 1\%, and $h^2_{\text{SNP}}$ was set to 0.5. 
The relative bias is the bias of estimated heritability contribution from causal bins $\hat{h}^2_{\text{causal}}$ 
relative to the true SNP heritability computed using the formula 
$(\hat{h}^2_{\text{causal}} - h^2_{\text{SNP}}) / h^2_{\text{SNP}} \times 100$. 
The values inside parentheses indicate standard errors of heritability estimates from 100 replications.
MAF: minor allele frequency; LD: linkage disequilibrium.}
\label{tab:cvc_mis_spec_partitioning_h2_0.5}
\resizebox{\ifdim\width>\linewidth\linewidth\else\width\fi}{!}{
\begin{tabular}[t]{llllll}
\toprule
\multicolumn{1}{c}{ } & \multicolumn{2}{c}{Genetic architecture} & \multicolumn{2}{c}{Heritability} & \multicolumn{1}{c}{ } \\
\cmidrule(l{3pt}r{3pt}){2-3} \cmidrule(l{3pt}r{3pt}){4-5}
CR & MAF of causal SNPs & MAF and LD coupling & Causal bin & Non-causal bin & Relative bias (\%)\\
\midrule
\multirow{8}{*}{0.2} & {}[0.01, 0.05] & a = 0, b = 0 & 0.505 (0.009) & -0.000 (0.005) & 0.963\\
 & {}[0.01, 0.05] & a = 0, b = 1 & 0.503 (0.008) & 0.000 (0.004) & 0.693\\
 & {}[0.01, 0.05] & a = 0.75, b = 0 & 0.504 (0.015) & -0.001 (0.009) & 0.808\\
 & {}[0.01, 0.05] & a = 0.75, b = 1 & 0.501 (0.019) & 0.001 (0.012) & 0.217\\
 & {}[0.05, 0.5] & a = 0, b = 0 & 0.501 (0.013) & -0.001 (0.007) & 0.261\\
 & {}[0.05, 0.5] & a = 0, b = 1 & 0.487 (0.011) & 0.002 (0.005) & -2.679\\
 & {}[0.05, 0.5] & a = 0.75, b = 0 & 0.512 (0.015) & -0.009 (0.008) & 2.327\\
 & {}[0.05, 0.5] & a = 0.75, b = 1 & 0.493 (0.016) & -0.002 (0.010) & -1.384\\
\midrule[0.3pt]
\multirow{8}{*}{0.5} & {}[0.01, 0.05] & a = 0, b = 0 & 0.505 (0.012) & -0.000 (0.011) & 1.072\\
 & {}[0.01, 0.05] & a = 0, b = 1 & 0.504 (0.013) & 0.001 (0.011) & 0.791\\
 & {}[0.01, 0.05] & a = 0.75, b = 0 & 0.509 (0.035) & 0.002 (0.037) & 1.740\\
 & {}[0.01, 0.05] & a = 0.75, b = 1 & 0.506 (0.049) & 0.001 (0.048) & 1.269\\
 & {}[0.05, 0.5] & a = 0, b = 0 & 0.498 (0.017) & 0.000 (0.012) & -0.451\\
 & {}[0.05, 0.5] & a = 0, b = 1 & 0.486 (0.016) & 0.004 (0.014) & -2.798\\
 & {}[0.05, 0.5] & a = 0.75, b = 0 & 0.511 (0.023) & -0.007 (0.018) & 2.161\\
 & {}[0.05, 0.5] & a = 0.75, b = 1 & 0.494 (0.031) & 0.004 (0.034) & -1.197\\
\midrule[0.3pt]
\multirow{8}{*}{0.7} & {}[0.01, 0.05] & a = 0, b = 0 & 0.502 (0.034) & 0.002 (0.037) & 0.488\\
 & {}[0.01, 0.05] & a = 0, b = 1 & 0.498 (0.036) & 0.006 (0.041) & -0.484\\
 & {}[0.01, 0.05] & a = 0.75, b = 0 & 0.494 (0.123) & 0.005 (0.140) & -1.107\\
 & {}[0.01, 0.05] & a = 0.75, b = 1 & 0.488 (0.165) & 0.017 (0.186) & -2.387\\
 & {}[0.05, 0.5] & a = 0, b = 0 & 0.499 (0.039) & 0.002 (0.037) & -0.156\\
 & {}[0.05, 0.5] & a = 0, b = 1 & 0.483 (0.044) & 0.008 (0.050) & -3.309\\
 & {}[0.05, 0.5] & a = 0.75, b = 0 & 0.507 (0.059) & 0.000 (0.061) & 1.468\\
 & {}[0.05, 0.5] & a = 0.75, b = 1 & 0.495 (0.103) & 0.013 (0.139) & -1.034\\
\midrule[0.3pt]
\multirow{8}{*}{0.8} & {}[0.01, 0.05] & a = 0, b = 0 & 0.495 (0.086) & 0.004 (0.091) & -0.995\\
 & {}[0.01, 0.05] & a = 0, b = 1 & 0.497 (0.087) & 0.008 (0.100) & -0.588\\
 & {}[0.01, 0.05] & a = 0.75, b = 0 & 0.516 (0.338) & 0.046 (0.369) & 3.195\\
 & {}[0.01, 0.05] & a = 0.75, b = 1 & 0.505 (0.443) & 0.063 (0.484) & 0.954\\
 & {}[0.05, 0.5] & a = 0, b = 0 & 0.503 (0.081) & -0.005 (0.094) & 0.520\\
 & {}[0.05, 0.5] & a = 0, b = 1 & 0.490 (0.099) & 0.001 (0.125) & -2.054\\
 & {}[0.05, 0.5] & a = 0.75, b = 0 & 0.522 (0.128) & -0.009 (0.152) & 4.341\\
 & {}[0.05, 0.5] & a = 0.75, b = 1 & 0.523 (0.240) & -0.002 (0.329) & 4.683\\
\bottomrule
\end{tabular}
}
\end{table}

\begin{table}[H]
\centering
\caption{Comparison of different methods for total heritability estimation ($N$ = 20000, $M$ = 20000, $C$ = 10, $K$ = 5). Relative bias is computed using the formula $(\hat{h}^2_{\text{SNP}} - h^2_{\text{SNP}}) / h^2_{\text{SNP}} \times 100$. MSE: Mean Squared Error; MAE: Mean Absolute Error. Values displayed as $<0.001$ indicate positive values smaller than the respective threshold.}
\label{tab:correct_spec_h2snp_comp_supp}
\resizebox{\ifdim\width>\linewidth\linewidth\else\width\fi}{!}{
\begin{tabular}[t]{cccrrrrrrrrr}
\toprule
 &  &  & \multicolumn{3}{c}{Normal} & \multicolumn{3}{c}{Gumbel} & \multicolumn{3}{c}{Logistic} \\
\cmidrule(l{3pt}r{3pt}){4-6} \cmidrule(l{3pt}r{3pt}){7-9} \cmidrule(l{3pt}r{3pt}){10-12}
$h^2_{\text{SNP}}$ & CR & Model & Rel. Bias (\%) & MSE & MAE & Rel. Bias (\%) & MSE & MAE & Rel. Bias (\%) & MSE & MAE\\
\midrule
\multirow{6}{*}{0.2} & \multirow{2}{*}{0.2} & CVC & 0.20 & $<0.001$ & 0.014 & 1.66 & $<0.001$ & 0.017 & 0.83 & $<0.001$ & 0.012\\
 &  & LT & 57.76 & 0.014 & 0.116 & 81.61 & 0.027 & 0.163 & 61.11 & 0.015 & 0.122\\ \cmidrule(l){2-12}
 & \multirow{2}{*}{0.5} & CVC & -2.78 & 0.002 & 0.035 & 4.81 & 0.005 & 0.054 & 3.69 & 0.002 & 0.036\\
 &  & LT & -49.13 & 0.010 & 0.098 & -42.99 & 0.008 & 0.086 & -48.46 & 0.010 & 0.097\\ \cmidrule(l){2-12}
 & \multirow{2}{*}{0.8} & CVC & -11.80 & 0.121 & 0.279 & -7.60 & 0.229 & 0.366 & 29.52 & 0.165 & 0.316\\
 &  & LT & -93.38 & 0.035 & 0.187 & -91.93 & 0.034 & 0.184 & -92.47 & 0.035 & 0.185\\ \midrule[1pt]
\multirow{6}{*}{0.5} & \multirow{2}{*}{0.2} & CVC & 0.15 & $<0.001$ & 0.015 & -0.39 & $<0.001$ & 0.015 & -0.24 & $<0.001$ & 0.016\\
 &  & LT & 56.58 & 0.081 & 0.283 & 66.02 & 0.110 & 0.330 & 56.49 & 0.081 & 0.282\\ \cmidrule(l){2-12}
 & \multirow{2}{*}{0.5} & CVC & -0.80 & 0.006 & 0.061 & -1.13 & 0.007 & 0.063 & -3.15 & 0.007 & 0.065\\
 &  & LT & -66.73 & 0.112 & 0.334 & -65.73 & 0.108 & 0.329 & -67.62 & 0.115 & 0.338\\ \cmidrule(l){2-12}
 & \multirow{2}{*}{0.8} & CVC & -7.28 & 0.272 & 0.420 & -22.49 & 0.476 & 0.466 & -12.54 & 0.376 & 0.464\\
 &  & LT & -94.82 & 0.225 & 0.474 & -95.39 & 0.228 & 0.477 & -94.21 & 0.222 & 0.471\\ \midrule[1pt]
\multirow{6}{*}{0.8} & \multirow{2}{*}{0.2} & CVC & -0.15 & $<0.001$ & 0.021 & -0.12 & $<0.001$ & 0.022 & -0.06 & $<0.001$ & 0.022\\
 &  & LT & 28.56 & 0.053 & 0.228 & 29.67 & 0.057 & 0.237 & 27.82 & 0.051 & 0.223\\ \cmidrule(l){2-12}
 & \multirow{2}{*}{0.5} & CVC & 0.58 & 0.006 & 0.062 & -0.54 & 0.005 & 0.058 & -0.40 & 0.008 & 0.070\\
 &  & LT & -66.06 & 0.280 & 0.528 & -65.85 & 0.278 & 0.527 & -66.33 & 0.282 & 0.531\\ \cmidrule(l){2-12}
 & \multirow{2}{*}{0.8} & CVC & 4.25 & 0.323 & 0.425 & -14.27 & 0.363 & 0.488 & -5.70 & 0.324 & 0.439\\
 &  & LT & -95.07 & 0.579 & 0.761 & -95.18 & 0.580 & 0.761 & -95.33 & 0.582 & 0.763\\
\bottomrule
\end{tabular}}
\end{table}

\begin{landscape}\begingroup\fontsize{10}{12}\selectfont
\begin{table}[H]
\centering
\caption{Comparison of CVC and BayesW for partitioned heritability estimation ($N$ = 20000, $M$ = 20000, $C$ = 10, $K$ = 5, CR = 0.2). Bin 1-5 represent genomic partitions, and bin e represents environmental component. $h^2_k$ values show the actual partitioned heritability used in simulation. Relative bias is computed using the formula $(\hat{h}_k^2 - h^2_k) / h^2_k \times 100$. MSE: Mean Squared Error; MAE: Mean Absolute Error. Values displayed as $<0.0001$ indicate positive values smaller than the respective threshold.}
\label{tab:correct_spec_h2k_comp}
\resizebox{\ifdim\width>\linewidth\linewidth\else\width\fi}{!}{
\begin{tabular}[t]{ccrrrrrrrrrrrrrrrrrrr}
\toprule
\multicolumn{3}{c}{ } & \multicolumn{6}{c}{Normal} & \multicolumn{6}{c}{Gumbel} & \multicolumn{6}{c}{Logistic} \\
\cmidrule(l{3pt}r{3pt}){4-9} \cmidrule(l{3pt}r{3pt}){10-15} \cmidrule(l{3pt}r{3pt}){16-21}
\multicolumn{3}{c}{ } & \multicolumn{2}{c}{Rel. bias (\%)} & \multicolumn{2}{c}{MSE} & \multicolumn{2}{c}{MAE} & \multicolumn{2}{c}{Rel. bias (\%)} & \multicolumn{2}{c}{MSE} & \multicolumn{2}{c}{MAE} & \multicolumn{2}{c}{Rel. bias (\%)} & \multicolumn{2}{c}{MSE} & \multicolumn{2}{c}{MAE} \\
\cmidrule(l{3pt}r{3pt}){4-5} \cmidrule(l{3pt}r{3pt}){6-7} \cmidrule(l{3pt}r{3pt}){8-9} \cmidrule(l{3pt}r{3pt}){10-11} \cmidrule(l{3pt}r{3pt}){12-13} \cmidrule(l{3pt}r{3pt}){14-15} \cmidrule(l{3pt}r{3pt}){16-17} \cmidrule(l{3pt}r{3pt}){18-19} \cmidrule(l{3pt}r{3pt}){20-21}
$h^2_{\text{SNP}}$ & Bin & $h_k^2$ & CVC & BayesW & CVC & BayesW & CVC & BayesW & CVC & BayesW & CVC & BayesW & CVC & BayesW & CVC & BayesW & CVC & BayesW & CVC & BayesW\\
\midrule
\multirow{6}{*}{0.2} & 1 & 0.003 & 32.64 & 973.80 & $<0.0001$ & 0.0008 & 0.005 & 0.027 & 48.87 & 1880.41 & $<0.0001$ & 0.0028 & 0.007 & 0.052 & 8.11 & 995.26 & $<0.0001$ & 0.0008 & 0.006 & 0.028\\
 & 2 & 0.026 & 3.67 & 99.41 & $<0.0001$ & 0.0007 & 0.005 & 0.026 & 1.97 & 185.06 & $<0.0001$ & 0.0024 & 0.006 & 0.048 & 4.88 & 106.40 & $<0.0001$ & 0.0008 & 0.005 & 0.028\\
 & 3 & 0.066 & -1.52 & 30.11 & $<0.0001$ & 0.0004 & 0.006 & 0.020 & 1.52 & 63.94 & 0.0001 & 0.0018 & 0.008 & 0.042 & -1.65 & 34.78 & $<0.0001$ & 0.0006 & 0.007 & 0.023\\
 & 4 & 0.044 & 0.17 & 53.62 & $<0.0001$ & 0.0006 & 0.006 & 0.024 & -0.74 & 104.94 & $<0.0001$ & 0.0022 & 0.008 & 0.046 & 0.65 & 59.59 & $<0.0001$ & 0.0007 & 0.006 & 0.026\\
 & 5 & 0.061 & -0.90 & 36.35 & $<0.0001$ & 0.0005 & 0.006 & 0.022 & 1.27 & 71.77 & 0.0001 & 0.0019 & 0.008 & 0.043 & 1.61 & 41.42 & $<0.0001$ & 0.0007 & 0.006 & 0.025\\
 & e & 0.800 & -0.05 & -14.85 & 0.0003 & 0.0142 & 0.014 & 0.119 & -0.42 & -29.13 & 0.0005 & 0.0544 & 0.017 & 0.233 & -0.21 & -16.26 & 0.0002 & 0.0170 & 0.012 & 0.130\\ \midrule[1pt]
\multirow{6}{*}{0.5} & 1 & 0.105 & 0.88 & 6.23 & $<0.0001$ & $<0.0001$ & 0.007 & 0.007 & -0.41 & 17.51 & $<0.0001$ & 0.0004 & 0.008 & 0.018 & -1.00 & 6.35 & $<0.0001$ & $<0.0001$ & 0.007 & 0.007\\
 & 2 & 0.141 & 0.78 & 1.09 & $<0.0001$ & $<0.0001$ & 0.008 & 0.005 & -0.27 & 9.42 & $<0.0001$ & 0.0002 & 0.008 & 0.013 & 0.73 & 2.37 & $<0.0001$ & $<0.0001$ & 0.008 & 0.006\\
 & 3 & 0.056 & -0.41 & 22.42 & $<0.0001$ & 0.0002 & 0.006 & 0.013 & 0.38 & 43.84 & $<0.0001$ & 0.0006 & 0.007 & 0.025 & -1.17 & 20.79 & $<0.0001$ & 0.0002 & 0.006 & 0.012\\
 & 4 & 0.069 & -0.21 & 15.73 & $<0.0001$ & 0.0001 & 0.007 & 0.011 & -1.90 & 31.54 & $<0.0001$ & 0.0005 & 0.007 & 0.022 & 0.04 & 15.84 & $<0.0001$ & 0.0001 & 0.007 & 0.011\\
 & 5 & 0.129 & -0.70 & 1.72 & 0.0001 & $<0.0001$ & 0.008 & 0.005 & -0.03 & 11.35 & 0.0001 & 0.0003 & 0.008 & 0.015 & -0.44 & 3.37 & 0.0001 & $<0.0001$ & 0.009 & 0.006\\
 & e & 0.500 & -0.15 & -6.75 & 0.0004 & 0.0012 & 0.015 & 0.034 & 0.39 & -18.54 & 0.0004 & 0.0087 & 0.015 & 0.093 & 0.24 & -7.40 & 0.0004 & 0.0014 & 0.016 & 0.037\\ \midrule[1pt]
\multirow{6}{*}{0.8} & 1 & 0.359 & 0.08 & -3.42 & 0.0003 & 0.0002 & 0.014 & 0.013 & 0.23 & -3.18 & 0.0003 & 0.0002 & 0.014 & 0.012 & 0.54 & -3.15 & 0.0003 & 0.0002 & 0.013 & 0.012\\
 & 2 & 0.258 & -0.11 & -3.74 & 0.0002 & 0.0001 & 0.010 & 0.010 & -1.19 & -2.75 & 0.0002 & $<0.0001$ & 0.010 & 0.008 & -0.26 & -3.88 & 0.0002 & 0.0001 & 0.011 & 0.010\\
 & 3 & 0.035 & -1.06 & 16.23 & $<0.0001$ & $<0.0001$ & 0.007 & 0.006 & 3.00 & 33.38 & 0.0001 & 0.0001 & 0.008 & 0.012 & -0.44 & 17.37 & $<0.0001$ & $<0.0001$ & 0.008 & 0.006\\
 & 4 & 0.108 & -1.21 & -0.57 & 0.0001 & $<0.0001$ & 0.009 & 0.004 & 0.11 & 4.40 & 0.0002 & $<0.0001$ & 0.010 & 0.005 & -0.36 & 0.27 & 0.0001 & $<0.0001$ & 0.009 & 0.004\\
 & 5 & 0.040 & 1.17 & 12.72 & $<0.0001$ & $<0.0001$ & 0.008 & 0.005 & 0.33 & 25.59 & 0.0001 & 0.0001 & 0.008 & 0.010 & -3.01 & 12.40 & $<0.0001$ & $<0.0001$ & 0.008 & 0.005\\
 & e & 0.200 & 0.61 & 5.89 & 0.0007 & 0.0002 & 0.021 & 0.012 & 0.47 & -4.08 & 0.0007 & $<0.0001$ & 0.022 & 0.008 & 0.25 & 4.99 & 0.0007 & 0.0001 & 0.022 & 0.010\\
\bottomrule
\end{tabular}}
\end{table}
\endgroup{}
\end{landscape}

\begin{table}[H]
    \centering
        \caption{Running times for various simulation settings ($h^2_{\text{SNP}} = 0.2$, $\Delta = 0.2$, $C = 10$, $J = 100$ and $B = 10$). Experiments were run on a compute node with up to 16 CPU cores, 160 GB RAM and 1 TB of temporary disk space for memory-mapped working arrays. The memory and disk space indicate peak RAM and temporary disk space usage observed while fitting the CVC model.}
    \label{tab:cvc_computational_benchmark}
    \resizebox{\textwidth}{!}{%
    \begin{tabular}{@{}r r r r r r r@{}}
      \toprule
      \multicolumn{3}{c}{\textbf{Parameters}} & 
      \textbf{Running time (h)} & \textbf{CPU cores} & \textbf{Memory (GB)} & \textbf{Disk space (GB)} \\
      \cmidrule(lr){1-3}
      $N$ & $M$ & $K$ & & & & \\
      \midrule
      50,000   & 50,000    & 10  & 0.11 & 2 & 1.6 & 3.8 \\
      50,000   & 100,000   & 10  & 0.14 & 2 & 1.7 & 3.8 \\
      100,000   & 100,000  & 10  & 0.19 & 2 & 2.7 & 7.5 \\
      100,000   & 100,000  & 25  & 0.25 & 2 & 3.6 & 19.0 \\
      100,000   & 100,000  & 50  & 0.66 & 2 & 5.2 & 38.0 \\
      250,000 & 500,000    & 50  & 1.53 & 8 & 11.3 & 94.0 \\
      1,000,000 & 1,000,000 & 100 & 8.39 & 16 & 86.7 & 746.0 \\
      \bottomrule
    \end{tabular}
    }
\end{table}

\begin{table}[H]
\centering
\caption{Impact of violation of marginal independence between 
$y_i$ and $r_i$ at varying degrees modulated by the variable 
$\xi$ ($N = 10,000$, $M = 10,000$, $C = 10$, $K = 10$, $J = 100$, 
$B = 10$, $h^2_{\text{SNP}} = 0.5$).}
\label{tab:conditional_independence}
\begin{tabular}[t]{ccccrrrr}
\toprule
CR & $\xi$ & $\rho$ & $\hat{h}_{\text{SNP}}^2$ & SE & JSE & SE of JSE & Rel. Bias of $\hat{h}_{\text{SNP}}^2$ (\%)\\
\midrule
 & -1.0 & -0.17 & 0.48 & 0.022 & 0.028 & 0.002 & -4.76\\

 & -0.5 & -0.09 & 0.49 & 0.022 & 0.028 & 0.002 & -2.99\\

 & 0.0 & 0.00 & 0.50 & 0.020 & 0.028 & 0.002 & 0.11\\

 & 0.5 & 0.09 & 0.52 & 0.025 & 0.029 & 0.002 & 3.97\\

\multirow{-5}{*}{\centering\arraybackslash 0.2} & 1.0 & 0.17 & 0.56 & 0.025 & 0.030 & 0.002 & 11.13\\
\cmidrule{1-8}
 & -1.0 & -0.17 & 0.45 & 0.055 & 0.057 & 0.006 & -9.42\\

 & -0.5 & -0.09 & 0.45 & 0.064 & 0.059 & 0.006 & -10.69\\

 & 0.0 & 0.00 & 0.50 & 0.072 & 0.069 & 0.010 & -0.56\\

 & 0.5 & 0.09 & 0.64 & 0.087 & 0.088 & 0.012 & 27.59\\

\multirow{-5}{*}{\centering\arraybackslash 0.5} & 1.0 & 0.17 & 1.07 & 0.145 & 0.135 & 0.018 & 114.43\\
\cmidrule{1-8}
 & -1.0 & -0.17 & 0.87 & 0.566 & 0.475 & 0.090 & 73.48\\

 & -0.5 & -0.09 & 0.52 & 0.302 & 0.331 & 0.077 & 3.93\\

 & 0.0 & 0.00 & 0.54 & 0.431 & 0.408 & 0.090 & 7.11\\

 & 0.5 & 0.09 & 0.74 & 18.070 & 5.119 & 9.108 & 47.52\\

\multirow{-5}{*}{\centering\arraybackslash 0.8} & 1.0 & 0.17 & -0.39 & 0.321 & 0.308 & 0.075 & -177.58\\
\bottomrule
\end{tabular}
\end{table}

\begin{table}[H]
\caption{Comparison of heritability estimates across different partitions ($K$) and models ($B = 10$). The number of jackknife blocks $J$ was set to 29 for $K = 96$, and to 100 for other $K$'s. The code indicates the UKB Data ID or Phecode. The models were  adjusted for sex, birth year, assessment center and 20 genetic PCs. The sex covariate was excluded for female-only traits. CR: censoring rate; JSE: jackknife standard error.}
\label{tab:ukb_age_at_dx_finer_bins}
\centering
\resizebox{\ifdim\width>\linewidth\linewidth\else\width\fi}{!}{
\begin{tabular}[t]{llrrrcrr}
\toprule
\textbf{Code} & \textbf{Description} & \textbf{N} & \textbf{M} & \textbf{K} & \textbf{Model} & $\mathbf{h^2_{\text{SNP}}}$ & \textbf{JSE}\\
\midrule
\multirow{6}{*}{2217} & \multirow{6}{*}{Started wearing glasses or contact lenses} & \multirow{6}{*}{276,169} & \multirow{6}{*}{592,454} & 24 & \multirow{3}{*}{CVC} & 0.101 & 0.004\\
 &  &   &   & 48 &  & 0.098 & 0.005\\
 &  &   &   & 96 &  & 0.098 & 0.005\\ \cmidrule(l){5-8}
 &  &   &   & 24 & \multirow{3}{*}{LT} & 0.249 & 0.008\\
 &  &   &   & 48 &  & 0.246 & 0.009\\
 &  &   &   & 96 &  & 0.245 & 0.011\\ \midrule[1pt]
\multirow{6}{*}{2754} & \multirow{6}{*}{First live birth} & \multirow{6}{*}{147,224} & \multirow{6}{*}{592,454} & 24 & \multirow{3}{*}{CVC} & 0.192 & 0.057\\
 &  &   &   & 48 &  & 0.199 & 0.046\\
 &  &   &   & 96 &  & 0.199 & 0.067\\ \cmidrule(l){5-8}
 &  &   &   & 24 & \multirow{3}{*}{LT} & 0.143 & 0.008\\
 &  &   &   & 48 &  & 0.143 & 0.007\\
 &  &   &   & 96 &  & 0.141 & 0.009\\ \midrule[1pt]
\multirow{6}{*}{3436.2867} & \multirow{6}{*}{Started smoking (current and former smokers)} & \multirow{6}{*}{276,169} & \multirow{6}{*}{592,454} & 24 & \multirow{3}{*}{CVC} & 0.307 & 0.050\\
 &  &   &   & 48 &  & 0.309 & 0.050\\
 &  &   &   & 96 &  & 0.300 & 0.051\\ \cmidrule(l){5-8}
 &  &   &   & 24 & \multirow{3}{*}{LT} & 0.225 & 0.007\\
 &  &   &   & 48 &  & 0.226 & 0.007\\
 &  &   &   & 96 &  & 0.226 & 0.009\\ \midrule[1pt]
\multirow{6}{*}{401} & \multirow{6}{*}{Hypertension} & \multirow{6}{*}{239,151} & \multirow{6}{*}{592,454} & 24 & \multirow{3}{*}{CVC} & -0.018 & 0.015\\
 &  &   &   & 48 &  & -0.020 & 0.016\\
 &  &   &   & 96 &  & -0.022 & 0.015\\ \cmidrule(l){5-8}
 &  &   &   & 24 & \multirow{3}{*}{LT} & -0.008 & 0.004\\
 &  &   &   & 48 &  & -0.010 & 0.004\\
 &  &   &   & 96 &  & -0.012 & 0.005\\
\bottomrule
\end{tabular}}
\end{table}

\newpage
\bibliographystyle{myapalike}
\bibliography{references}